\documentclass[reprint,superscriptaddress,twocolumn,floatfix,secnumarabic,amssymb, nobibnotes, aps, prc,floatfix]{revtex4-1}

\usepackage{graphicx}
\usepackage{epstopdf}
\usepackage{color}
\usepackage{longtable}
\usepackage{dcolumn}
\usepackage{amsmath}
\usepackage{hyperref}

\begin{document}
\title{Single and double spin asymmetries for deeply virtual Compton scattering measured with CLAS and a longitudinally polarized proton target}%
\newcommand*{\INFNFR}{INFN, Laboratori Nazionali di Frascati, 00044 Frascati, Italy}
\newcommand*{\INFNFRindex}{4}
\affiliation{\INFNFR}
\newcommand*{\ORSAY}{Institut de Physique Nucl\'eaire Orsay, 91406 Orsay, France}
\newcommand*{\ORSAYindex}{5}
\affiliation{\ORSAY}
\newcommand*{\FAIRFIELD}{Fairfield University, Fairfield, Connecticut 06824}
\newcommand*{\FAIRFIELDindex}{3}
\affiliation{\FAIRFIELD}
\newcommand*{\UCONN}{University of Connecticut, Storrs, Connecticut 06269}
\newcommand*{\UCONNindex}{1}
\affiliation{\UCONN}
\newcommand*{\SACLAY}{CEA, Centre de Saclay, Irfu/Service de Physique Nucl\'eaire, 91191 Gif-sur-Yvette, France}
\newcommand*{\SACLAYindex}{2}
\affiliation{\SACLAY}
\newcommand*{\ODU}{Old Dominion University, Norfolk, Virginia 23529}
\newcommand*{\ODUindex}{8}
\affiliation{\ODU}
\newcommand*{\GLASGOW}{University of Glasgow, Glasgow G12 8QQ, United Kingdom}
\newcommand*{\GLASGOWindex}{6}
\affiliation{\GLASGOW}
\newcommand*{\JLAB}{Thomas Jefferson National Accelerator Facility, Newport News, Virginia 23606}
\newcommand*{\JLABindex}{9}
\affiliation{\JLAB}
\newcommand*{\INFNGE}{INFN, Sezione di Genova, 16146 Genova, Italy}
\newcommand*{\INFNGEindex}{10}
\affiliation{\INFNGE}
\newcommand*{\ITEP}{Institute of Theoretical and Experimental Physics, Moscow, 117259, Russia}
\newcommand*{\ITEPindex}{11}
\affiliation{\ITEP}
\newcommand*{\WM}{College of William and Mary, Williamsburg, Virginia 23187-8795}
\newcommand*{\WMindex}{13}
\affiliation{\WM}
\newcommand*{\GWUI}{The George Washington University, Washington, D.C. 20052}
\newcommand*{\GWUIindex}{14}
\affiliation{\GWUI}
\newcommand*{\UTFSM}{Universidad T\'{e}cnica Federico Santa Mar\'{i}a, Casilla 110-V Valpara\'{i}so, Chile}
\newcommand*{\UTFSMindex}{15}
\affiliation{\UTFSM}
\newcommand*{\OHIOU}{Ohio University, Athens, Ohio  45701}
\newcommand*{\OHIOUindex}{16}
\affiliation{\OHIOU}
\newcommand*{\INFNRO}{INFN, Sezione di Roma Tor Vergata, 00133 Roma, Italy}
\newcommand*{\INFNROindex}{17}
\affiliation{\INFNRO}
\newcommand*{\ROMAII}{Universit\`a di Roma Tor Vergata, 00133 Roma, Italy}
\newcommand*{\ROMAIIindex}{22}
\affiliation{\ROMAII}
\newcommand*{\ISU}{Idaho State University, Pocatello, Idaho 83209}
\newcommand*{\ISUindex}{18}
\affiliation{\ISU}
\newcommand*{\INFNFE}{INFN, Sezione di Ferrara, 44100 Ferrara, Italy}
\newcommand*{\INFNFEindex}{19}
\affiliation{\INFNFE}
\newcommand*{\VIRGINIA}{University of Virginia, Charlottesville, Virginia 22901}
\newcommand*{\VIRGINIAindex}{20}
\affiliation{\VIRGINIA}
\newcommand*{\FSU}{Florida State University, Tallahassee, Florida 32306}
\newcommand*{\FSUindex}{21}
\affiliation{\FSU}
\newcommand*{\SCAROLINA}{University of South Carolina, Columbia, South Carolina 29208}
\newcommand*{\SCAROLINAindex}{24}
\affiliation{\SCAROLINA}
\newcommand*{\ANL}{Argonne National Laboratory, Argonne, Illinois 60439}
\newcommand*{\ANLindex}{26}
\affiliation{\ANL}
\newcommand*{\LPSC}{LPSC, Universit\'{e} Grenoble-Alps, CNRS/IN2P3, Grenoble, France}
\newcommand*{\LPSCindex}{40}
\affiliation{\LPSC}
\newcommand*{\MSU}{Skobeltsyn Institute of Nuclear Physics, Lomonosov Moscow State University, 119234 Moscow, Russia}
\newcommand*{\MSUindex}{27}
\affiliation{\MSU}
\newcommand*{\INFNTUR}{INFN, Sezione di Torino, 10125 Torino, Italy}
\newcommand*{\INFNTURindex}{28}
\affiliation{\INFNTUR}
\newcommand*{\EDINBURGH}{Edinburgh University, Edinburgh EH9 3JZ, United Kingdom}
\newcommand*{\EDINBURGHindex}{7}
\affiliation{\EDINBURGH}
\newcommand*{\YEREVAN}{Yerevan Physics Institute, 375036 Yerevan, Armenia}
\newcommand*{\YEREVANindex}{23}
\affiliation{\YEREVAN}
\newcommand*{\URICH}{University of Richmond, Richmond, Virginia 23173}
\newcommand*{\URICHindex}{27}
\affiliation{\URICH}
\newcommand*{\JMU}{James Madison University, Harrisonburg, Virginia 22807}
\newcommand*{\JMUindex}{29}
\affiliation{\JMU}
\newcommand*{\FIU}{Florida International University, Miami, Florida 33199}
\newcommand*{\FIUindex}{12}
\affiliation{\FIU}
\newcommand*{\UNH}{University of New Hampshire, Durham, New Hampshire 03824-3568}
\newcommand*{\UNHindex}{31}
\affiliation{\UNH}
\newcommand*{\VT}{Virginia Tech, Blacksburg, Virginia   24061-0435}
\newcommand*{\VTindex}{36}
\affiliation{\VT}
\newcommand*{\LOSLM}{Los Alamos National Laboratory, Los Alamos, NM 87545}
\newcommand*{\LOSLMindex}{42}
\affiliation{\LOSLM}
\newcommand*{\TEMPLE}{Temple University,  Philadelphia, Pennsylvania 19122 }
\newcommand*{\TEMPLEindex}{32}
\affiliation{\TEMPLE}
\newcommand*{\KNU}{Kyungpook National University, Daegu 702-701, Republic of Korea}
\newcommand*{\KNUindex}{34}
\affiliation{\KNU}
\newcommand*{\CUA}{Catholic University of America, Washington, D.C. 20064}
\newcommand*{\CUAindex}{35}
\affiliation{\CUA}
\newcommand*{\CMU}{Carnegie Mellon University, Pittsburgh, Pennsylvania 15213}
\newcommand*{\CMUindex}{38}
\affiliation{\CMU}
\newcommand*{\CSUDH}{California State University, Dominguez Hills, Carson, California 90747}
\newcommand*{\CSUDHindex}{37}
\affiliation{\CSUDH}
\newcommand*{\NSU}{Norfolk State University, Norfolk, Virginia 23504}
\newcommand*{\NSUindex}{33}
\affiliation{\NSU}
\newcommand*{\RPI}{Rensselaer Polytechnic Institute, Troy, New York 12180-3590}
\newcommand*{\RPIindex}{30}
\affiliation{\RPI}
\newcommand*{\CANISIUS}{Canisius College, Buffalo, New York 14208}
\newcommand*{\CANISIUSindex}{41}
\affiliation{\CANISIUS}

\newcommand*{\NOWUK}{University of Kentucky, Lexington, Kentucky 40506}
\newcommand*{\NOWJLAB}{Thomas Jefferson National Accelerator Facility, Newport News, Virginia 23606}
\newcommand*{\NOWUCONN}{University of Connecticut, Storrs, Connecticut 06269}
\newcommand*{\NOWODU}{Old Dominion University, Norfolk, Virginia 23529}
\newcommand*{\NOWGLASGOW}{University of Glasgow, Glasgow G12 8QQ, United Kingdom}
\newcommand*{\NOWINFNGE}{INFN, Sezione di Genova, 16146 Genova, Italy}
\newcommand*{\NOWUMASS}{University of Massachusetts, Amherst, Massachusetts  01003}
\newcommand*{\NOWMISSI}{Mississippi State University, Mississippi State, MS 39762}


\author{S. Pisano}
\email[contact author: ]{pisanos@jlab.org}
\affiliation{\INFNFR}
\affiliation{\ORSAY}
\author{A. Biselli}
\affiliation{\FAIRFIELD}
\author{S. Niccolai}
\affiliation{\ORSAY}
\author{E. Seder}
\affiliation{\UCONN}
\affiliation{\SACLAY}
\author{M. Guidal}
\affiliation{\ORSAY}
\author{M. Mirazita}
\affiliation{\INFNFR}

\author {K.P. ~Adhikari} 
\affiliation{\ODU}
\author {D.~Adikaram} 
\affiliation{\ODU}
\author {M.J.~Amaryan} 
\affiliation{\ODU}
\author {M.D.~Anderson} 
\affiliation{\GLASGOW}
% ...

\collaboration{The CLAS Collaboration}
     \noaffiliation
\date{\today}%

\begin{abstract}
\noindent Single-beam, single-target, and double-spin asymmetries for hard exclusive photon production on the proton $\vec{e}\vec{p} \to e' p'\gamma$  are presented. The data were taken at Jefferson Lab using the CLAS detector and a longitudinally polarized ${}^{14}$NH$_3$ target. 
The three asymmetries were measured in 165 4-dimensional kinematic bins, covering the widest kinematic range ever explored simultaneously for beam and target-polarization observables in the valence quark region. 
The kinematic dependences of the obtained asymmetries are discussed and compared to the predictions of models of Generalized Parton Distributions. The measurement of three DVCS spin observables at the same kinematic points allows a quasi-model-independent extraction of the imaginary parts of the $H$ and $\tilde{H}$ Compton Form Factors, which give insight into the electric and axial charge distributions of valence quarks in the proton. 
\end{abstract}

\maketitle
\section{Generalized Parton Distributions and deeply virtual Compton scattering}\label{sec_intro}
It is well known that the fundamental particles that form hadronic matter are the quarks and the gluons, whose interactions are described by the Lagrangian of Quantum Chromo-Dynamics (QCD). However, exact QCD-based calculations cannot yet be performed to explain the properties of hadrons in terms of their constituents. One has to resort to phenomenological functions to interpret experimental measurements in order to understand how QCD works at the ``long distances'' at play when partons are confined in nucleons. Typical examples of such functions include form factors and parton distributions. 
Generalized Parton Distributions (GPDs), which unify and extend the information carried by both form factors and parton distributions, are nowadays the object of an intense effort of research,
for a more complete understanding of the structure of the nucleon. 
The GPDs describe the correlations between the longitudinal momentum and the transverse position of the partons inside the nucleon, they give access to the contribution of the orbital momentum of the quarks to the nucleon spin, and they are sensitive to the correlated $q\bar{q}$ components. The original articles, general reviews on GPDs and details on the formalism can be found in Refs.~\cite{muller,ji,rady,collins,goeke,revdiehl,revrady}.

\begin{figure}
\begin{center}
\hbox{\hspace{-4.5ex}\includegraphics[scale=0.6]{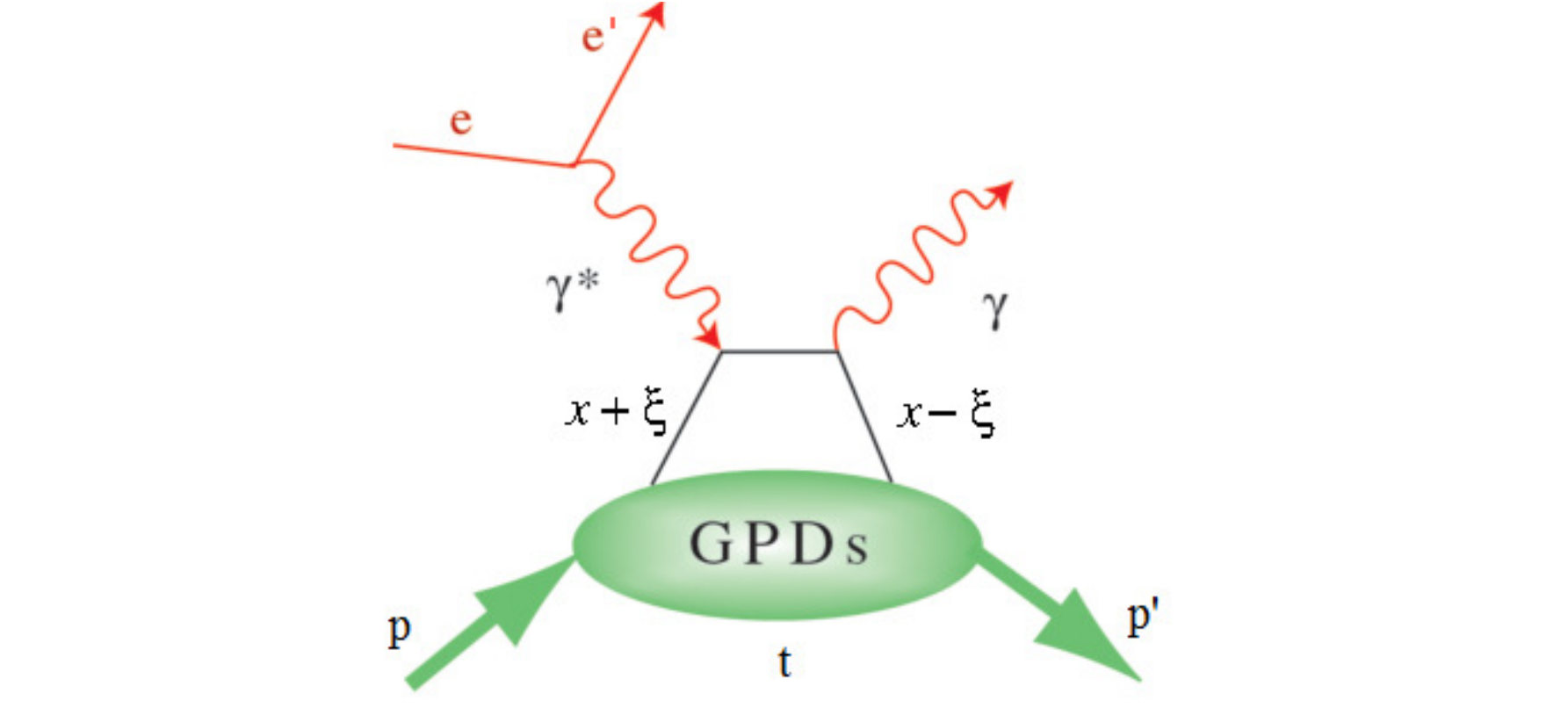}}
\caption{(Color online) The handbag diagram for the DVCS process on the proton $ep\to e'p'\gamma'$.}
\label{fig:dvcs}
\end{center}
\end{figure}

The GPDs are universal nucleon-structure functions, which can be accessed experimentally via the exclusive leptoproduction of a photon (DVCS, deeply virtual Compton scattering) or of a meson from the nucleon at high momentum transfer. More precisely, the virtuality $Q^2$ of the photon exchanged with the nucleon by the initial lepton, defined as 
\begin{equation}
Q^2=-(k-k')^2, 
\end{equation}
where $k$ and $k'$ are the four momenta of, respectively, the incoming ($e$) and scattered electron ($e'$), must be sufficiently large for the reaction to happen at the quark level. Figure~\ref{fig:dvcs} illustrates the leading-twist \cite{jaffe} process for DVCS, also called the ``handbag mechanism'', on a proton target. The virtual photon interacts with one of the quarks of the proton, which propagates radiating a real photon. 
\begin{equation}
t=(p-p')^2
\end{equation}
is the squared four-momentum transfer between the initial ($p$) and final proton ($p'$). 
 $x+\xi$ and $x-\xi$ are the longitudinal momentum fractions of the quark before and after radiating the real photon, respectively. $\xi$ is defined as \cite{belitski}: 
\begin{equation}\label{xi_def}
\xi = x_B\frac{1+\frac{t}{Q^2}}{2-x_B+x_B\frac{t}{Q^2}},
\end{equation}
which at leading twist ($-t<<Q^2$) becomes
\begin{equation}\label{xi_def2}
\xi\simeq\frac{x_B}{2-x_B}, 
\end{equation}
where $x_B$ is the Bjorken scaling variable
\begin{equation}
x_B = \frac{Q^2}{2M\nu},
\end{equation}
$M$ is the proton mass, and
\begin{equation}
\nu = E_e-E_{e'}.
\end{equation}
%
%
In the Bjorken limit, defined by high $Q^2$, high $\nu$ and fixed $x_B$, the DVCS process can be factorized into a hard-scattering part ($\gamma^* q \to \gamma q'$) that can be treated perturbatively, and a soft nucleon-structure part, described by the GPDs.
At leading-order QCD and at leading twist, considering only quark-helicity conserving quantities and the quark sector, the DVCS process is described by four GPDs, $H, \tilde H, E, \tilde E$, which account for the possible combinations of relative orientations of nucleon spin and quark helicity between the initial and final state. 

The GPDs depend upon the variables $x$, $\xi$ and $t$. 
The Fourier transform, at $\xi=0$, of the $t$ dependence of a GPD provides the spatial distribution in the transverse plane for partons having a longitudinal momentum fraction $x$ \cite{burkhardt}.

Model-independent sum rules link the first moment in $x$ of the GPDs to the elastic form factors (FFs) \cite{ji}: 
\begin{eqnarray}
   \int_{-1}^{1}dx H(x,\xi,t)=F_{1}(t) & ; & \int_{-1}^{1}dx  E(x,\xi,t)=F_{2}(t) \nonumber \\
   \int_{-1}^{1}dx \widetilde{H}(x,\xi,t)=G_{A}(t) & ; & 	\int_{-1}^{1}dx  \widetilde{E}(x,\xi,t)=G_{P}(t), 
   \label{EqGPDsFFlink}
\end{eqnarray}
where $F_{1}(t)$ and $F_{2}(t)$ are the Dirac and Pauli FFs, and $G_{A}(t)$ and $G_{P}(t)$ are the axial and pseudoscalar FFs. 

Among the three variables $x$, $\xi$ and $t$, which appear in the DVCS formalism, only two, $\xi$ and $t$, are experimentally accessible, since $x$ appears only in the quark loop and is integrated over. Indeed, the DVCS amplitude is proportional to sums of integrals over $x$ of the form: 
\begin{equation}\label{dvcs-ampl}
\int_{-1}^{1} d x F(\mp x,\xi,t)\left[\frac{1}{x - \xi + i \epsilon}\pm\frac{1}{x + \xi - i \epsilon}\right]
\end{equation}
where $F$ represents a generic GPD and the top and bottom signs apply, respectively, to the quark-helicity independent, or unpolarized, GPDs ($H, E$) and to the quark-helicity dependent, or polarized, GPDs ($\widetilde {H}, \widetilde {E}$). Each of these 4 integrals, which are called Compton Form Factors (CFFs), can be decomposed into their real and imaginary parts, as
\begin{eqnarray}\label{def_cffs}
\Re{\rm e}{\cal F} &=& {\cal P}\int_{-1}^{1}dx\left[\frac{1}{x-\xi}\mp\frac{1}{x+\xi}\right]F(x,\xi,t) \\
\Im{\rm m}{\cal F}(\xi,t)&=& -\pi [F(\xi,\xi,t)\mp F(-\xi,\xi,t)], 
\end{eqnarray}
where ${\cal P}$ is Cauchy's principal value integral. The information that can be extracted from the experimental data at a given ($\xi,t$) point depends on the observable involved. When measuring observables sensitive to the real part of the DVCS amplitude, such as double-spin asymmetries, beam-charge asymmetries or unpolarized cross sections, the real part of the CFF, $\Re{\rm e}{\cal F}$, is accessed. When measuring observables sensitive to the imaginary part of the DVCS amplitude, such as single-spin asymmetries or cross-section differences, the imaginary part of the CFF, $\Im{\rm m}{\cal F}$, can be obtained. 
However, knowing the CFFs 
does not define the GPDs uniquely. A model input is necessary to deconvolute their $x$ dependence.

The DVCS process is accompanied by the Bethe-Heitler 
(BH) process, in which the final-state photon is radiated by the incoming 
or scattered electron and not by the nucleon itself. The BH process, 
which is not sensitive to the GPDs, is experimentally indistinguishable from 
DVCS and interferes with it. However, considering that the nucleon form factors are well known at small $t$, the BH process is precisely calculable.

It is clearly a non-trivial task to measure the GPDs. It calls for a long-term experimental program comprising the measurement of different observables. 
%
Such a dedicated experimental program, concentrating on a proton target, has started worldwide in the past few years. Jefferson Lab (JLab) has provided the first measurement, in the valence region, of beam-polarized and unpolarized DVCS cross sections at Hall A~\cite{carlos}, providing a $Q^2$-scaling test that supports the validity of the leading-order, leading-twist handbag mechanism starting at values of $Q^2$ of 1-2 (GeV/$c$)$^2$.
Hall B provided pioneering measurements of beam~\cite{stepan} and target~\cite{shifeng} spin asymmetries with the CLAS detector~\cite{clas}, and afterwards obtained beam-spin asymmetries (BSA) over a large kinematic range and with high statistics~\cite{fx}. Beam-charge asymmetries, longitudinally and transversely polarized target-spin asymmetries (TSAs), as well as double-spin asymmetries (DSAs), have also been measured by the HERMES Collaboration~\cite{hermes}.

This paper is focused on the extraction of longitudinal TSAs, BSAs and DSAs for proton DVCS over a wide phase space using dedicated data taken at Jefferson Lab with the CLAS detector.

\section{DVCS on a longitudinally polarized proton target}
\begin{figure}
\begin{center}
\includegraphics[width=86mm]{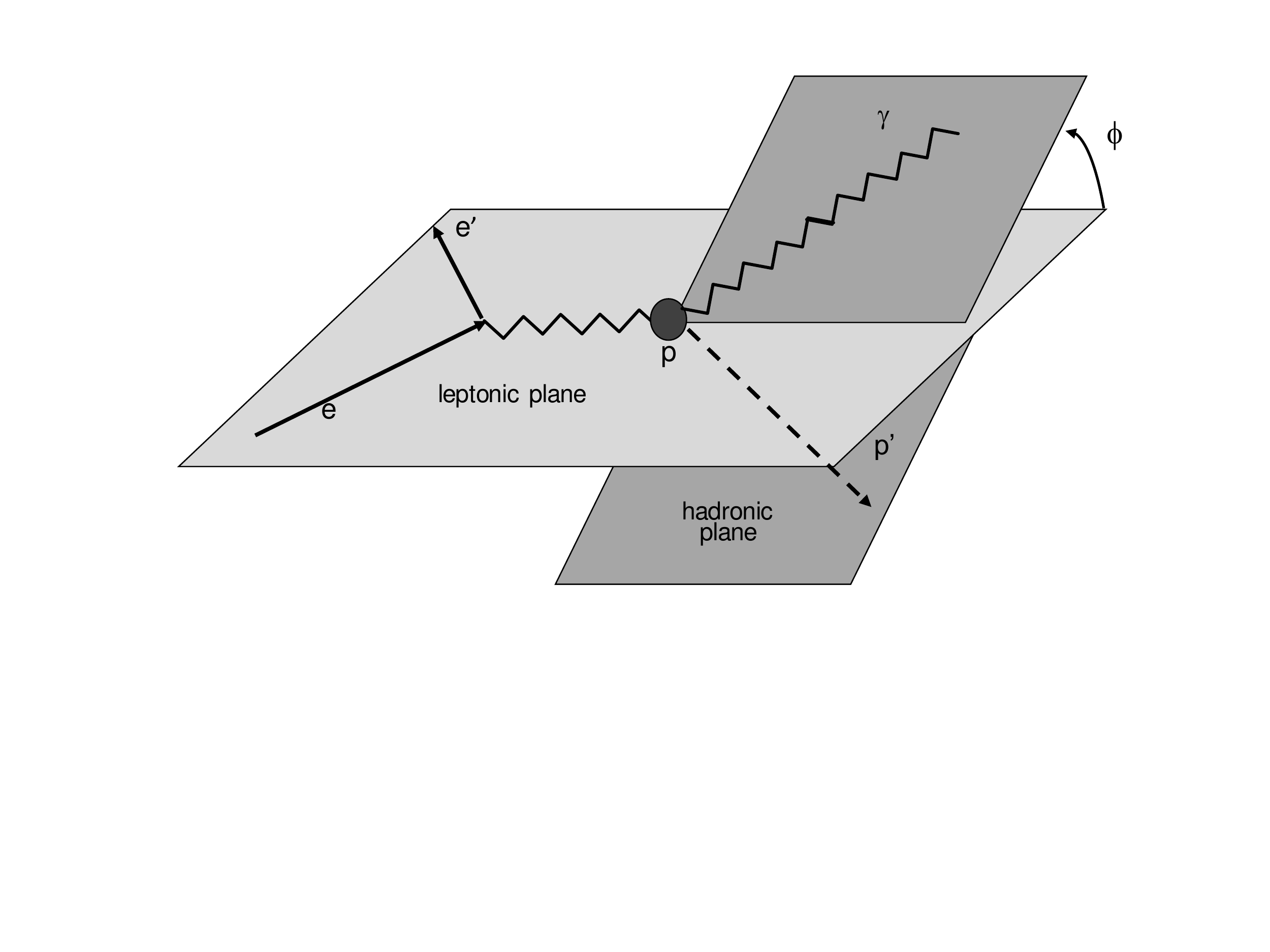}
\vspace{-2.5cm}
\caption{Scheme to illustrate the definition of the angle $\phi$, formed by the leptonic and hadronic planes.}
\label{fig:dvcs_phi}
\end{center}
\end{figure}

An analysis of DVCS observables, including the asymmetries of interest in this work, in terms of Fourier moments with respect to the azimuthal angle, was carried out by Belitsky {\it et al.} \cite{belitski}, up to a twist-3 approximation. 
These asymmetries allow one to extract separate components of the angular dependence of the $ep\to e'p'\gamma$ cross section, that are related to the Compton Form Factors, defined in Eq.~\eqref{def_cffs}. The five-fold cross section for exclusive real photon electroproduction \cite{belitski}
\begin{equation}
\label{WQ}
\frac{d\sigma}{d x_B dy dt d\phi d\varphi_e}=\frac{\alpha^3  x_B y } {16 \, \pi^2 \,  Q^2 \sqrt{1 + \epsilon^2}}
\left| \frac{\cal T}{e^3} \right|^2 \, 
\end{equation}
depends on $x_B$, $t$, the lepton energy fraction $y= p\cdot q_1/p\cdot k$, with $q_1 = k - k'$, $\epsilon=2 x_B \frac{M}{Q}$, and, in general, two azimuthal angles $\phi$ and $\varphi_e$. The observables of interest here do not depend on $\varphi_e$, the azimuthal angle of the scattered electron in the laboratory frame.
The angle $\phi$ is formed by the leptonic and hadronic planes, as shown in Fig.~\ref{fig:dvcs_phi}.
The charge of the electron is denoted with $e$ and $\alpha$ represents the fine-structure constant.
The amplitude ${\cal T}$ is the coherent sum of the DVCS  and Bethe-Heitler  amplitudes (${\cal T}_{\rm DVCS}$, ${\cal T}_{\rm BH}$):
\begin{equation}\label{eq_amplitude}
{\cal T}^2= |{\cal T}_{\rm BH}|^2 + |{\cal T}_{\rm DVCS}|^2 + {\cal I}
\, ,
\end{equation}
with the interference term defined as
\begin{equation}
{\cal I}
= {\cal T}_{\rm DVCS} {\cal T}_{\rm BH}^\ast
+ {\cal T}_{\rm DVCS}^\ast {\cal T}_{\rm BH}
\, .
\end{equation}
The azimuthal angular dependence of each of the three terms in Eq.~\eqref{eq_amplitude} is given by \cite{belitski}:
\begin{eqnarray}
\label{Par-BH}
|{\cal T}_{\rm BH}|^2&=& \frac{e^6} {x_B^2 y^2 (1 + \epsilon^2)^2 t\, {\cal P}_1 (\phi) {\cal P}_2 (\phi)} [ c^{\rm BH}_0 +  \nonumber\\ 
&+&\sum_{n=1}^2 c^{\rm BH}_n \, \cos{n\phi} + s^{\rm BH}_1 \, \sin{\phi} ],
\end{eqnarray}
\begin{eqnarray}
\label{AmplitudesSquared}
|{\cal T}_{\rm DVCS}|^2
&=& \frac{e^6}{y^2 {\cal Q}^2}\{c^{\rm DVCS}_0 + \sum_{n=1}^2 [c^{\rm DVCS}_n \cos {n\phi} + \nonumber\\
&+& s^{\rm DVCS}_n \sin {n\phi} ]\} \, ,
\end{eqnarray}
\begin{eqnarray}
\label{InterferenceTerm}
{\cal I}&=& \frac{e^6}{x_B y^3 t {\cal P}_1 (\phi) {\cal P}_2 (\phi)}\{c_0^{\cal I}+ \sum_{n = 1}^3[c_n^{\cal I} \cos{n\phi} +\nonumber\\
&+&  s_n^{\cal I} \sin{n\phi} ] \} \, ,
\end{eqnarray}
where ${\cal P}_1$ and ${\cal P}_2$ are intermediate lepton propagators 
(for more details and definitions, see \cite{belitski}).
Among the Fourier coefficients $c^{\rm P}_i$, $s^{\rm P}_i$ appearing in the previous expansions, the ones appearing in $|{\cal T}_{\rm BH}|^2$ (P = BH) depend on the well-known electromagnetic FFs,
while the ones appearing in $|{\cal T}_{\rm DVCS}|^2$ (P = DVCS) and ${\cal I}$ (P = ${\cal I}$) depend on the Compton Form Factors, the latter linearly. 
In the next sections, the sensitivity to the various CFFs of the DVCS spin observables presented in this paper will be outlined. 

\subsection{Target-spin asymmetry}
The use of a longitudinally polarized (LP) target allows the extraction of the target-spin asymmetry $A_{\rm UL}$, where the first letter in the subscript refers to the beam polarization (``Unpolarized'', in this case) and the second to the target polarization
(``Longitudinally polarized'', in this case), and which is given at twist-2 by:
\begin{equation}\label{eq_tsa}
A_{\rm UL}(\phi) \sim \frac{s_{1,{\rm LP}}^{\cal I}\sin\phi}{c_{0,{\rm unp}}^{\rm BH}+(c_{1,{\rm unp}}^{\rm BH}+c_{1,{\rm unp}}^{\cal I}+...)\cos\phi+...}\ ~,
\end{equation}
where the ellipses in the denominator represent smaller, higher-twist terms, and the $\rm unp$ subscript stands for ``unpolarized''. The $\sin\phi$ coefficient $s_{1,{\rm LP}}$, originating from the DVCS/BH interference term, at leading twist is proportional to a linear combination of the imaginary parts of the four CFFs (Eq.~\eqref{def_cffs}), 
\begin{eqnarray}\label{eq_tsa_cffs}
s_{1,{\rm LP}} &\propto&  \Im{\rm m}[ F_1\widetilde{\cal H}+\xi(F_1+F_2)({\cal H}+\frac{x_B}{2}{\cal E})+\nonumber \\
&-& \xi(\frac{x_B}{2} F_1+ \frac{t}{4M^2}F_2)\widetilde{\cal E}] ~.
\end{eqnarray}
\noindent
Due to the relative values of the proton form factors $F_1$ and $F_2$, and given the typical JLab kinematics, the coefficients in Eq.~\eqref{eq_tsa_cffs} enhance the contribution to $A_{\rm UL}$ coming from $\Im{\rm m} \widetilde{\cal H}$ over the ones from other CFFs. However, given that $\Im{\rm m}{\cal H}$ is expected to be about twice as big as $\Im{\rm m}\widetilde{\cal H}$, $A_{\rm UL}$ will also be sensitive to $\Im{\rm m}{\cal H}$.
In the kinematical range of the data presented here, higher-twist effects would appear in Eq.~\eqref{eq_tsa} as additional $\phi$-dependent terms, the dominant of which is a $\sin2\phi$ term in the numerator \cite{belitski}. 

\subsection{Beam-spin asymmetry}
The expression at twist-two of the beam-spin asymmetry is
\begin{eqnarray}\label{eq_bsa}
A_{\rm LU}(\phi) \sim \frac{s_{1,{\rm unp}}^{\cal I}\sin\phi}{c_{0,{\rm unp}}^{\rm BH}+(c_{1,{\rm unp}}^{\rm BH}+c_{1,{\rm unp}}^{\cal I}+...)\cos\phi...}\  ~,
\end{eqnarray}
where
\begin{equation}
s_{1,{\rm unp}}^{\cal I}\propto \Im{\rm m}[F_1 { \cal H}+\xi (F_1 + F_2)\widetilde{\cal H}-\frac{t}{4M^2}F_2{\cal E}].
\end{equation}
Thus, the beam-spin asymmetry for DVCS on a proton target is particularly sensitive to the imaginary part of the CFF of the unpolarized GPD $H$. As for the TSA, more terms must be added if one goes beyond the leading-twist approximation, the larger of which is a $\sin 2\phi$ term in the numerator. 
As mentioned in Section~\ref{sec_intro}, the slope in $t$ of the GPDs is related via a Fourier transform to the transverse position of the struck parton. Therefore, a measurement of $\Im{\rm m} \widetilde{\cal H}$ and $\Im{\rm m}{\cal H}$ provides, respectively,
information on the axial and electric charge distributions in the nucleon as a function of $x_B$ (see Eq. \eqref{EqGPDsFFlink}). A measurement of both the TSA and BSA at the same kinematics is needed to truly distinguish between the two contributions.

\subsection{Double-spin asymmetry}
The use of a polarized electron beam along with a longitudinally polarized target allows also the determination of the double-spin asymmetry $A_{\rm LL}$. Unlike $A_{\rm UL}$, the Bethe-Heitler process alone can generate a double-spin asymmetry. At twist-2 level, this observable takes the form: 

\begin{equation}
A_{\rm LL}(\phi) \sim \frac{c_{0,{\rm LP}}^{\rm BH}+c_{0,{\rm LP}}^{\cal I}
	 +(c_{1,{\rm LP}}^{\rm BH}+c_{1,{\rm LP}}^{\cal I})\cos\phi}
	{c_{0,{\rm unp}}^{\rm BH}+(c_{1,{\rm unp}}^{\rm BH}+c_{1,{\rm unp}}^{\cal I}+...)\cos\phi...} ~,
\end{equation}
with
\begin{eqnarray}\label{eq_dsa}
c_{0,{\rm LP}}^{\cal I}, c_{1,{\rm LP}}^{\cal I}
&\propto& \Re{\rm e} [F_1 \widetilde { \cal H} + \xi (F_1 +  F_2)({\cal H}+\frac{x_B}{2}{\cal E})+\nonumber\\
&-& \xi(\frac{x_B}{2} F_1+\frac{t}{4M^2}F_2)\widetilde{\cal E} ],
\end{eqnarray}
where terms depending on powers of $\frac{t}{Q^2}$ were neglected.
In this expression, the interference terms, related to the GPDs, are expected to be smaller than the known BH terms \cite{belitski}. Moreover, both the constant and the $\cos\phi$-dependent terms contain contributions from both BH and the DVCS/BH interference. Nonetheless, given the fast variation of BH depending on the kinematics, it is important to sample $A_{\rm LL}$ over a wide phase space to find possible regions of sensitivity to $\Re{\rm e}{\cal H}$ and $\Re{\rm e} \widetilde{\cal H}$, which dominate in Eq.~\eqref{eq_dsa}.

\section{The experiment}\label{sec_experiment}
\begin{figure*}
\begin{center}
\includegraphics[width=172mm]{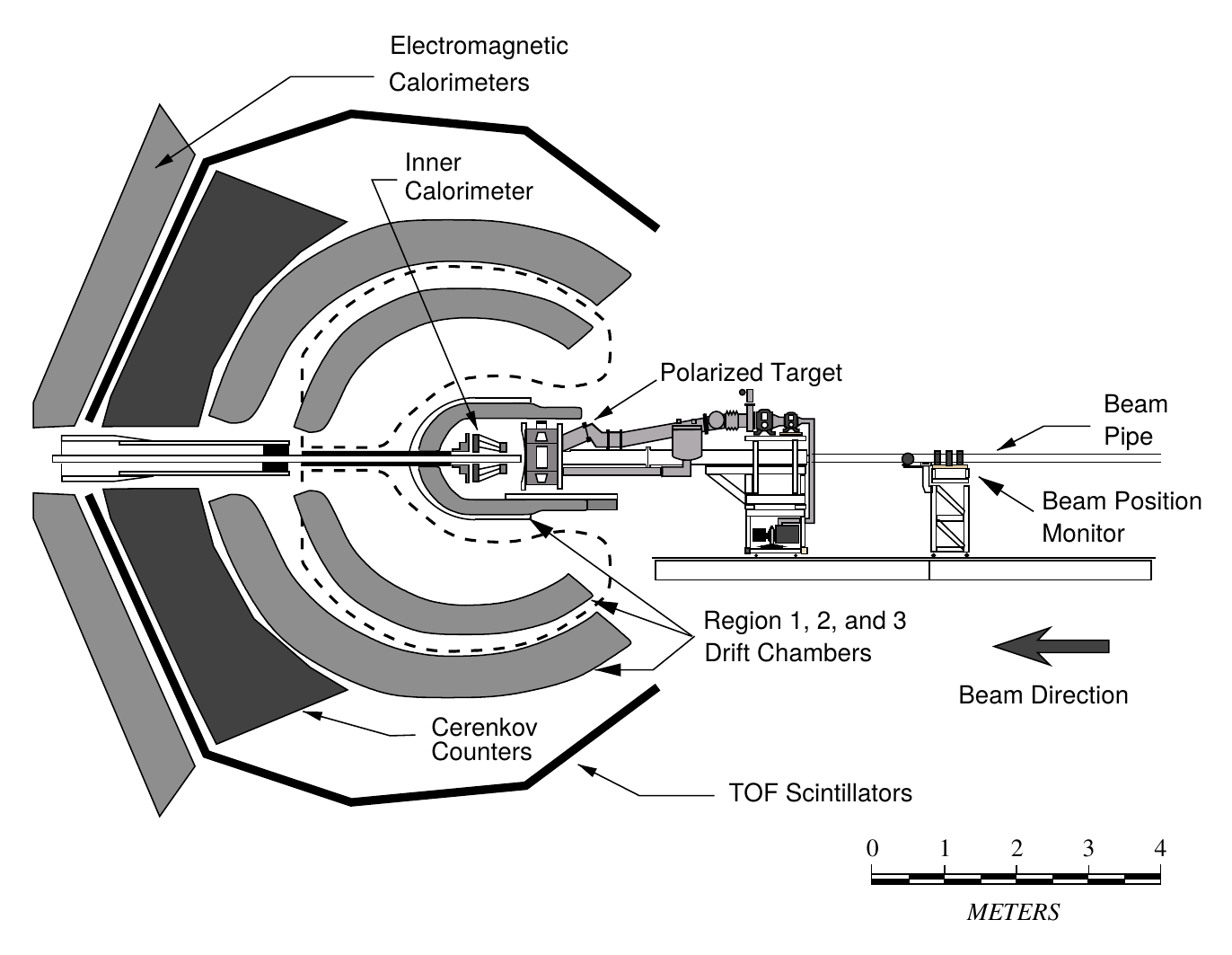}
\caption{Drawing of the experimental setup, including the CLAS detector with its components (DC, EC, CC, TOF), the Inner Calorimeter and the polarized target.}
\label{setup}
\end{center}
\end{figure*}

The data were taken in Hall B at Jefferson Lab from February to September 2009, for a total of 129 days. 
A continuous polarized electron beam was delivered by the CEBAF accelerator onto a solid ammonia target, polarized along the beam direction. Frozen beads of paramagnetically doped $^{14}$NH$_{3}$, kept at temperatures of about 1K and in a 5T magnetic field, were dynamically polarized by microwave irradiation \cite{chris_nim}. The target was 1.45 cm long and 1.5 cm in diameter. The target system included a $^4$He evaporation refrigerator and a superconducting split-coil magnet. The magnet, which provided a uniform polarizing field
($\Delta B/B = 10^{-4}$) and at the same time focused the low-energy M{\o}ller electrons toward the beam line, was inserted in the center of the CLAS detector \cite{clas}. CLAS was a spectrometer, based on a toroidal magnetic field, providing a wide acceptance. The CLAS magnetic field was generated by six superconducting coils arranged around the beamline to produce a field pointing primarily in the azimuthal 
direction. The particle detection system consisted of drift chambers (DC) to determine the trajectories and the momenta of charged particles curved by the magnetic field, gas {\v C}erenkov counters (CC) for electron identification, scintillation counters for measuring time-of-flight (TOF) and electromagnetic calorimeters (EC) to detect showering particles (electrons and photons) and neutrons. The segments were individually instrumented to form six independent magnetic spectrometers with a common target, trigger, and data-acquisition system.
An additional detector, the Inner Calorimeter, constructed for a previous DVCS-dedicated experiment \cite{fx} to complete the photon acceptance at low polar angles (from 4${}^{o}$ to 15${}^{o}$), was placed at the center of CLAS.
Figure~\ref{setup} shows the whole setup, including the polarized target, CLAS and the IC. 
A totally absorbing Faraday cup (FC), downstream of CLAS, was used to determine the integrated beam charge passing through the target. 

The data presented here come from the first two of the three parts in which the experiment was divided, and for which a ${}^{14}$NH$_3$ target was used. The main differences between these two parts (referred to as A and B) were the target position with respect to the center of CLAS ($z=-57.95$ cm for part A and $z=-67.97$ cm for part B) and the electron-beam energy ($E=5.886 \pm 0.005$ GeV for part A and $E=5.952 \pm 0.005$ GeV for part B). The beam energy values were obtained from elastic-scattering analysis on these data, using the ${}^{14}$NH$_3$ target.

The beam was rastered over the target in a spiral motion in order to assure a homogeneous depolarization over the whole volume of the target. 
The beam polarization was frequently monitored in M{\o}ller runs, via the measurement of the asymmetry of elastic electron-electron scattering. The target polarization was continuously monitored by a Nuclear Magnetic Resonance (NMR) system. 
Runs on a carbon target were taken periodically throughout the duration of the experiment for unpolarized-background studies. The selection of the data to be analyzed was done by monitoring the stability over time of Faraday-cup normalized yields per sector. 
For the results presented here, runs taken with the ${}^{14}$NH$_3$ and carbon targets from parts A and B were used. The two parts were analyzed separately and the final results were found by combining the results from the two parts, as will be described in Section~\ref{sec_merge}.

\section{Definitions of the asymmetries}
This paper reports on the extraction of three kinds of asymmetries, the experimental definitions of which are given here. In all of the formulae below, the first sign in the superscript on the number of normalized DVCS/BH events $N$ is the beam helicity ($b$) and the second sign
is the target polarization ($t$). 
$N$ is obtained from $ep\gamma$ events ($N_{ep\gamma}$), normalized by the corresponding Faraday-cup charge ($FC^{bt}$) after subtraction of the $\pi^0$ background 
as follows:
\begin{equation}
N^{bt}= (1-B_{\pi^0}^{bt})\cdot \frac{N^{bt}_{ep\gamma}}{FC^{bt}},
\end{equation}

where $B_{\pi^0}$ is the relative $\pi^0$ contamination, outlined in Section \ref{sec_pi0_back}. 

The  beam-spin asymmetry is calculated as:

\begin{equation}\label{def_bsa}
A_{\rm LU} = \frac{P_t^-(N^{++}-N^{-+})+P_t^+(N^{+-}-N^{--})}{P_b(P_t^-(N^{++}+N^{-+})+P_t^+(N^{+-}+N^{--}))},
\end{equation}
where $P_b$ is the polarization of the beam and $P_t^{+(-)}$ are the values of the polarization of the target for its two polarities.

The target-spin asymmetry is computed as:
\begin{equation}
\label{def_tsa_1}
A_{\rm UL} = A_{\rm UL}^{\rm lab}  + c_{A_{\rm UT}},
\end{equation}
where
\begin{equation}\label{def_tsa}
A_{\rm UL}^{\rm lab} = \frac{N^{++}+N^{-+}-N^{+-}-N^{--}}{D_f (P_t^-(N^{++}+N^{-+})+P_t^+(N^{+-}+N^{--}))}~.
\end{equation}
 $D_f$ is the dilution factor to account for the contribution of the unpolarized background (Section \ref{sec_dilution}) and $c_{A_{UT}}$ represents a correction applied to define the target-spin asymmetry with respect to the virtual-photon direction (Section \ref{transverse_sec}). 

The double (beam-target) spin asymmetry is obtained as:

\begin{equation}
\label{def_dsa_1}
A_{\rm LL} = A_{\rm LL}^{\rm lab}  + c_{A_{\rm LT}},
\end{equation}
where
\begin{equation}\label{def_dsa}
A_{\rm LL}^{\rm lab} = \frac{N^{++}+N^{--}-N^{+-}-N^{-+}}{P_b\cdot D_f (P_t^-(N^{++}+N^{-+})+P_t^+(N^{+-}+N^{--}))}
\end{equation}
and $c_{A_{\rm LT}}$ is the analog of $c_{A_{\rm UT}}$ for the double-spin asymmetry (Section \ref{transverse_sec}).\\
In the following, the steps leading to the extraction from the data of all the terms composing these asymmetries will be presented. 

\section{Data analysis}

\subsection{Particle identification}

The final state was selected requiring the detection of exactly one electron and one proton, and at least one photon. 

\begin{figure}
\begin{center}
\includegraphics[width=86mm]{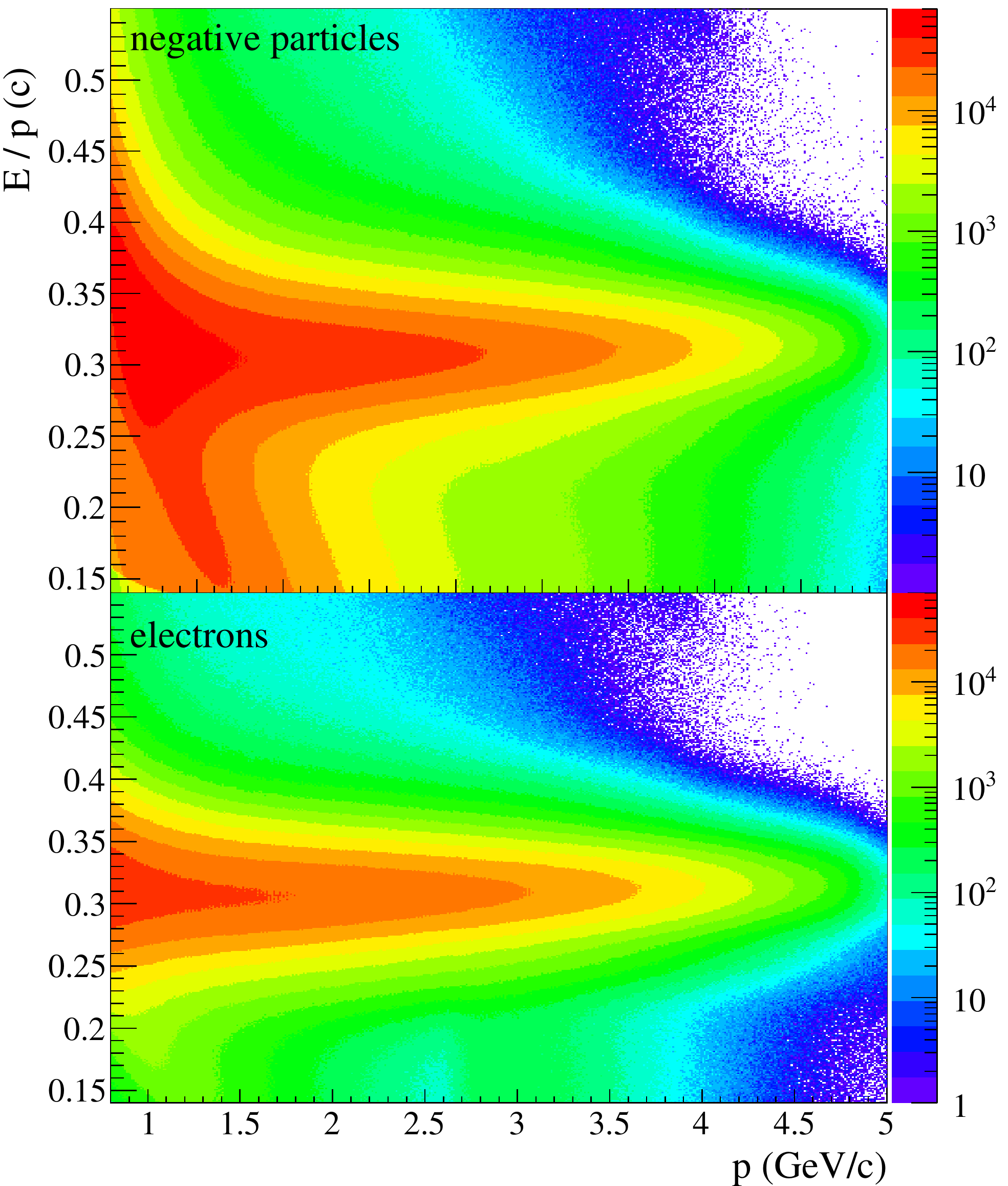}
\caption{(Color online)  Total energy deposited in the EC (inner+outer layers), $E$, divided by the particle momentum $p$ as a function of $p$ for all the negative tracks. Top: negative charged particles, before cuts. Bottom: after minimum-momentum, $EC_{in}$, fiducial and timing cuts.} 
\label{etot_p_before_after}
\end{center}
\end{figure}

\begin{figure}
\begin{center}
\includegraphics[width=86mm]{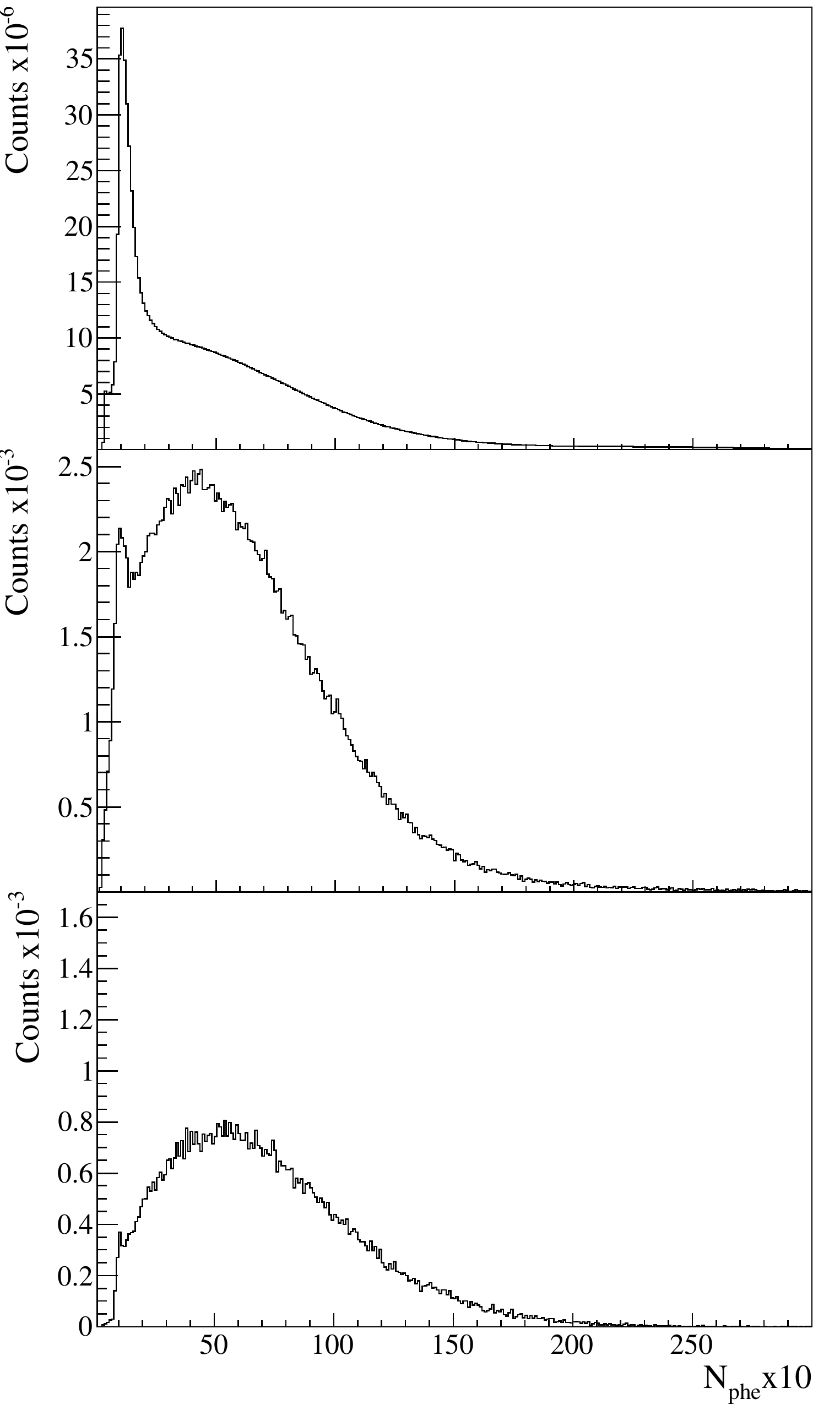}
\caption {Number of CC photoelectrons (times 10) for all negative tracks (top), after applying all PID and fiducial cuts for electrons (middle), and after $ep\gamma$ exclusivity cuts.}
\label{cc_nphe}
\end{center}
\end{figure}

The electrons were selected among all the negative tracks with momenta above 0.8~GeV/$c$, but requiring that their energy deposited in the inner layers of the EC \cite{ec_nim} be greater than 0.06 GeV, in order to reject negative pions, and that their hits in the CC and in the SC be matched in time. Fiducial cuts were also applied to eliminate the events at the edges of the EC (where the energy of a particle cannot be correctly reconstructed because a large part of the induced electromagnetic shower is lost), and to remove the ``shadow'' of the IC, which limits the CLAS acceptance for charged particles.
The effect of these cuts on the distribution of the total energy deposited in the EC divided by the momentum is shown as a function of the momentum in Fig.~\ref{etot_p_before_after}. The comparison of the top and middle plots of Fig.~\ref{cc_nphe} shows the effect of these same cuts on the number of CC photoelectrons. Most of the events in the single-photoelectron peak, which come from either pions or electronic noise in the PMTs, are removed by our particle identification (PID) cuts for electrons. The rest of the background is eliminated by the exclusivity cuts applied afterwards as explained in Section~\ref{sec_excl_cuts} (Fig.~\ref{cc_nphe}, bottom). 

\begin{figure}
\begin{center}
\includegraphics[width=86mm]{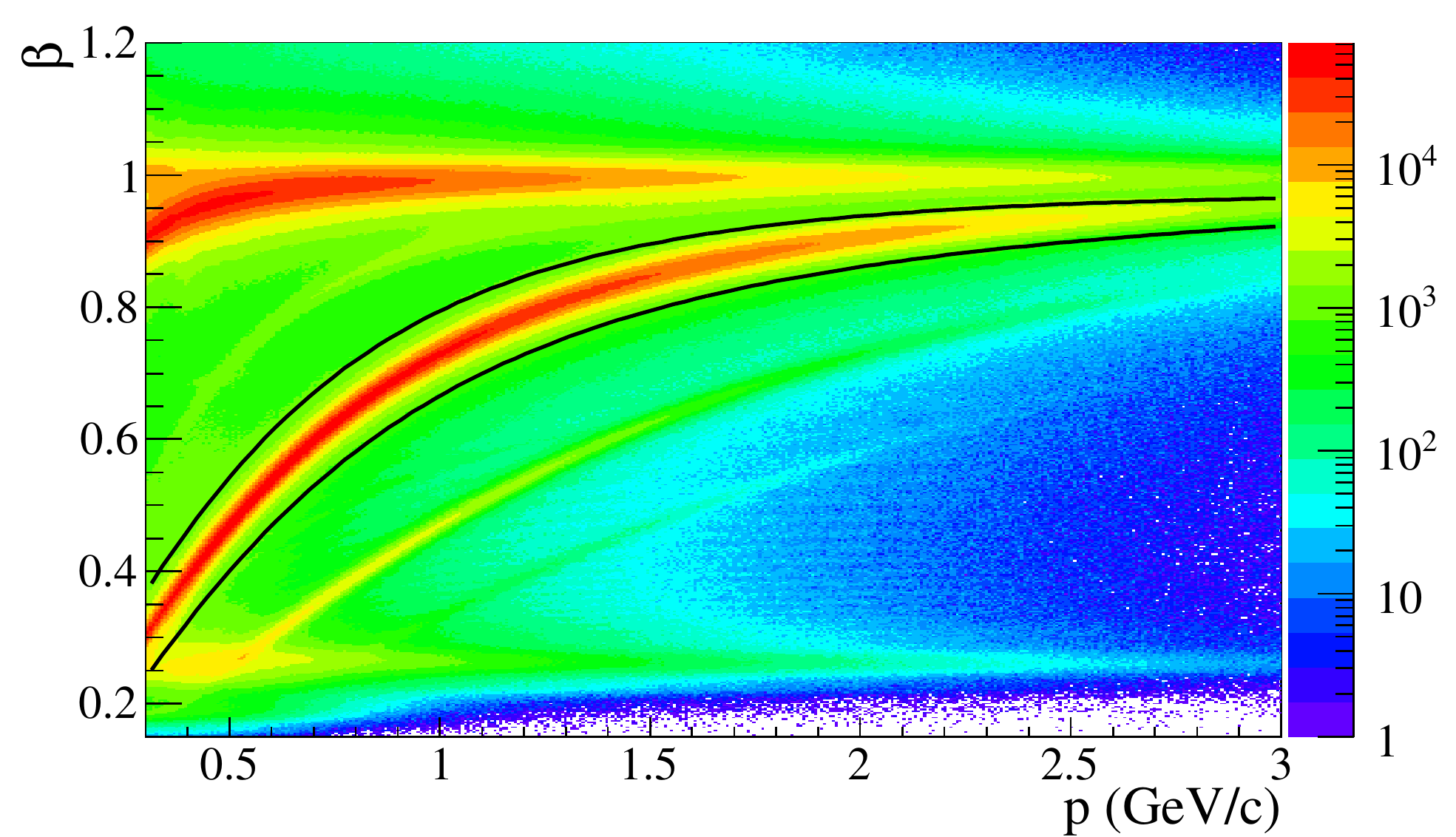}
\caption{(Color online)  $\beta$ as a function of $p$ for positively charged particles. The black lines represent the cut applied to select protons.}
\label{p_beta}
\end{center}
\end{figure}

The protons were selected from the correlation between their velocity, deduced from the time-of-flight measurement using the SC, and the proton momentum as measured by the DCs (Fig.~\ref{p_beta}).  
Fiducial cuts on $\theta$ and $\phi$  were also applied in order to remove the shadow of the IC.

\begin{figure}
\begin{center}
\includegraphics[width=86mm]{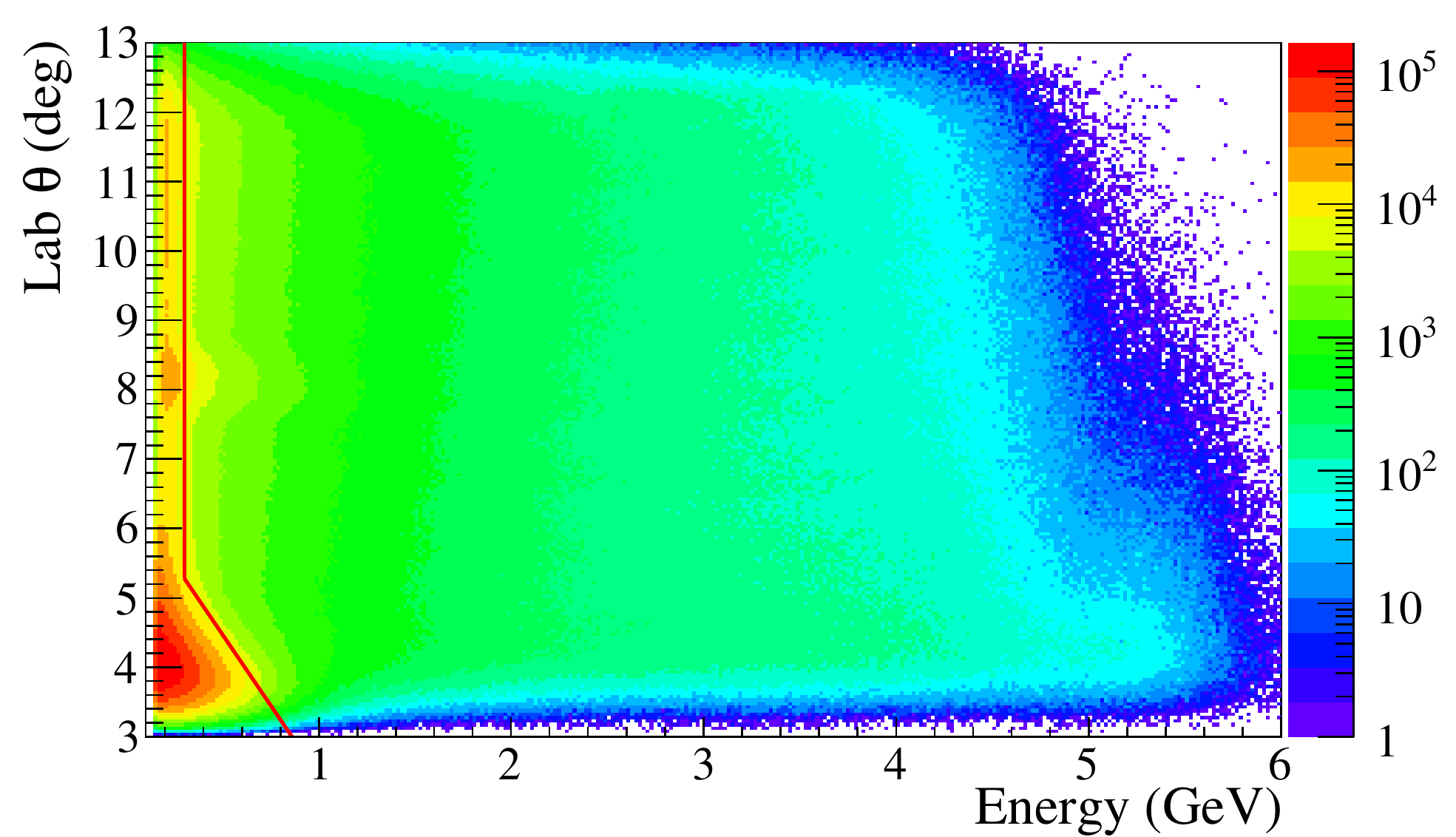}
\caption{(Color online)  Polar angle $\theta$ as a function of the reconstructed energy for IC hits, showing the cut on the minimum energy at 0.25 GeV, as well as the ``triangular'' cut to remove the low-energy/low-$\theta$ accidental background,
applied to select photons.}
\label{ic_photon_en}
\end{center}
\end{figure}

\begin{figure}
\begin{center}
\includegraphics[width=86mm]{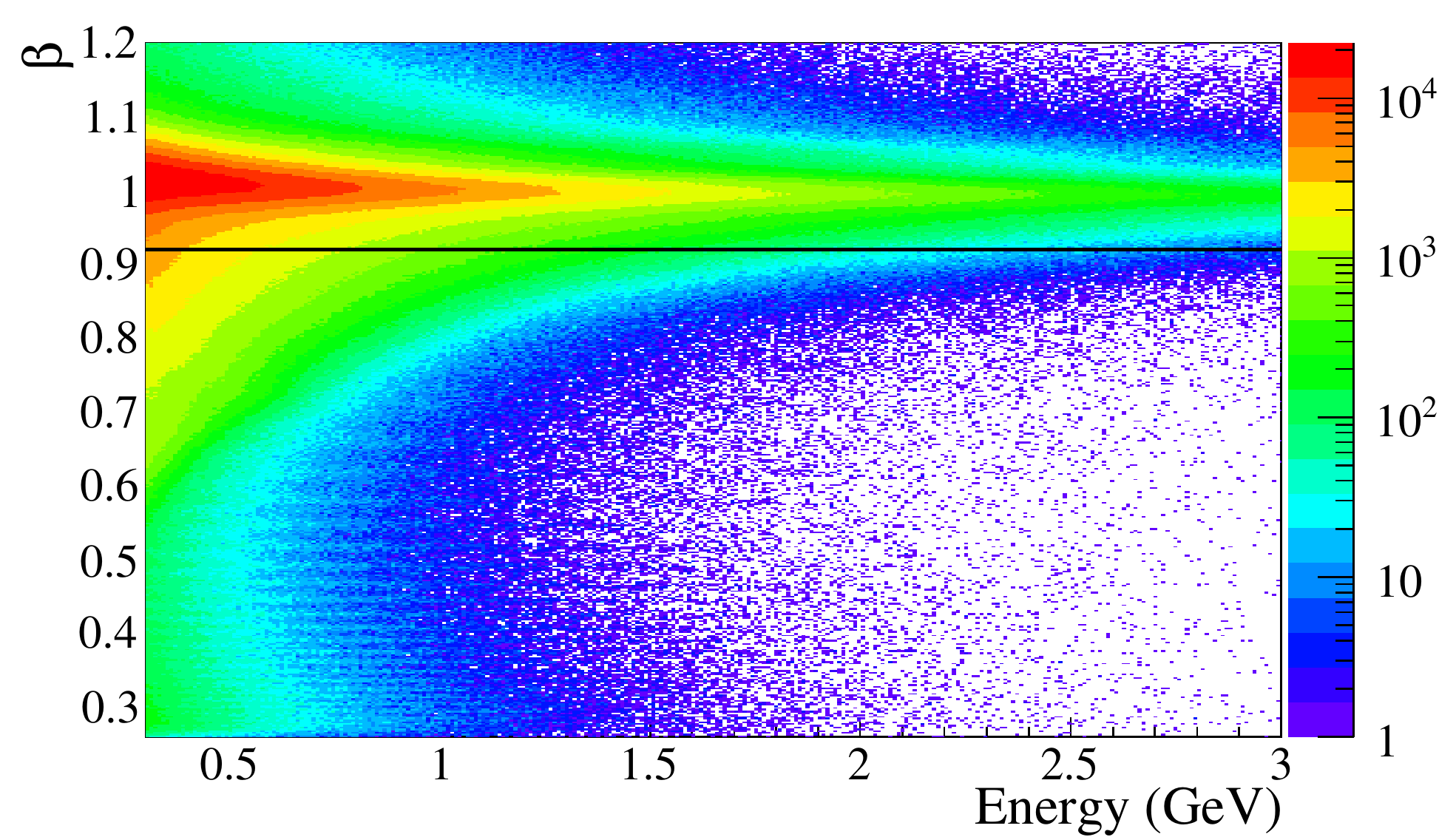}
\caption{(Color online)  Distribution of $\beta$ vs energy for neutrals measured by time-of-flight with EC. The events for which $\beta>0.92$ (black line) were retained as photon candidates.}
\label{ph_ec_beta}
\end{center}
\end{figure}

For the photons, two different sets of cuts were adopted, depending on whether the photon was detected in the IC or in the EC. 
For the IC case, a low-energy threshold of 0.25 GeV and a cut on $\theta$ versus $E_{\gamma}$ to remove the background coming from M{\o}ller electrons (Fig.~\ref{ic_photon_en}) were applied, as well as fiducial cuts on $x$ and $y$, to remove the outer and inner edges of the detector, where clusters could not be fully reconstructed.
%
For the EC case, all neutrals passing a low-energy threshold of 0.25 GeV and having $\beta>0.92$ (Fig.~\ref{ph_ec_beta}) to select the in-time photons were retained. Fiducial cuts as for the electrons were also adopted to remove the edges of the detector
and IC-frame cuts were applied to remove the shadow of the IC over the EC.

\subsection{Monte-Carlo simulations}
Monte-Carlo simulations were used in this analysis for the three following tasks:
\begin{itemize}
\item{determination of angular and momentum corrections for electrons and protons to compensate for multiple scattering and energy losses in the target and detector materials, as well as for imperfections in the trajectory reconstruction;}
\item{optimization of the DVCS and $ep\pi^0$ exclusivity cuts (Sections~\ref{sec_excl_cuts} and \ref{sec_pi0_back}, respectively);}
\item{evaluation of the $ep\pi^0$ contamination in the $ep\gamma$ event sample (Section~\ref{sec_pi0_back}).}
\end{itemize}
The two sets of generated events, DVCS/BH and $ep\pi^0$ (Section~\ref{sec_dvcs_gen} and \ref{sec_pi0_gen}), were fed to the CLAS GEANT3-based simulation package (GSIM) to simulate the response of the CLAS detector. The output of GSIM was then fed to a post-processing code (GPP) that simulates dead or inefficient DC wires and smears the DC and TOF resolutions to more closely agree with the experiment. 
The output of GPP was finally fed to the CLAS reconstruction package, and reduced ntuples and root files were obtained from the reconstructed files in the same way as was done in the data processing. 

\subsubsection{DVCS/BH simulation}\label{sec_dvcs_gen}

\begin{figure}
\begin{center}
\includegraphics[width=86mm]{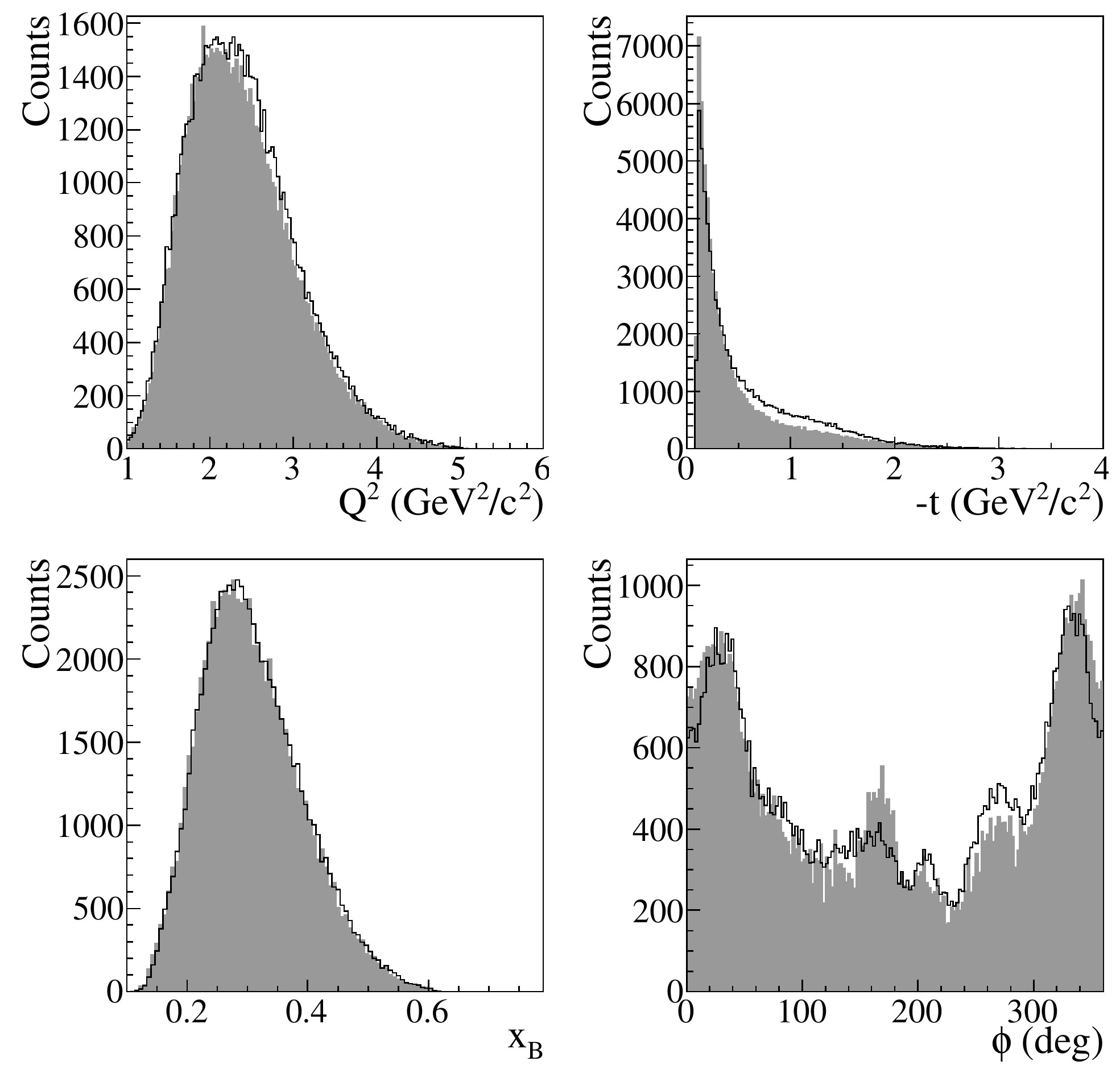}
\caption {DVCS channel. Comparison of data (black lines) and Monte-Carlo simulation (shaded areas). From the top left, $Q^2$, $-t$, $x_B$ and $\phi$ are plotted. The histograms are normalized to each other via the ratio of their integrals.}
\label{data_mc_dvcs_kinvars}
\end{center}
\end{figure}

A DVCS/BH event-generator code, which produced $ep\gamma$ events according to the formalism of Belitsky {\it et al.} \cite{belitski}, was used for the Monte-Carlo simulation. 
Figure \ref{data_mc_dvcs_kinvars} shows a comparison of the distributions of the relevant kinematic variables for the data (black lines) and the Monte-Carlo simulation (shaded areas). Here, PID and $ep\gamma$ exclusivity cuts, which will be described in Section \ref{sec_excl_cuts}, were included for both the data and the Monte Carlo. The agreement between data and simulation is quite good, especially given the purposes of the simulations in these analyses: they are not used for absolute acceptance corrections, but only to help in the determination of cut widths and for background subtraction. Slight differences between data and MC, especially visible in the high-$t$ and central-$\phi$ region, are coming from events in the EC topology. This is probably due to the fact that the data, unlike the MC, are not only pure $ep\gamma$ events, but are contaminated by exclusive $\pi^0$ events. The fact that the $ep\pi^0$ contamination is larger in the EC topology, as will be reiterated in Section~\ref{sec_excl_cuts}, can explain the data/MC discrepancies. 

\subsubsection{Exclusive $\pi^0$ simulation}\label{sec_pi0_gen}

\begin{figure}
\begin{center}
\includegraphics[width=86mm]{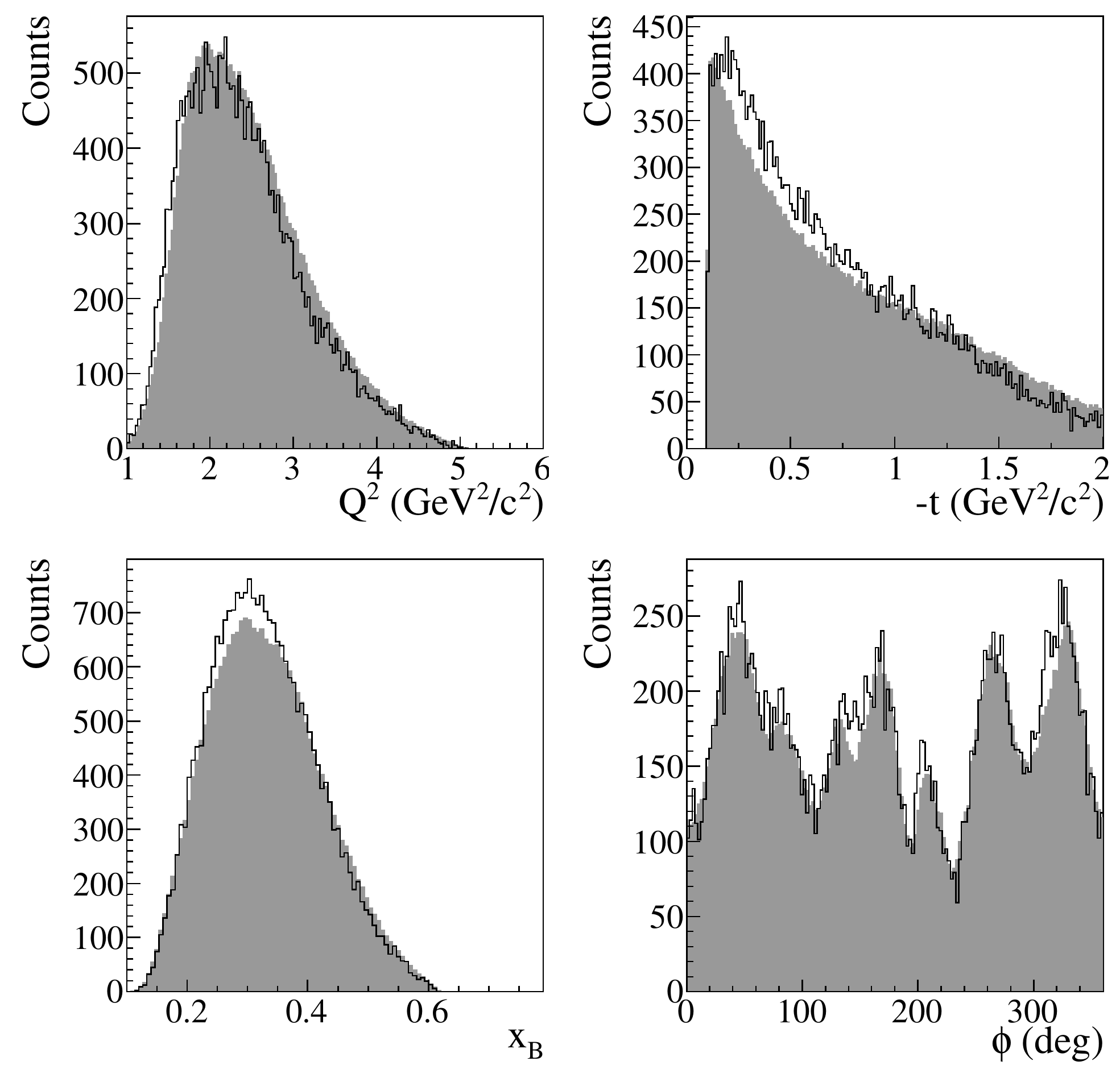}
\caption {Exclusive $\pi^0$ channel. Comparison of data (black lines) and Monte-Carlo simulation (shaded areas). Starting from the top left, $Q^2$, $-t$, $x_B$ and $\phi$ are plotted. The histograms are normalized to each other via the ratio of their integrals.}
\label{data_mc_pi0_kinvars}
\end{center}
\end{figure}

Exclusive $ep\pi^0$ events were generated using a code for meson electroproduction that included a parametrization of the $ep\pi^0$ differential cross sections that have recently been measured by CLAS \cite{ivan}. 
Figure \ref{data_mc_pi0_kinvars} shows a comparison of the distributions of the relevant kinematic variables for the data (black) and the Monte-Carlo simulation (shaded areas). Here, PID and $ep\pi^0$-exclusivity cuts, which will be described in Section \ref{sec_pi0_back}, were included for both the data and the Monte Carlo. The agreement between data and simulation is quite good. 

\subsection{DVCS channel selection}\label{sec_excl_cuts}
\begin{figure*}
\begin{center}
\includegraphics[width=172mm]{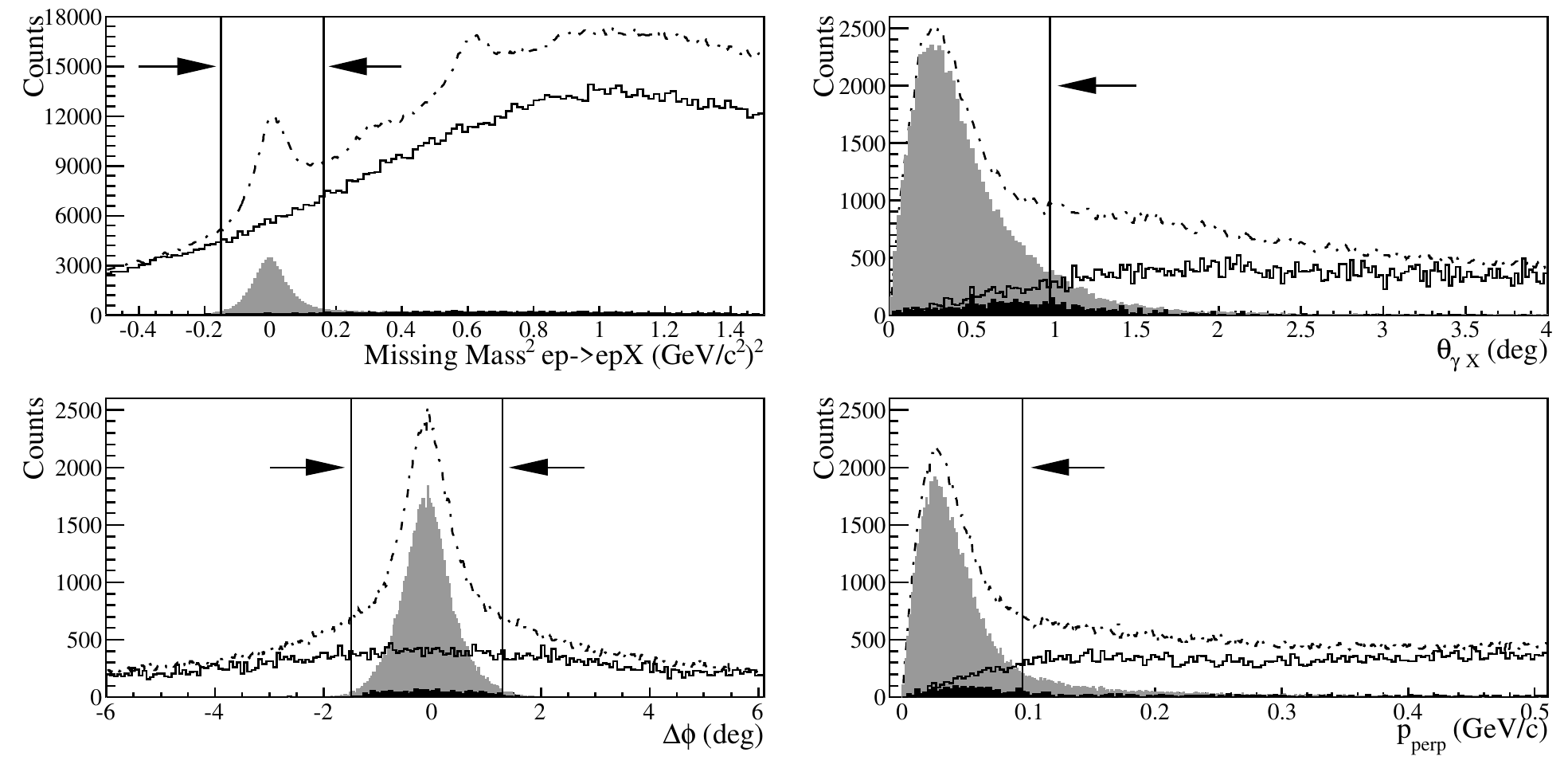}
\caption {IC topology. Effects of the exclusivity cuts. Top left: squared missing mass of the $ep$ system; top right: the angle $\theta_{\gamma X}$ between the measured photon and the calculated photon from $ep \to e'p' X$; bottom left: difference between two ways to compute the $\phi$ angle; bottom right: missing transverse momentum $p_{perp}$ calculated from $ep \to e'p' \gamma X$. The dot-dashed and solid lines show the events {\it before} exclusivity cuts for, respectively, $^{14}$NH$_{3}$ and $^{12}$C data, while the gray and black shaded areas represent the events {\it after} all exclusivity cuts except for the one on the plotted variable for, respectively, $^{14}$NH$_{3}$ and $^{12}$C data. The lines and arrows show the limits of the selection cuts. The plots for $^{14}$NH$_{3}$ and $^{12}$C data are normalized to each other via their relative luminosities.}
\label{ic_cuts}
\end{center}
\end{figure*}

\begin{figure*}
\begin{center}
\includegraphics[width=172mm]{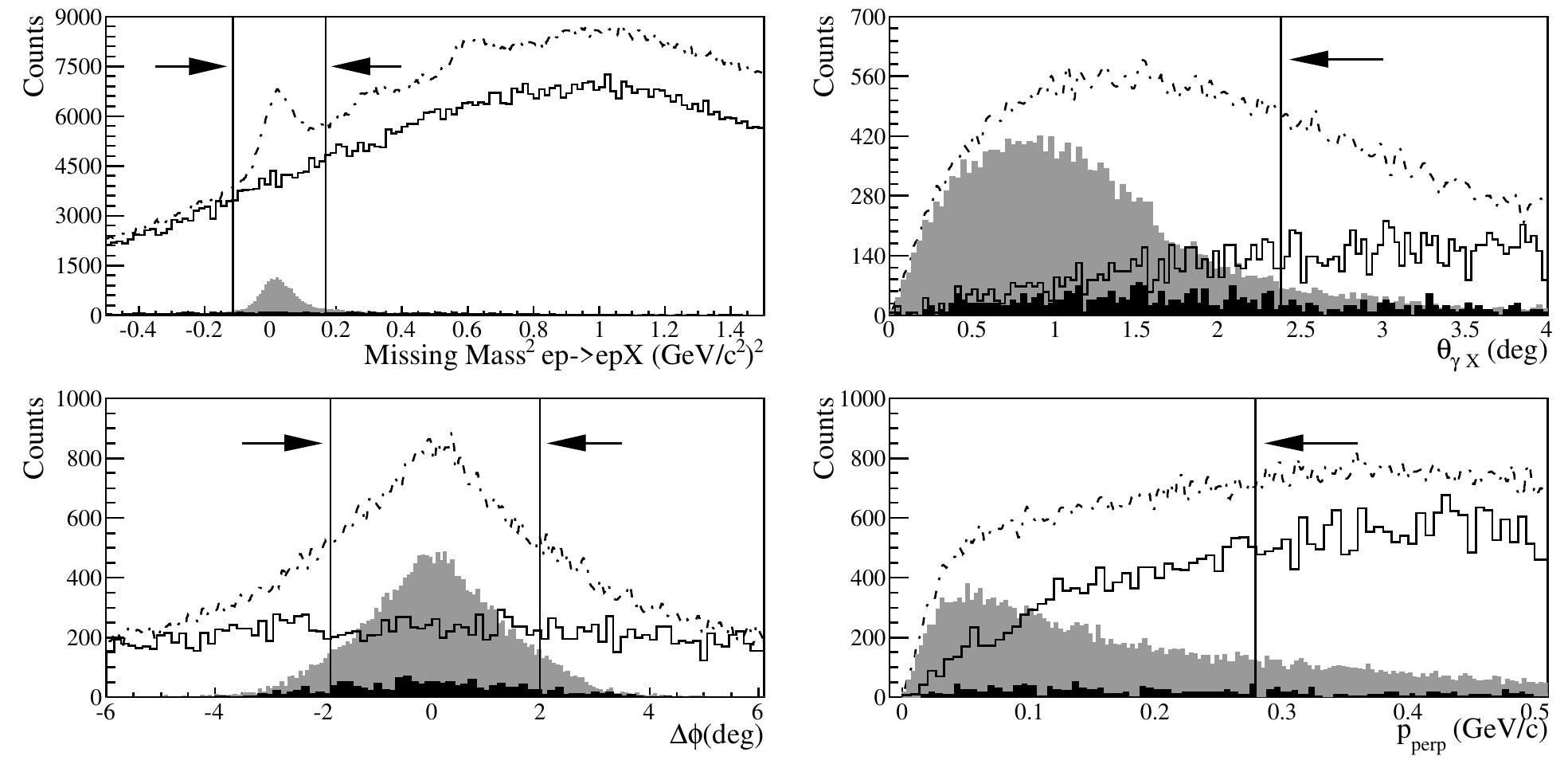}
\caption {EC topology. See caption of Fig.~\ref{ic_cuts}.}
\label{ec_cuts}
\end{center}
\end{figure*}

After selecting events with exactly one electron and one proton, and at least one photon, and having applied the momentum corrections, further cuts need to be applied to ensure the exclusivity of the $ep\gamma$ final state. Two kinds of backgrounds need to be minimized: the nuclear background coming from scattering off the nitrogen of the ${}^{14}$NH$_3$ target, and the background coming from other channels containing an electron, a proton and at least one photon in the final state. Having measured the four-momenta of the three final-state particles, one can construct several observables, hereafter referred to as ``exclusivity variables'', on which cuts can be applied to select the $ep\gamma$ channel. Here, the following quantities were studied: 
\begin{itemize}
\item{the squared missing mass of the $ep$ system, MM$^2(ep)$;}
\item{the angle $\theta_{\gamma X}$ between the measured photon and the calculated photon, using the detected electron and proton together with $ep \to e'p'X$ kinematics;}
\item{$\Delta\phi$, the difference between two ways to compute the angle $\phi$ between the leptonic and the hadronic plane. The two ways concern the definition of the normal vector to the hadronic plane: one is via the cross product of the momentum vectors of the proton and the real photon, and the other one is via the cross product of the momentum vectors of the proton and the virtual photon. For the $ep\gamma$ final state, $\Delta\phi$ should be distributed as a gaussian centered at zero, with width determined by the experimental resolution;}
\item{$p_{perp}$, the transverse component of the missing momentum of the reaction $ep\to e'p'\gamma X$, given by $p_{perp}=\sqrt{p_x(X)^2+p_y(X)^2}$, in the laboratory frame.}
\end{itemize}
The definition of the exclusivity cuts is quite a delicate step in the DVCS analysis. 
It is important that the cuts are determined in a consistent way for the data and for the Monte-Carlo simulation, because the latter will be used to evaluate the $\pi^0$ background contamination. 
In this analysis it was chosen to define each exclusivity cut by fitting the corresponding variable with a gaussian and cutting at $\pm 3\sigma$ around the fitted mean. This procedure was done separately for the data and for the simulation. This way, the same fraction of events was kept, for both data and simulation. 
The exclusivity variables to be fitted were plotted after applying {\it preliminary} cuts that included:
\begin{itemize}
\item{Kinematic cuts to be above the region of the nucleon resonances and to approach the regime of applicability of the leading-twist GPD formalism: $Q^2>1$ (GeV/$c$)$^2$, $-t<Q^2$, and $W>2$ GeV/$c^2$;} (where $W=\sqrt{M^2 + 2M\nu - Q^2}$)
\item{$E_{\gamma}>1$ GeV, since the real photons of interest are expected to have high energy;}
\item{$3\sigma$ cut around the mean of MM$^2(ep)$ to eliminate from the experimental data the background from channels other than $ep\gamma$ or $ep\pi^0$ (visible in the top left plot of Figs.~\ref{ic_cuts} and \ref{ec_cuts}, where peaks from $\eta$ and $\omega$/$\rho$ are evident).}
\end{itemize}
In order to eliminate broadening on the widths of the peaks due to events from electron scattering on the nitrogen, the fits to the exclusivity variables were done on the spectra obtained after subtracting carbon data from the ${}^{14}$NH$_3$ data. The two datasets were normalized to each other via the ratio of their Faraday-cup counts multiplied by a constant that accounts for different densities of materials for the two target types (see Section \ref{sec_dilution}). 

The method to define the cuts described above was adopted for the topology where the photon was detected in the IC, since the comparison with Monte Carlo showed that these data are strongly dominated by the DVCS/BH channel. Figure~\ref{ic_cuts} shows, for the IC topology, the effects of the exclusivity cuts, which appear successful both in extracting quite a clean $ep\gamma$ final state (shaded areas) and in minimizing the background originating from the nitrogen part of the target (black areas). 

A different strategy was found to be necessary for the EC case, which displayed, before cuts, a larger contribution from $ep\pi^0$ events.
The peaks in the exclusivity variables for the data in this topology are very broad, when visible, and not necessarily produced by DVCS/BH candidates. In the first plot of Fig.~\ref{ec_cuts}, for example, the distribution of the squared missing mass of the $ep$
system is shown for the EC case. As is clearly visible, the peak of the distribution is not centered at zero but around the squared $\pi^0$ mass, indicating a significant contamination from the exclusive $\pi^0$ events that will be subtracted later through the
procedure described in Section~\ref{sec_pi0_back}.

Thus it was decided, for the EC topology, not to fit the distributions of the exclusivity variables to extract cut means and widths. Instead it was chosen to fit only the peaks of the DVCS/BH Monte-Carlo simulations. To correct for the discrepancies in resolutions between data and simulation, the widths of the various exclusivity variables obtained from the fits were then multiplied by appropriate scaling factors. These factors were obtained from the comparisons of data/MC for the $ep\pi^0$ channel in the EC-EC topology. 
The cuts and their effects are shown in Fig.~\ref{ec_cuts}. 

\subsection{Four-dimensional binning and central kinematics}\label{sec_binning}
\begin{figure}
\begin{center}
\includegraphics[width=86mm]{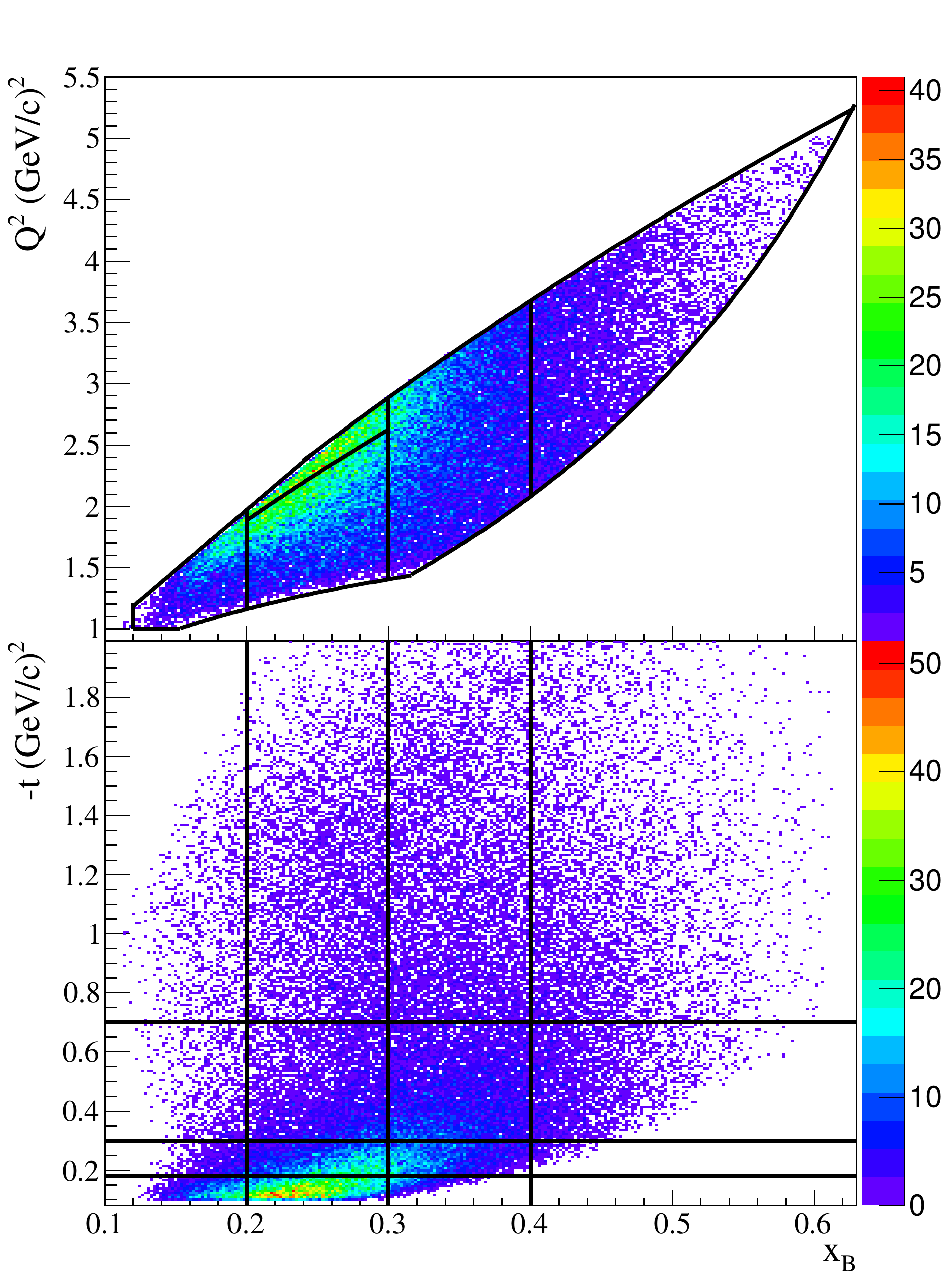}
\caption{(Color online) Grid showing the binning in the $Q^2$-$x_{B}$ space (top) and in the $-t$-$x_B$ space (bottom).}
\label{binning_q2_x_t}
\end{center}
\end{figure}

The DVCS reaction can be described by four independent kinematic variables. 
The typical variables used to interpret the results in terms of Generalized Parton Distributions are $Q^2$, $x_B$, $-t$ and $\phi$. In accordance with the choice made in previous DVCS analyses \cite{fx}, the binning of the data in the $Q^2$-$x_B$ plane was done making 5 slices in the polar angle $\theta_e$ of the electron and in $x_B$.
The limits of the slices are given in Table \ref{tab_binning}, as well as the bin averages for $Q^2$ and $x_B$, defined as the weighted average over the distribution of events in each bin. 
The size of the bins was optimized to have comparable statistics. The top plot of Fig.~\ref{binning_q2_x_t} shows the chosen grid in the $Q^2$-$x_B$ plane. 
Ten equally spaced bins in $\phi$ and 4 bins in $-t$ were adopted. The bin limits and data-averaged bin centers are summarized in Table~\ref{tab_binning_t}. The bottom plot of Fig.~\ref{binning_q2_x_t} shows the binning in the $t$-$x_B$ plane. 
\begin{table}
   \centering
   \begin{tabular}{|c|c|c|c|c|} 
\hline
      Bin   & $x_B$ bin      & $\theta_e$ bin    &   $\langle x_B\rangle$   &   $\langle Q^2\rangle$ ((GeV/$c$)$^2$) \\
\hline
      1      &  $0.1<x_B<0.2$     & $15^o<\theta_e<48^o$ & 0.179 & 1.52\\
      2      &  $0.2<x_B<0.3$     & $15^o<\theta_e<34^o$ & 0.255 & 1.97 \\
      3      &  $0.2<x_B<0.3$     & $34^o<\theta_e<48^o$ & 0.255 & 2.41\\
      4      &  $0.3<x_B<0.4$     & $15^o<\theta_e<45^o$ & 0.345 & 2.60 \\
      5      &  $x_B>0.4$     & $15^o<\theta_e<45^o$ & 0.453 & 3.31 \\
\hline
   \end{tabular}
   \caption{Definition of the bins in $x_B$ and $\theta_e$ ($Q^2$), and average kinematics for $x_B$ and $Q^2$ ((GeV/$c$)$^2$) for each bin.}
   \label{tab_binning}
\end{table}
\begin{table}[tbph]
   \centering
   \begin{tabular}{|c|c|c|} 
\hline
      Bin    & $-t$ range (GeV/$c$)$^2$    & $\langle -t\rangle$ (GeV/$c$)$^2$\\
\hline
      1      &  $0.08<-t<0.18$     & 0.137\\
      2      &  $0.18<-t<0.3$       & 0.234\\
      3      &  $0.3<-t<0.7$       & 0.467\\
      4      &  $0.7<-t<2.0$       & 1.175\\
\hline
   \end{tabular}
   \caption{Definition of the bins in $-t$ and average kinematics for each bin.}
   \label{tab_binning_t}
\end{table}

The central kinematics in this analysis were defined as the average value from the data of each of the 4 kinematic variables for each bin. In fact, at first order the uncertainties on an asymmetry induced by taking bins of finite size are minimized when the central kinematics are chosen to be the weighted average over the distribution of events in that bin. 
The procedure to compute second-order bin-centering corrections is reported in Section \ref{sec_bcc}.
The grid of bins was applied to both parts A and B of the experiment. In order to establish whether the asymmetries obtained from the two sets of data could eventually be combined, the central values of the various bins into which the available phase space was divided were computed, and compared for the two parts. For each bin in $Q^2$ ($\theta_e$)-$x_B$ and $-t$, the central kinematics for parts A and B were found compatible well within their standard deviations.



\subsection{Dilution factors}\label{sec_dilution}
\begin{figure}
\begin{center}
\includegraphics[width=86mm]{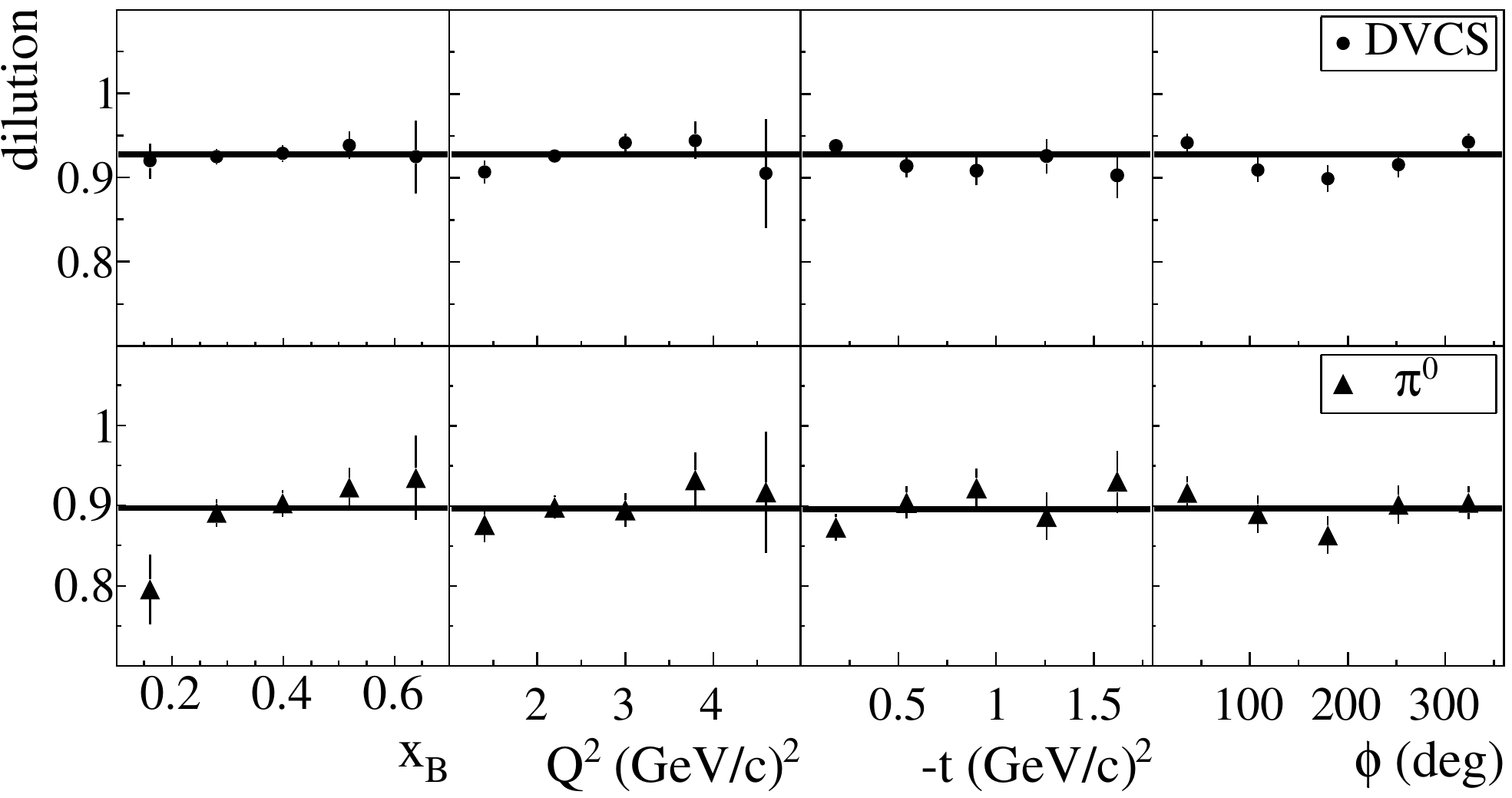}
\caption {Dilution factor as a function of (from left to right) $x_B$, $Q^2$, $-t$, $\phi$, for the DVCS analysis (top) and for the $ep\pi^0$ analysis (bottom). Part B.}
\label{dil_q2}
\end{center}
\end{figure}

For both the DVCS and $ep\pi^0$ final states, dilution factors to correct the experimental yields for the contribution from the scattering on the unpolarized nitrogen of ${}^{14}$NH$_3$ were determined using data taken on ${}^{14}$NH$_3$ and on ${}^{12}$C targets. The dilution factor is defined as
\begin{equation}
D_f = 1-c\cdot\frac{N_{{}^{12}{\rm C}}}{N_{{}^{14}{\rm NH}_3}} \label{eq_dilution}.
\end{equation}
Here, $N_{{}^{12}{\rm C}}$ is the number of events, normalized by the corresponding Faraday-cup counts, taken on carbon and surviving all of the DVCS (or $ep\pi^0$) selection cuts, while $N_{{}^{14}{\rm NH}_3}$ is the number of events, normalized by the corresponding Faraday-cup counts, taken on ${}^{14}$NH$_3$ passing the DVCS (or $ep\pi^0$) selection cuts. The factor $c$ accounts for the different luminosities of the two sets of data, which are in turn related to the ratio of the areal densities of the materials 
present at the target level for the two kinds of runs (${}^{14}$NH$_3$ in the numerator, ${}^{12}$C in the denominator). 
As the statistics acquired on carbon during the experiment was much smaller than for the ${}^{14}$NH$_3$ data, it was not possible to perform the dilution factor analyses for each 4-dimensional bin. The dependence of the dilution factor on each of the 4 kinematic variables was checked by integrating over the other three. The results for part B, as an example, are shown as functions of $x_B$, $Q^2$, $-t$, and $\phi$ in Fig.~\ref{dil_q2}. Both dilution factors show an approximately flat dependence in each of the 4 kinematic variables. There may be a small $x_B$ dependence, although the dilution factor is not inconsistent with a constant behavior.
The fit results for each variable are consistent with each other within error bars. Therefore, a constant value of $D_f$ for all the kinematic bins, for both the $ep\gamma$ and $ep\pi^0$ analysis, was assumed. The following values were adopted:
\begin{itemize}
\item{part A: $D_f(ep\gamma)= 0.912 \pm 0.009$; \\
part B: $D_f(ep\gamma)=0.928 \pm 0.006$;}
\item{part A: $D_f(ep\pi^0)=0.921 \pm 0.016$; \\
part B $D_f(ep\pi^0)=0.896 \pm 0.010$.}
\end{itemize}
 
\subsection{Beam and target polarization}\label{sec_pbpt}
The product of the polarizations of beam and target, $P_b P_t$, was extracted for part A and part B separately. Exclusive elastic events were used for this analysis, since the elastic asymmetry is well known \cite{donnelly} and $P_b P_t$ is the proportionality factor between the experimental asymmetry ($A_{exp}$) and the expected one ($A$):
\begin{equation}\label{el_asym_eq_th}
A=\frac{A_{exp}}{P_bP_t},
\end{equation}
where
\begin{equation}\label{el_asym_eq}
A_{exp}=\frac{\bar{N}^+-\bar{N}^-}{D_f(\bar{N}^++\bar{N}^-)} ~.
\end{equation}
$\bar{N}^{+(-)}$ refers to the number of elastic events for positive (negative) beam helicity for a given target polarization, normalized to the corresponding Faraday-cup counts. $D_f$ is the dilution factor ($\sim 98\%$) determined in the same way as for the DVCS and $ep \to e'p'\pi^0$ events (Section~\ref{sec_dilution}).
The expected asymmetry $A$ is computed according to \cite{donnelly}, and for the elastic form factors a parametrization, obtained by fitting the results from a JLab polarization-transfer experiment \cite{puckett}, was used. 
Inverting Eq.~\eqref{el_asym_eq_th} $P_b P_t$ was calculated for each run part as the average over 7 different $Q^2$ bins. 
\begin{table*}
   \centering
   \begin{tabular}{|c|c|c|c|c|c|}
\hline
      Run part    & $P_bP_t^+$ & $P_bP_t^-$ & $P_b$ & $P_t^+$ & $P_t^-$\\
\hline
      Part A  &  $0.648 \pm 0.018$   &    $0.625 \pm 0.016$  &  $0.87 \pm 0.04$ & $0.75 \pm 0.04$ & $0.72 \pm 0.04$ \\ 
      Part B &  $0.674 \pm 0.011$   &    $0.625 \pm 0.010$ & $0.837 \pm 0.017$ & $0.81 \pm 0.03$ & $0.75 \pm 0.03$\\
\hline
   \end{tabular}
   \caption{Obtained values of $P_bP_t$, $P_b$, and $P_t$, for each part of the experiment, and for both positive and negative target polarizations.}
   \label{table_pbpt}
\end{table*}
The beam polarization alone ($P_b$) was obtained from the analysis of the various M{\o}ller runs that were taken during the experiment. Event-weighted averages of $P_b$ were computed for the two parts of the experiment. The target polarization values were thus deduced by dividing the measured $P_b P_t$ by $P_b$.
The results for the product of the polarizations of beam and target for each polarization sign, and of the beam and target polarizations alone for the two parts of the experiment, are summarized with their uncertainties in Table~\ref{table_pbpt}.

\subsection{$\pi^0$ background subtraction}\label{sec_pi0_back}
The procedure adopted to compute the $\pi^0$ contamination to the $ep\gamma$ event sample is described here. All the following was done separately for each of the 4-dimensional ($Q^2$, $x_B$, $-t$, $\phi$) bins described in Section~\ref{sec_binning} and for each beam and target polarization state. To keep the notation simpler, in the following formulae we omit the dependence on ($Q^2$, $x_B$, $-t$, $\phi$) of each quantity.
After applying the $ep\gamma$ event selection cuts and the DVCS exclusivity cuts, one is left with $N_{ep\gamma}$ events that are not only DVCS or Bethe-Heitler candidate events, but can also contain some $ep\pi^0$ events in which one of the two $\pi^0$-decay photons has escaped detection. 
The number of DVCS/BH events can be obtained as
\begin{equation}
N_{\rm DVCS/BH} = N_{ep\gamma}( 1 - B_{\pi^0}),
\end{equation}
where the contamination fraction is given by 
\begin{equation}\label{eq_background_fraction}
B_{\pi^0}= \frac{D_f(ep\pi^0) N_{ep\pi^0 (1\gamma)}}{D_f(ep\gamma) N_{ep\gamma}}.
\end{equation}
To calculate $N_{ep\pi^0 (1\gamma)}$, we use both the MC simulation and exclusive $ep\pi^0$ events ($N_{ep\pi^0}^{\rm DATA}$) from the real data: 
\begin{equation}
N_{ep\pi^0 (1\gamma)} = N_{ep\pi^0}^{\rm DATA} * R_{\rm Acc}(1\gamma/2\gamma).
\end{equation}
The ratio of acceptances $R_{\rm Acc}(1\gamma / 2\gamma)$ is obtained from simulated $ep\pi^0$ events, and is defined as:
\begin{equation}\label{eq_acc_ratio}
R_{\rm Acc}(1\gamma/2\gamma)=\frac{N_{ep\pi^0 (1\gamma)}^{\rm MC}}{N_{ep\pi^0 (2\gamma)}^{\rm MC}},
\end{equation}
where $N_{ep\pi^0 (1\gamma)}^{\rm MC}$ is the number of reconstructed events obtained by applying the DVCS selection cuts to the $ep\pi^0$ simulation, and $N_{ep\pi^0 (2\gamma)}^{\rm MC}$ is the number of events obtained by applying the $ep\pi^0$ selection cuts to the same $ep\pi^0$ simulation and for the same number of generated events.

\begin{figure}
\begin{center}
\includegraphics[width=86mm]{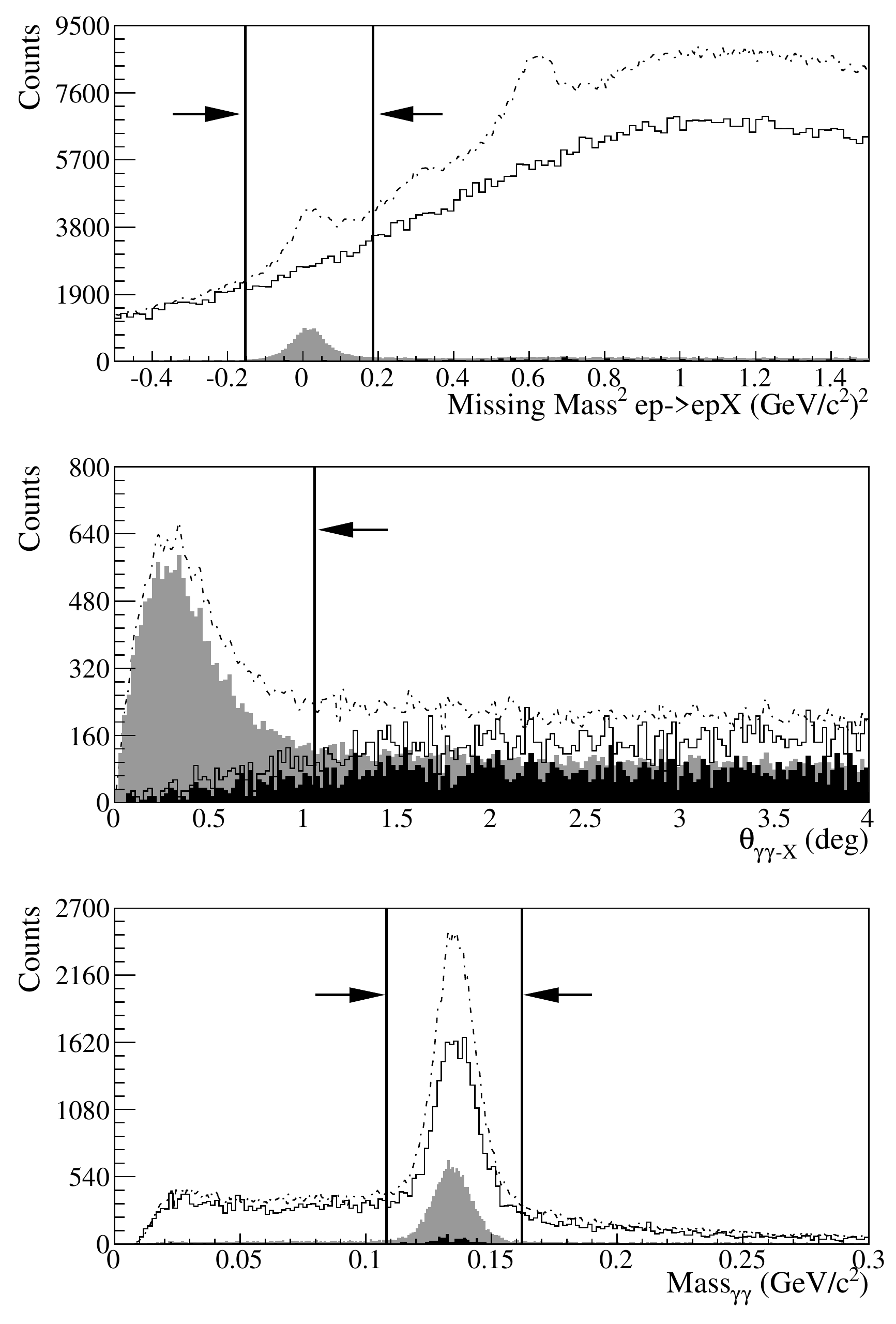}
\caption {Exclusive $\pi^0$ analysis, IC-IC topology. Effects of the $ep\pi^0$ exclusivity cuts on MM$^2(ep)$ (top), $\theta_{\pi^0 X}$ (middle), and two-photons invariant mass (bottom). The dot-dashed and solid lines show the events {\it before} exclusivity cuts for, respectively, $^{14}$NH$_{3}$ and $^{12}$C data, while the gray and black shaded plots represent the events {\it after} all exclusivity cuts but the one on the plotted variable for, respectively, $^{14}$NH$_{3}$ and $^{12}$C data. The lines and arrows show the limits of the selection cuts.}
\label{ic_pi0_cuts}
\end{center}
\end{figure}

\begin{figure}
\begin{center}
\includegraphics[width=86mm]{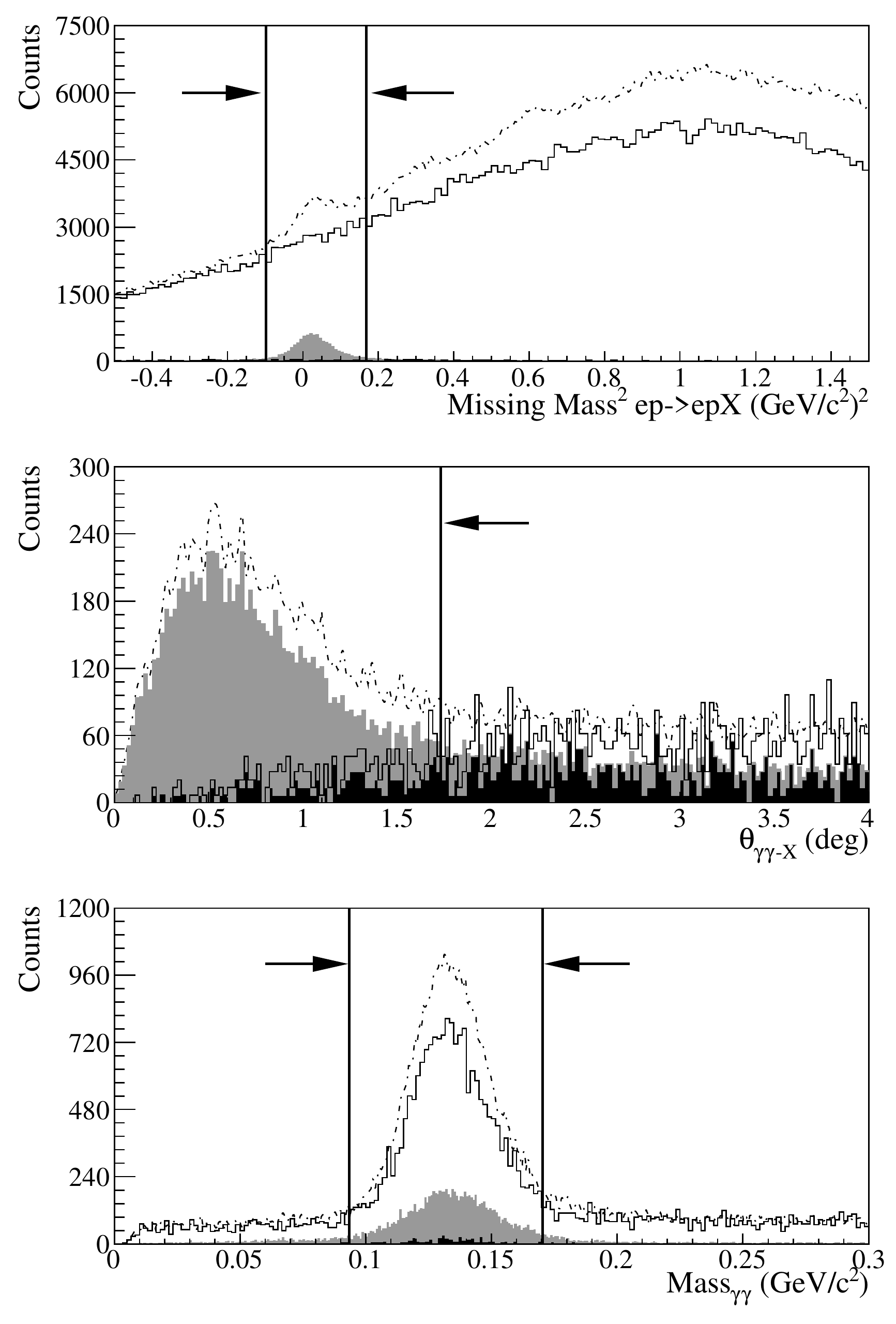}
\caption {Exclusive $\pi^0$ analysis, EC-EC topology. See caption of Fig.~\ref{ic_pi0_cuts}.}
\label{ec_pi0_cuts}
\end{center}
\end{figure}

$N_{ep\pi^0}^{\rm DATA}$ was extracted by selecting events with one electron, one proton and at least two photons using the same PID cuts as for the DVCS channel. The chosen photon pair would then be the one whose invariant mass was the closest to the nominal $\pi^0$ mass. Then, the same method as the one adopted to define the DVCS selection cuts (Section~\ref{sec_excl_cuts}) was adopted to obtain the exclusivity of the $ep\pi^0$ final state. Three ``exclusivity variables'' were used:
\begin{itemize}
\item{the missing mass squared of the $ep$ system, MM$^2(ep)$;}
\item{the angle between the measured and calculated $\pi^0$ direction, $\theta_{\pi^0 X}$;}
\item{the two-photon invariant mass, M$(\gamma\gamma)$.}
\end{itemize}
The carbon data were subtracted from the ${}^{14}$NH$_3$ data to remove possible smearing effects on the fitted peaks coming from background events due to electron scattering on ${}^{14}$N. When possible, data and Monte-Carlo distributions were fitted separately, at the ``preliminary cuts'' stage (the preliminary cuts being the same as for the DVCS selection, excluding the $E_{\gamma}>1$ GeV cut). The obtained widths were compared to the widths of the distributions after all cuts were applied to check for possible correlation effects. The three different photon-detection topologies were treated differently, as in the case of the DVCS selection. Figures~\ref{ic_pi0_cuts} and \ref{ec_pi0_cuts} show the results of the $ep\pi^0$ exclusivity cuts for the data when both photons were detected in the IC and in the EC, respectively. 

The acceptance ratio $R_{acc}$ varies for each kinematic bin, and is, on average, around 30\%. 
In absolute value, the background/signal ratios range from 1\% (at low $t$, low-mid $x_B$, high $Q^2$) to about 50\% (in the highest $t$, highest $x_B$, central $\phi$ range). However, what will impact the final asymmetries is not the size of the contamination itself, but the point-by-point difference of contamination for positive and negative target (or beam, or beam-target) polarization. 
The background subtraction was done for each individual 4-dimensional bin. The average impact of the background subtraction on each of the 3 asymmetries, relative to their average value at $90^{\circ}$, is $\sim 8\%$ for the BSA, $\sim 11\%$ for the TSA, and $\sim 6\%$ for the DSA. 

\subsection{Nuclear background effects on BSA}
The beam-spin asymmetry extracted using the ${}^{14}$NH$_3$ data contains a contribution from the nitrogen nuclei. The dilution factor analysis (Section \ref{sec_dilution}) shows that, thanks to the tight exclusivity cuts adopted, less than 10\% of the final DVCS/BH event candidates come from the nitrogen. However, given the low statistics collected in the carbon runs, it is not possible to directly subtract for each kinematic bin and helicity state the nitrogen contribution, calculated on the basis of the $^{12}$C measurement, from the ${}^{14}$NH$_3$ events. 
The raw BSA, without $\pi^{0}$ subtraction, was extracted from the carbon-target data, applying all of the DVCS/BH selection cuts and integrating, for statistics reasons, over all kinematic variables but $\phi$. The resulting asymmetry was compared to the corresponding observable obtained on the ${}^{14}$NH$_3$ target.
A Student's $t$-test was performed, and it was found that the two DVCS/BH BSAs for a free proton and quasi-free proton in carbon  are compatible at the $3\sigma$ level. 
This result points to the fact that given the limits imposed by our statistics, our experiment is not sensitive to possible medium-modification effects for DVCS, going from proton to carbon target. Moreover, our results for the dilution factor show a contamination of events from nitrogen in the $ep\gamma$ sample of less than 10\%, and this contamination comes from bound protons at quasi free kinematics due to the tight exclusive cuts. All this supports the assumption that the contribution to the final beam-spin asymmetries of the incoherent DVCS on nitrogen does not impact sizeably our result, within the limits of our statistical accuracy. 

\subsection{Combination of asymmetries from part A and part B}\label{sec_merge}
Having verified that the bin centers for parts A and B are equal within uncertainties (Section~\ref{sec_binning}), the final asymmetries, after $ep\pi^0$ background subtraction, have been combined via uncertainty-weighted averages for each four-dimensional bin in ($Q^2$, $x_B$, $-t$, $\phi$). 
In order to verify that the two sets of asymmetries are statistically compatible with each other, $t$-tests were perfomed for the three asymmetries. The tests proved that the fluctuations between part A and B are purely statistical, and the two datasets can therefore be combined. The combined asymmetries are shown in Section \ref{sec_results}.

The combined beam energy, 5.932 GeV, was computed as the weighted average over all $ep\gamma$ events of the values of the energies for the two parts, which were separately deduced from elastic measurements (Section~\ref{sec_experiment}).

\subsection{Transverse corrections}\label{transverse_sec}
In this experiment, the target polarization was parallel to the electron beam direction. However, for the theoretical interpretations of the asymmetries, the longitudinal polarization with respect to the virtual photon direction is usually adopted. 
In order to be consistent in the comparisons with the theoretical models, a model-dependent correction was computed to finally obtain the TSA (and DSA) with respect to the virtual photon direction. 
The purely longitudinal asymmetry referred to the virtual photon, $A_{\rm UL}$, is linked to the $x$-component of the transverse one, $A_{\rm UT}(0)$, and to what is measured in the lab, $A_{\rm UL}^{\rm lab}$, by the relation \cite{diehl}: 
\begin{equation}\label{eq_transverse}
A_{\rm UL} = \frac{A_{\rm UL}^{\rm lab}}{\cos\theta^*} + \tan\theta^* A_{\rm UT}(0), 
\end{equation}
where $\theta^*$ is the angle formed by the virtual photon and the beam direction. An analogous relation holds for the double-spin asymmetry,
\begin{equation}\label{eq_transverse_dsa}
A_{\rm LL} = \frac{A_{\rm LL}^{\rm lab}}{\cos\theta^*} + \tan\theta^* A_{\rm LT}(0).
\end{equation}
The idea adopted here is to compute a set of model-dependent bin-by-bin corrections using the predictions of the GPD model VGG \cite{vgg} (described in Section~\ref{sec_gpd_models}) for $A_{UT}(0)$, and the average value of $\theta^*$ over the selected DVCS events for each bin. The VGG model was run for all kinematic bins to obtain $A_{UT}(0)$, with three different sets of options for systematic studies of the model dependence of the correction. 
The size of the transverse-target spin asymmetry varies with $-t$, from a few percent at low $-t$ to up to 30\% for the highest $-t$ bins. Also the difference between the results obtained for the three sets of VGG options is more sizeable at high $-t$, where changes of sign for the $\phi$ distribution are observed. As far as the double-transverse spin asymmetry is concerned, it is higher at low $-t$, mostly positive, and never greater than 30\%. 
The angle $\theta^*$ between the virtual photon and the beam direction is, on average, around $7^{\circ}$. 
Thanks to the combination of small values for $\theta^*$ and relatively small values for $A_{\rm UT}(0)$, the corrections, which are defined as 
\begin{eqnarray}
c_{A_{\rm UT}} & = & A_{\rm UL} - A_{\rm UL}^{\rm lab}\\
c_{A_{\rm LT}} & = & A_{\rm LL} -  A_{\rm LL}^{\rm lab}
\end{eqnarray}
and obtained using Eqs.~\eqref{eq_transverse} and~\eqref{eq_transverse_dsa}, are very small. 
The standard deviation of the differences between the three versions of ``corrected'' TSAs (or DSAs) and the measured one was adopted as a systematic uncertainty. 
In general, the correction brings a slight increase to both kinds of asymmetries, and it is always smaller than the statistical uncertainties. The same holds for the associated systematic uncertainty, the averages of which for both TSA and DSA are reported in Table~\ref{table_syst}. The transverse correction was applied to each 4-dimensional bin. The average impact of this correction on each of the 2 asymmetries, relative to their average value at $90^{\circ}$, is $\sim 4\%$ for the TSA and $\sim 2\%$ for the DSA. The values of the corrections for each 4-dimensional bin are reported in Table~\ref{table_all_data} along with the values of the corrected asymmetries.

\subsection{Bin-centering corrections}\label{sec_bcc}
Second-order bin-centering corrections were applied to the single-spin asymmetries. As stated in Section~\ref{sec_binning}, the asymmetries are correctly defined at first order if one takes the average value of each kinematic variable over all the events of each bin. 
To apply second-order bin-centering corrections, one needs to make assumptions on the kinematic dependences of the asymmetry. Based on the information from the present data, only the effect of the $-t$ dependence was studied, since the asymmetries do not exhibit strong variations in $Q^2$ and $x_B$ (see Section~\ref{sec_results}). 
The bin-centering corrections were evaluated for each 4-dimensional bin as
\begin{equation}
c_{\rm BCC}=\frac{\alpha(t)\sin\phi}{1+\beta(t)\cos\phi}-\frac{\alpha(\langle t\rangle )\sin \langle \phi\rangle }{1+\beta(\langle t\rangle )\cos \langle \phi\rangle },
\end{equation} 
where the first term is computed event-by-event while the second term is evaluated at the central values $\langle t \rangle$ and $\langle \phi \rangle$ of the bin. The $\alpha(t)$ and $\beta(t)$ functions were determined by fitting the data.

For the BSA and the TSA, the corrections are on average at the level of 4\%, relative to the average value of the asymmetries at $90^{\circ}$.
Since the extrema in the DSA are much less pronounced, the corrections are expected to be even smaller and they are therefore neglected here. 

\subsection{Systematic uncertainties}\label{sec_systematics}
Systematic checks were performed to evaluate the stability of the measured observables against the variation of the terms that compose them. 
For each of the three asymmetries extracted, and for each kind of systematic effect studied, one of the quantities in the definition of the asymmetry was varied two or more times. Then, the asymmetries were produced for each variation, for parts A and B separately. The two parts were merged as described previously.  
Finally, the systematic uncertainty for each bin was computed as
\begin{equation}\label{eq_syst}
\sigma_{syst}= \sqrt{\frac{\sum_{i=1}^n(A_i - A_0)^2}{n}},
\end{equation}
where $A_i$ are the asymmetries corresponding to each variation $i$ out of $n$ variations performed, while $A_0$ is the standard asymmetry. 
Each systematic check was performed as follows:
\begin{itemize}
\item{Systematics on the exclusivity cuts: the ``standard'' analysis was performed with $3\sigma$ cut widths (see Section~\ref{sec_excl_cuts}), and four others were done, using cut widths of 2.25, 2.5, 3.25 and 3.5 $\sigma$. The bin-by-bin systematics were computed using Eq.~\eqref{eq_syst}. This source is the biggest contributor to the overall systematic uncertainty (see Table~\ref{table_syst}), and it encompasses also effects due to variations of the dilution factor.}
\item{Systematics on $P_b$, $P_t$, $P_bP_t$: the beam-spin asymmetry was computed two more times, taking two different values of $P_b$: $P_b+\Delta(P_b)$ and $P_b-\Delta(P_b)$, where $\Delta(P_b)$ is the statistical uncertainty that was estimated on this quantity (see Section~\ref{sec_pbpt}). An equivalent treatment was adopted for the TSA and the DSA. 
There is no major kinematic dependence for this source of uncertainty, and its contribution to the overall systematic uncertainty is very small (see Table~\ref{table_syst}). }
\item{Systematics on $ep\pi^0$ background subtraction:
the three asymmetries (BSA, TSA and DSA) were computed each three times, taking three different values of the $R_{Acc}$ factor (defined in Eq.~\eqref{eq_acc_ratio}) that was used to compute the $ep\pi^0$ background. Specifically, the background was computed using the ``real'' value $R_{Acc}$ increased and decreased by 30\%. There is some kinematic dependence for this source of uncertainty, and its contribution to the overall systematic uncertainty is smaller than that from the exclusivity cuts, but larger than those from the polarizations or the dilution factors.}
\end{itemize}
Table~\ref{table_syst} reports the averages of each kind of systematic uncertainty for three three asymmetries. The biggest contribution to the systematic uncertainties comes from the exclusivity cuts which contributes to $\sim80\%$ of the total systematic error. 
The total systematic uncertainty was computed as a quadratic sum of all contributions. For all bins and for the three kinds of asymmetries, the statistical uncertainty is bigger than the total systematic uncertainty. Both kinds of uncertainties are listed, along with the values of the asymmetries, in Table~\ref{table_all_data}.
\begin{table}
\begin{tabular}{|c|c|c|c|} 
\hline
Source & BSA & TSA & DSA \\
\hline
$\pi^0$ background & 0.005 & 0.005 & 0.009 \\
Polarization & 0.003 & 0.004 & 0.008 \\
Exclusivity cuts & 0.021 & 0.019 & 0.027  \\
Transverse correction & N.A. & 0.006 & 0.006 \\
\hline
\end{tabular}
\caption{Average systematic uncertainties  for each source of uncertainty and for each asymmetry type. ``Polarization'' stands for $P_b$, $P_t$, $P_bP_t$ for BSA, TSA and DSA, respectively.
}
\label{table_syst}
\end{table}
\subsubsection{Radiative corrections}
Afanasev {\it et al.} \cite{afanasev} have computed the radiative corrections for the DVCS and BH processes for CLAS kinematics. It was found that, given the strict kinematic cuts adopted to select the final state, the undetected radiated photon can only have small energies. In this case, therefore, the main contribution to the radiative correction comes from spin-independent soft-photon emission that does not affect the polarization observables - while instead it can affect unpolarized cross sections even up to the 20\% level. The approximation of negligible contribution from the radiative corrections to the BSA, TSA and DSA, compared to the size of the asymmetries, is valid at CLAS kinematics at the 0.1\% level \cite{afanasev}. Given the statistical uncertainties and the larger size of other systematic effects, it was chosen to neglect this contribution. 

\section{Models of GPDs}\label{sec_gpd_models}
In the following sections the experimental asymmetries are compared to the predictions of four GPD models:  Vanderhaeghen-Guichon-Guidal (VGG) \cite{vgg}, Goloskokov-Kroll (GK) \cite{kroll}, Kumericki-M\"uller (KMM12) \cite{kreso}, and Goldstein-Gonzalez-Liuti (GLL) \cite{liuti}. 

Both the VGG and GK models are based on double distributions \cite{radyuskin2,muller} to parametrize the $(x,\xi)$ dependence of the GPDs, and on Regge phenomenology for their $t$ dependence. The main differences between these two models are in the parametrization of the high-$t$ part of the electromagnetic form factors and in the fact that the parameters of the GK model are tuned using low-$x_B$ DVMP data from HERA, which are particularly sensitive to gluon and sea-quark GPDs. Therefore, the GK model is also suited for gluonic GPDs that are not accounted for in VGG. However, given the $x_B$ range of the results presented here, the description of the valence-quark GPDs is sufficient. 

KMM12 is a hybrid model designed for global fitting, in which sea-quark GPDs are modeled in a Mellin-Barnes integral representation; valence quarks are modeled in terms of these GPDs on the cross-over line $\xi = x \sim x_B/(2-x_B)$. The parameters of the model were fixed using unpolarized-proton DVCS data from CLAS and HERMES, as well as the polarized-proton HERMES data. The kinematic range of applicability of this model is defined by the relation $-t<\frac{Q^2}{4}$.

Finally, the GGL model provides a diquark model based parametrization of GPDs that incorporates Regge behavior by introducing a spectral function for the spectator diquark's invariant mass distribution. The parameters of the model were obtained by fitting both deep inelastic scattering (DIS) structure functions and the recently available flavor-separated nucleon form factors data \cite{cates}.

\section{Results}\label{sec_results}

Hereafter, the results for the three asymmetries are presented, discussed and compared to the GPD models described in Section~\ref{sec_gpd_models}. The values of the asymmetries for each 4-dimensional bin, along with their uncertainties, are listed in Table~\ref{table_all_data} and in Ref.~\cite{clasdb}.

The harmonic structure in $\phi$ of the asymmetries versus ($Q^2$-$x_B$) and $-t$ was studied by fitting their $\phi$ distributions. For this goal, the fact that the three asymmetries have the same denominator has been exploited.
In fact, by performing a simultaneaous fit of the 3 asymmetries, the common denominator can be constrained to be the same for the three different observables.
Thus BSA, TSA and DSA were fitted, respectively, with three functions that shared a common denominator $(1+\beta\cos\phi$). In the highest $-t$ bin of the third ($Q^2$-$x_B$) bin, $\beta$ was set to zero due to the limited $\phi$ coverage.

\subsection{Beam-spin asymmetry}\label{sec_results_bsa}

\begin{figure*}
\begin{center}
\includegraphics[width=172mm]{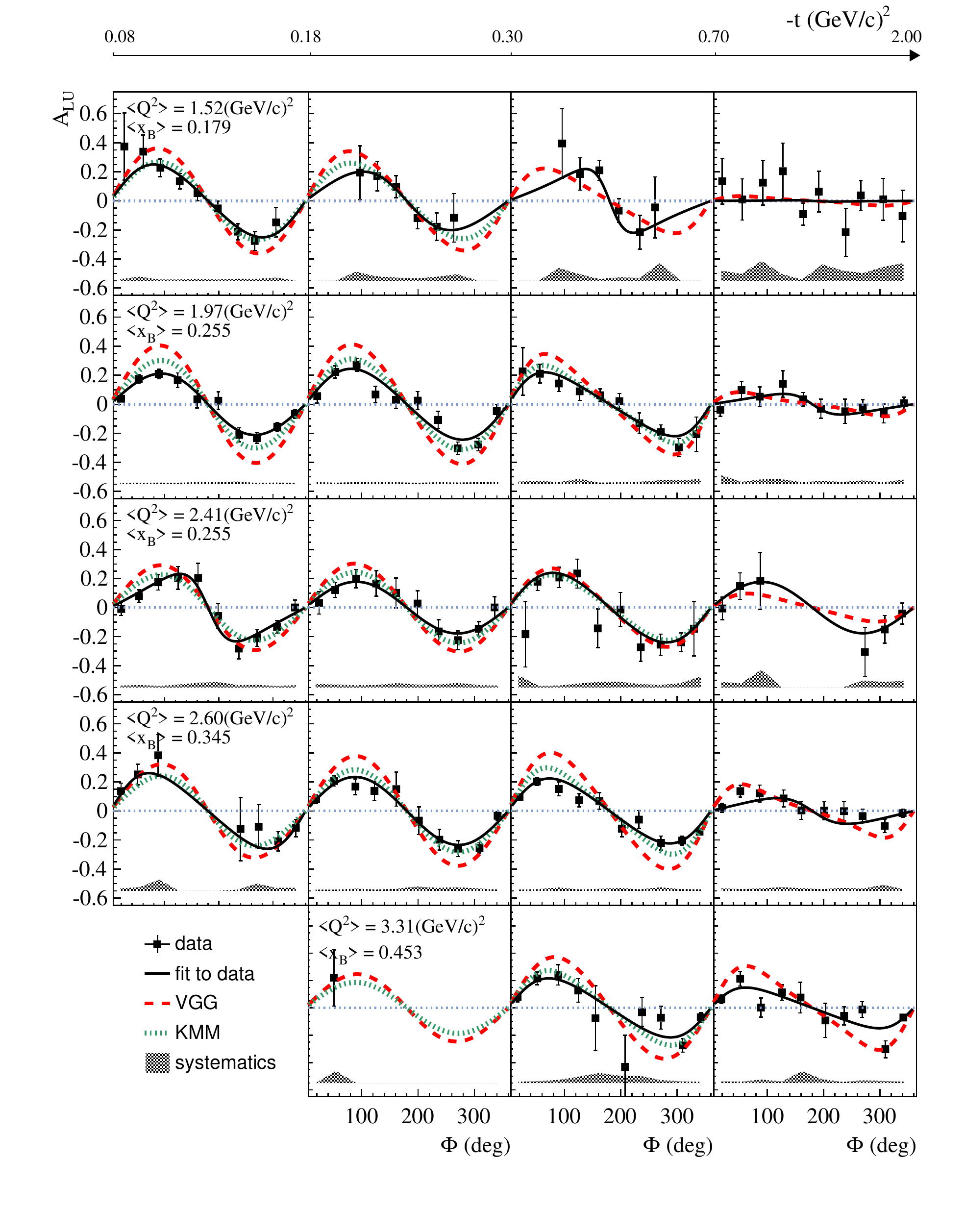}
\caption {(Color online) Beam-spin asymmetry for the reaction $ep \to e'p'\gamma$ as a function of $\phi$ for the various $Q^2$-$x_B$ (rows) and $-t$ (columns) bins. The point-by-point systematic uncertainties are represented by the shaded bands. The solid black curve is the fit with the function in Eq.~\eqref{bsa_fit_func}. In the highest $-t$ bin of the third ($Q^2$-$x_B$) bin, $\beta$ was set to zero due to the limited $\phi$ coverage, while no fit is performed on the
first $-t$ bin of the highest ($Q^2$-$x_B$) bin, where only one data point is present. The curves show the predictions of the VGG \cite{vgg} (red-dashed) and KMM12 \cite{kreso} (cyan-dotted) models.}
\label{bsa_phi}
\end{center}
\end{figure*}

\begin{figure}
\begin{center}
\includegraphics[width=86mm]{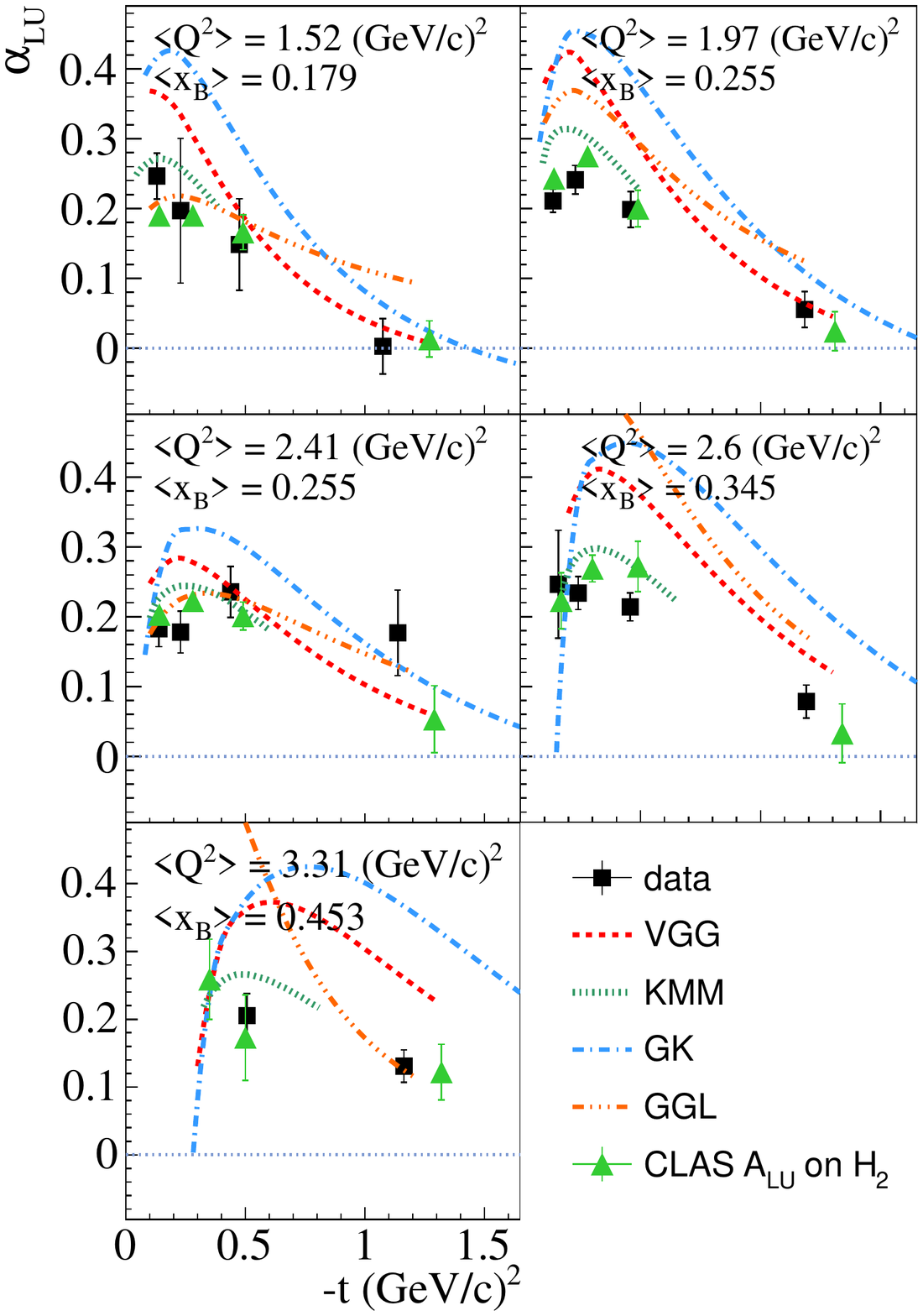}
\caption {(Color online) $t$ dependence for each $Q^2$-$x_B$ bin of the $\alpha_{\rm LU}$ term of the beam-spin asymmetry. The curves show the predictions of four GPD models for the BSA at $\phi=90^o$: i) VGG \cite{vgg} (red dashed), ii) KMM12 \cite{kreso} (blue dotted), iii) GK \cite{kreso} (blue dash-dotted), and iv) GGL \cite{liuti} (orange dashed-three-dotted). The square black points are the results obtained from the present analysis, the triangular green
data come from the previous CLAS experiment with unpolarized proton target \cite{fx}.}
\label{bsa_tdep}
\end{center}
\end{figure}

Figure \ref{bsa_phi} shows the beam-spin asymmetry as a function of $\phi$ for each slice in the $Q^2$-$x_B$ space (rows) and for each bin in $-t$ (columns). Each asymmetry is fitted with the function
%
%
\begin{equation}\label{bsa_fit_func}
\frac{\alpha_{\rm LU}\sin\phi}{(1+\beta\cos\phi)}
\end{equation}
and shows a clear $\sin\phi$-like modulation, with a decreasing amplitude as $-t$ increases, ranging from $\sim 0.25$ down to $\sim 0$. The dependence in the other kinematic variables appears less marked, although a slight decrease in the $-t$ slope seems
to happen at the highest $Q^2$ and $x_B$ values. This is confirmed by Fig.~\ref{bsa_tdep}, which shows the beam-spin asymmetry at $90^{\circ}$ (i.e. the $\alpha_{\rm LU}$ coefficient in Eq.~\eqref{bsa_fit_func})
as a function of $-t$ for each $Q^2$-$x_B$ bin. The choice of the fitting function was motivated by the physics (see Eq.~\eqref{eq_bsa}).
%
The data are compared to the predictions of the VGG, GK, KMM12 and GGL models. 
As expected, the VGG and GK models don't show strong differences between each other. With respect to the data, they overestimate the amplitude of the experimental asymmetries especially at low $-t$ and at low $Q^2$-$x_B$. At the highest $-t$ values, the VGG model gets closer to the data, while the GK model is systematically higher. Both models expect a steeper $-t$ slope than the data display. This can be due to the fact that these models are based on Double-Distributions, where the $-t$ dependence is
factorized with respect to the $(x, \xi)$ dependence. The data, instead, seem to point to more complex correlations between the three variables.
The GGL model is in good agreement with the data at low $-t$ especially for the first and third $Q^2$-$x_B$ bin, while it diverges away from the data in the high-$x_B$ bins. The discrepancy observed for larger $x_B$ values is an indication that using only DIS and form factor data one can only provide a loose constraint on the $\xi$ dependence of the model. The best fit to the data is provided by the KMM12 model, which however does not cover our whole set of kinematics. 
For consistency, our beam-spin asymmetries were also compared to those obtained from previous CLAS data \cite{fx} (e1-dvcs experiment). For this task, the results for the $\alpha_{\rm LU}$ coefficient were used, taking the kinematic bins from the e1-dvcs data that were closest to our own. The comparison is shown in Fig.~\ref{bsa_tdep}, where the e1-dvcs results are represented by the (green online) triangles. The agreement is good, especially considering the imperfect kinematical overlap.

\subsection{Target-spin asymmetry}\label{sec_results_tsa}
\begin{figure*}
\begin{center}
\includegraphics[width=172mm]{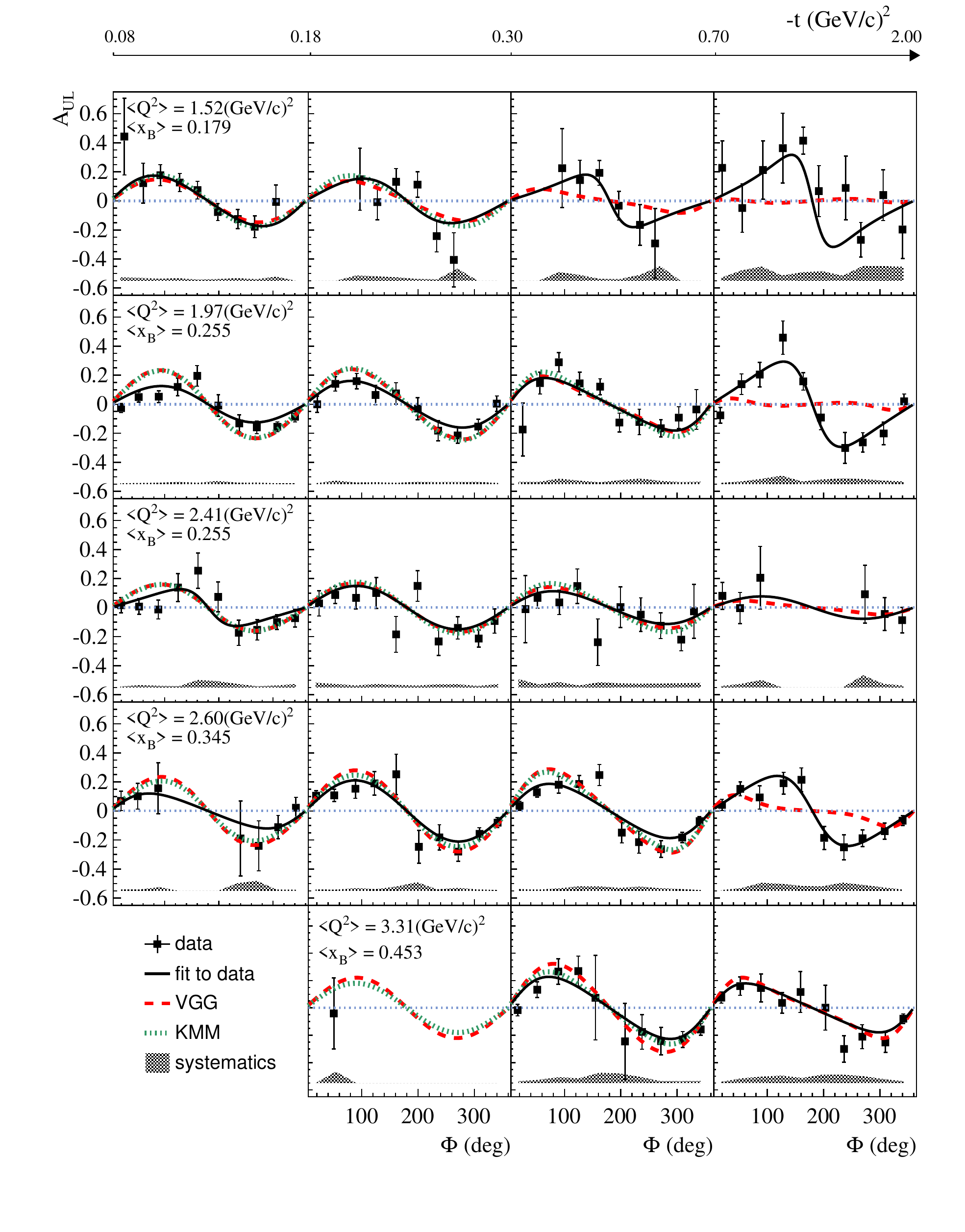}
\caption {(Color online) Target-spin asymmetry for the reaction $ep \to e'p'\gamma$ as a function of $\phi$ for the various $Q^2$-$x_B$ (rows) and $-t$ (columns) bins. The point-by-point systematic uncertainties are represented by the shaded bands. The solid black curve is the fit with the function in Eq.~\eqref{tsa_fit_func}. In the highest $-t$ bin of the third ($Q^2$-$x_B$) bin, $\beta$ was set to zero due to the limited $\phi$ coverage, while no fit is performed on the
first $-t$ bin of the highest ($Q^2$-$x_B$) bin, where only one data point is present. The curves show the predictions of the VGG \cite{vgg} (red-dashed) and KMM12 \cite{kreso} (blue-dotted) models.}
\label{tsa_phi}
\end{center}
\end{figure*}

\begin{figure}
\begin{center}
\includegraphics[width=86mm]{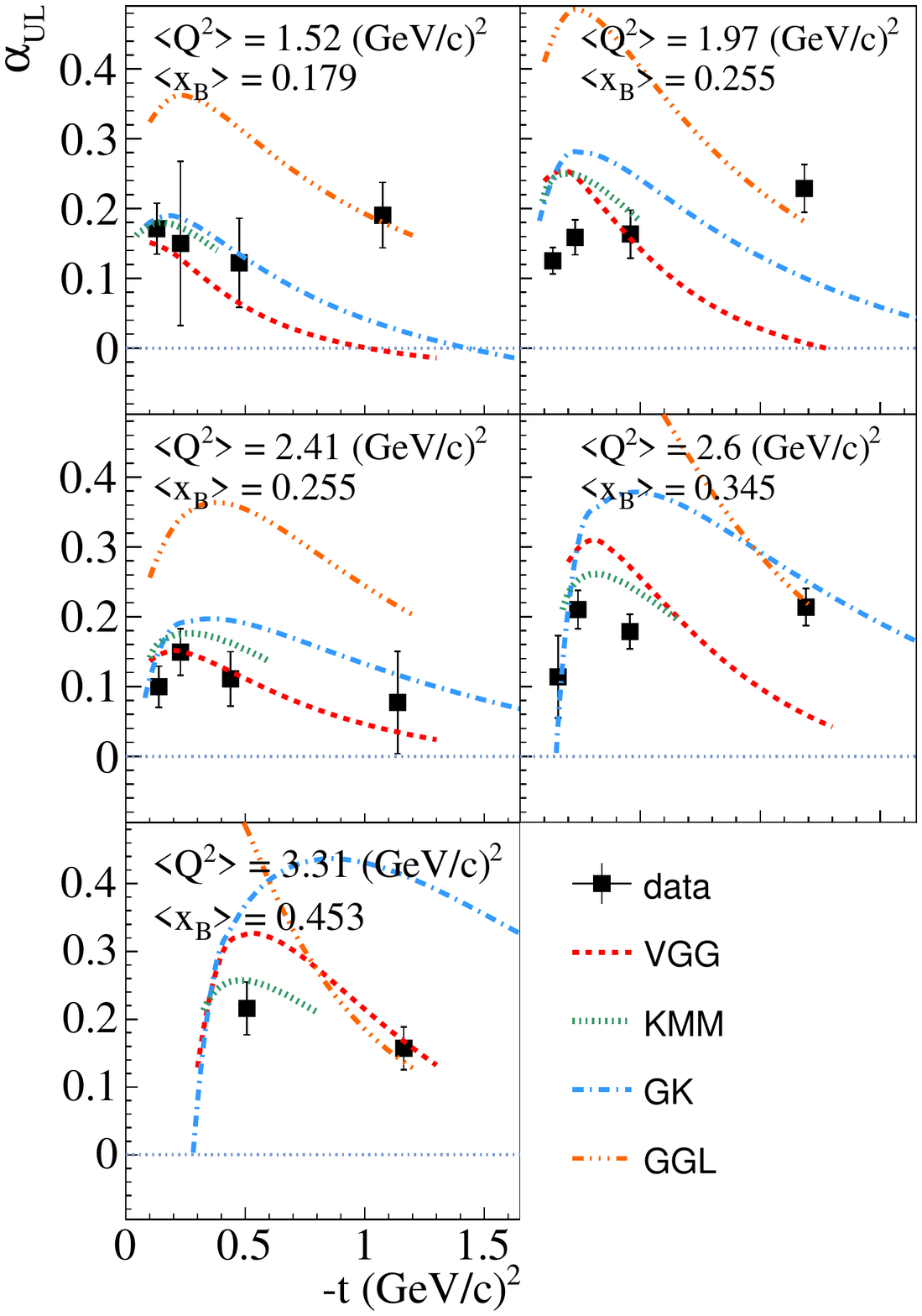}
\caption {(Color online) $t$ dependence, for each $Q^2$-$x_B$ bin, of the $\alpha_{\rm UL}$ term of the target-spin asymmetry. The curves show the predictions of four GPD models for the TSA at $\phi=90^o$: i) VGG \cite{vgg} (red dashed), ii) KMM12 \cite{kreso} (cyan dotted), iii) GK \cite{kreso} (blue dash-dotted), and iv) GGL \cite{liuti} (orange dashed-three-dotted).}
\label{tsa_tdep}
\end{center}
\end{figure}

The results for the target-spin asymmetry \cite{prl_erin} are presented in Fig.~\ref{tsa_phi} as a function of $\phi$ for each slice in the $Q^2$-$x_B$ space (rows) and for each bin in $-t$ (columns). As for the BSA, it is fitted with the function
%
%
\begin{equation}\label{tsa_fit_func}
\frac{\alpha_{\rm UL}\sin\phi}{1+\beta\cos\phi}
\end{equation}
and shows the typical $\sin\phi$-like dependence, with amplitudes ranging from 0.1 to 0.3, but its evolution with $-t$ is quite different from the BSA, in shape and magnitude. In fact, the amplitude of the target-spin asymmetry seems rather constant as a function of all kinematic variables, $-t$ included, apart from the expected systematic drop towards $t\sim t_{min}$. 
Figure~\ref{tsa_tdep} shows the $t$-dependence for each bin in $Q^2$-$x_B$ of the $\sin\phi$ fitting coefficient $\alpha_{\rm UL}$ (Eq.~\eqref{tsa_fit_func}), which appears fairly constant, unlike what was observed for the beam-spin asymmetry. As mentioned
above, the variable $t$ yields the Fourier conjugate of the impact parameter, describing the transverse position of the partons in the reference frame where the proton goes at the speed of light. Therefore, a steep $t$ slope is equivalent to a rather flat spatial distribution, and vice versa. Equations~\eqref{eq_tsa} and \eqref{eq_bsa} point to the proportionality between, respectively, TSA and $\Im{\rm m}\widetilde{\cal {H}}$ and BSA and $\Im{\rm m}\cal{H}$. Thus, the $t$ behavior of the TSA compared to that of the BSA suggests that the axial charge (linked to $\tilde{H}$) is more concentrated in the center of the proton than the electric charge (linked to $H$). This fact was already observed in a paper \cite{fitmick2} devoted to the extraction of the CFFs $\Im{\rm m}\cal{H}$ and $\Im{\rm m}\widetilde{\cal {H}}$ from the HERMES data. 
This finding is clearly not predicted by the VGG or GK models, which instead display a similar drop with $t$ for the TSA as what was computed for the BSA. These models approximately reproduce the low-$t$ magnitude of the asymmetry in some kinematics (namely, in $Q^2$-$x_B$ bins 1 and 3), with a slightly better fit of the data for VGG. GK predicts an increase of the TSA with $x_B$ that is not observed in the experiment. 
By comparing the behavior of these two models for the two single-spin asymmetries, it can be observed how both reach good agreement with the data at high $-t$ for the BSA and a low $-t$ for the TSA. These data can therefore provide strong guidance to correct the $t$ dependence of the parametrizations of both $H$ and $\widetilde{H}$.
The GGL model also predicts a drop in $-t$ not confirmed by the data, and moreover it overestimates the magnitude of the asymmetry by at least a factor of 2. 
The best fit to the data is provided, in the bins where it applies, by the KMM12 model. 

\begin{figure}
\begin{center}
\includegraphics[width=86mm]{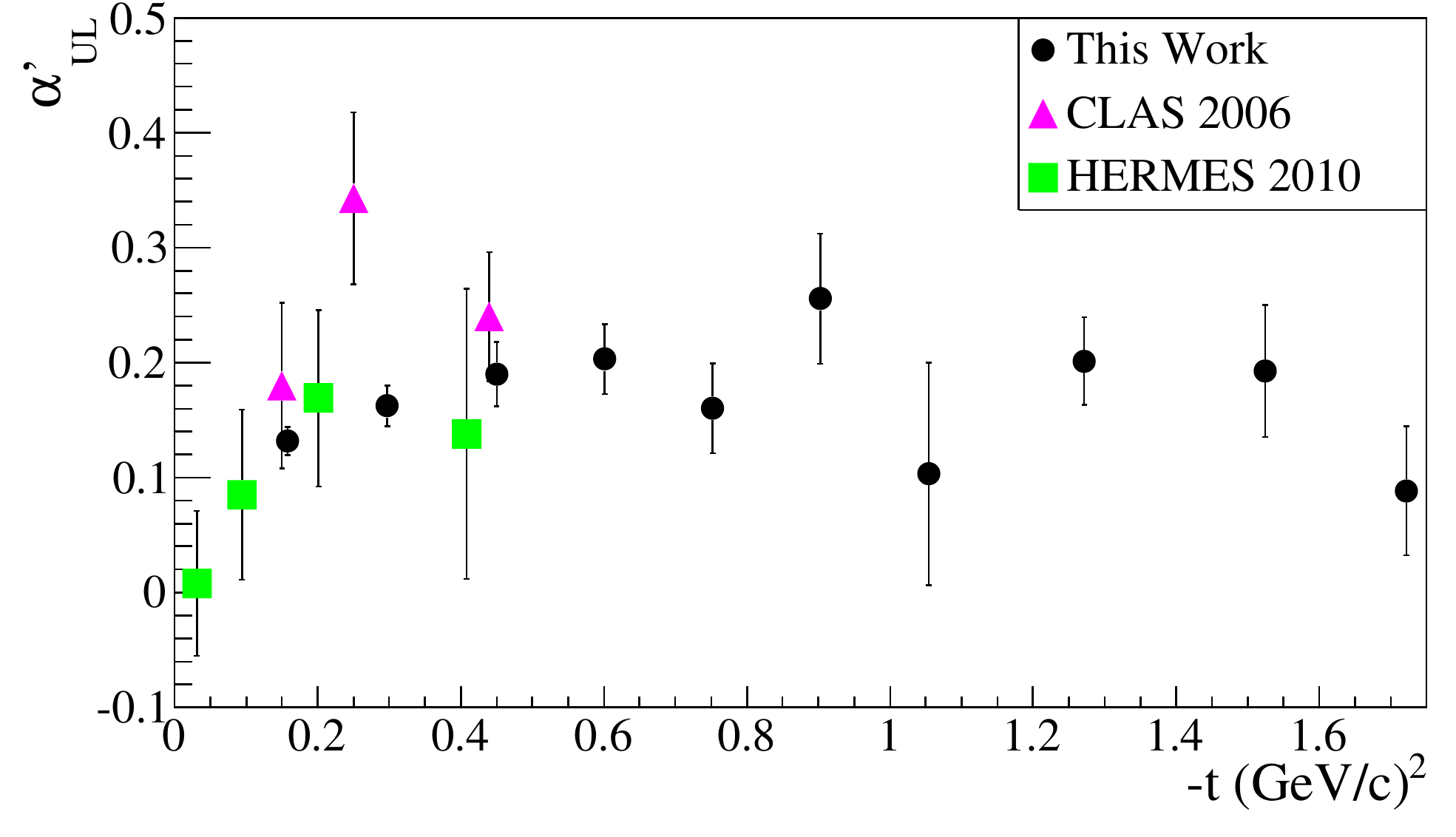}
\caption {(Color online) Comparisons of the $t$ dependences of the $\sin\phi$ term of the $ep\gamma$ target-spin asymmetries for the present data, integrated over $Q^2$ and $x_B$ (black circles), the previous CLAS experiment \cite{shifeng} (magenta triangles), and HERMES \cite{hermes} (green squares).}
\label{tsa_comparison}
\end{center}
\end{figure}

Our target-spin asymmetries were compared to those obtained from HERMES and to the results of a previous exploratory CLAS experiment \cite{shifeng}. Given the different kinematical coverages of the three experiments, it was decided to integrate the data over all values of $Q^2$-$x_B$ of the phase space, and obtain a set of TSAs only as a function of $\phi$ and $-t$, with a finer binning in $-t$ than for our four-fold asymmetries. For this comparison, each $\phi$ distribution for a given $-t$ bin was fitted with the function $\alpha'_{\rm UL}\sin\phi+\gamma_{\rm UL}\sin(2\phi)$, which was the one used for both the HERMES and the old CLAS data.
The comparison of the $\alpha'_{\rm UL}$ coefficients is shown in Fig.~\ref{tsa_comparison}. 
Our data have at least a factor 5 smaller error bars than the previously published data, and extend the $-t$ range up to 1.7 (GeV/$c$)$^2$. 

A further test was made including in the fit an additional $\sin(2\phi)$ term for the TSA. Thus the adopted fitting function was
%
%
\begin{equation}\label{tsa_fit_func_sin2phi}
\frac{\alpha_{\rm UL}\sin\phi+\gamma_{\rm UL}\sin(2\phi)}{1+\beta\cos\phi} ~.
\end{equation}
The data from HERMES \cite{hermes}, in fact, show a non-negligible contribution of the $\sin 2 \phi$ moment to the TSA. It must be pointed out, however, that in the expansion in $\sin\phi$ moments done by the HERMES Collaboration, the contribution of the denominator could mix with the $\sin n \phi$ terms, while in our analysis it is treated with its own parameter. In the Belitsky-M\"uller-Kirchner formalism, the parts of the DVCS/BH interference term depending on $\sin 2 \phi$ appear only at twist-three level \cite{belitski}. Therefore, observing a sizeable $\sin 2\phi$ component in the target-spin asymmetry would mean having some sensitivity to twist-three CFFs. A recent paper \cite{courtoy} pointed to a possible way to access the quarks' orbital angular momentum via the measurement of twist-three GPDs. 
First of all, the stability of $\alpha_{\rm UL}$ was verified by comparing the $\sin\phi$ parameter obtained with and without $\sin 2 \phi$ term in the fitting function. The $\sin 2\phi$ term appears to be much smaller, at least in the low-$t$ region, than the $\sin\phi$ term, often compatible with zero, with a slight tendency to increase at high $-t$ in some kinematic bins towards negative values. The $\sin\phi$ component is always dominant. However, given the limited statistics and the relatively small number of $\phi$ bins, the uncertaintes on the denominator parameter highly affect the extraction of such a potentially small higher-twist $\sin 2\phi$ modulation in the numerator, so no reliable  $\sin 2\phi$-extraction can be achieved with the present fitting procedure. 

\subsection{Double-spin asymmetry}\label{sec_results_dsa}
\begin{figure*}
\begin{center}
\includegraphics[width=172mm]{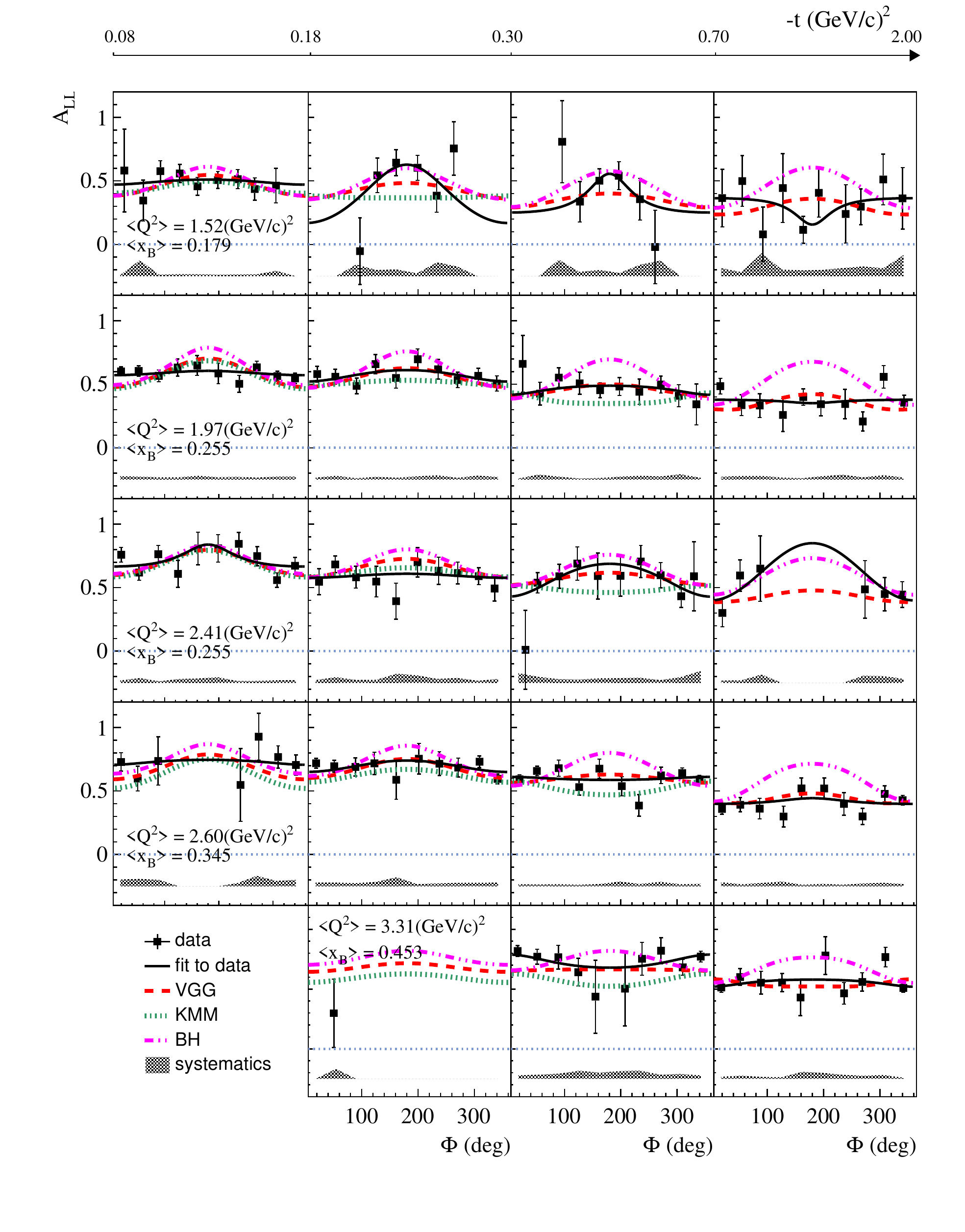}
\caption {(Color online) Double-spin asymmetry for the reaction $ep \to e'p'\gamma$ as a function of $\phi$ for the various $Q^2$-$x_B$ (rows) and $-t$ (columns) bins. The point-by-point systematic uncertainties are represented by the shaded bands. The solid black curve is the fit with the function in Eq.~\eqref{dsa_fit_func}. In the highest $-t$ bin of the third ($Q^2$-$x_B$) bin $\beta$, was set to zero due to the limited $\phi$ coverage, while no fit is performed on the
first $-t$ bin of the highest ($Q^2$-$x_B$) bin, where only one data point is present. The red-dashed and cyan-dotted curves are predictions of the VGG and KMM12 models, respectively. The pink two-dot-dashed curves are the calculations for the Bethe-Heitler process.}
\label{dsa_phi}
\end{center}
\end{figure*}

\begin{figure}
\begin{center}
\includegraphics[width=86mm]{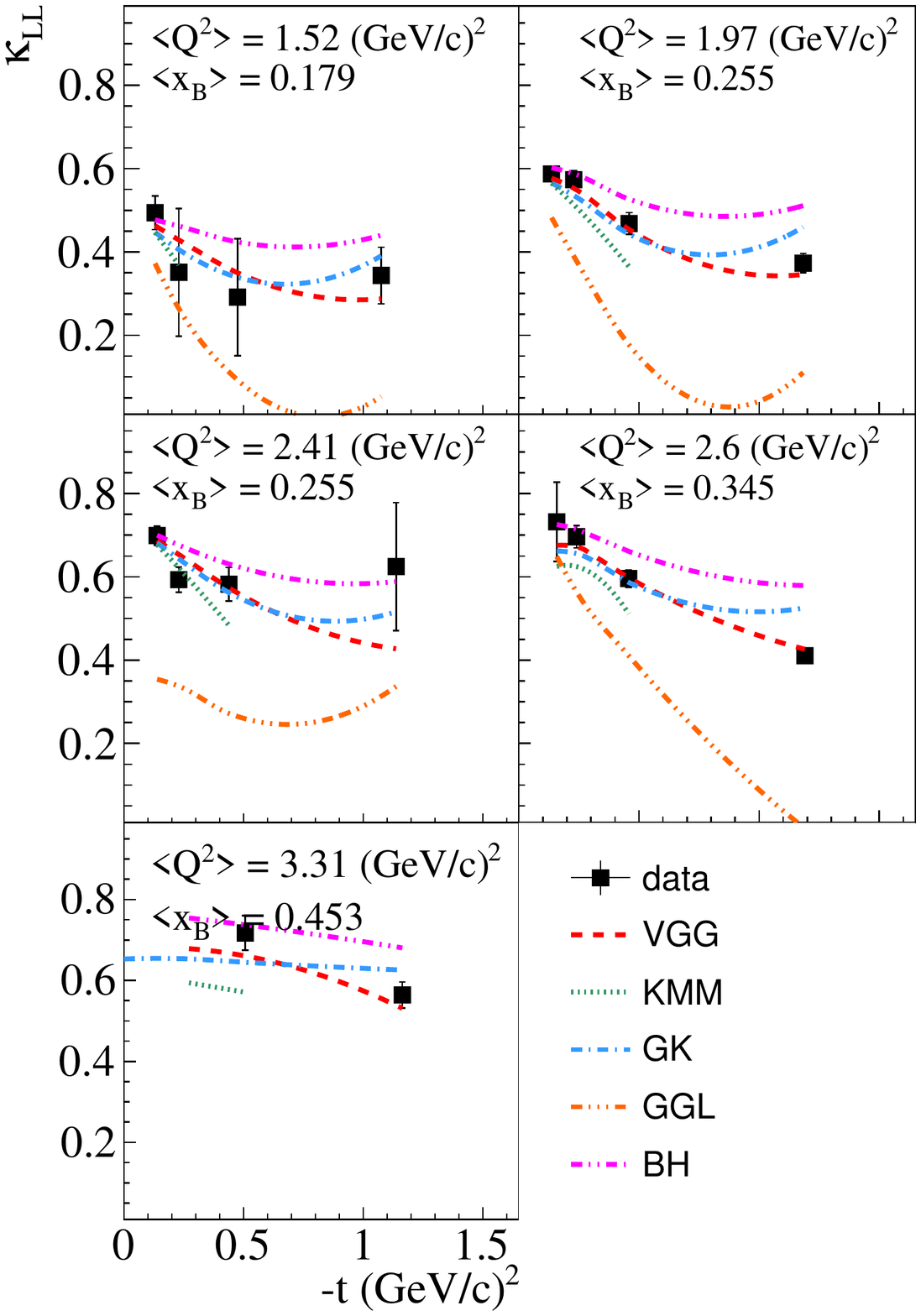}
\caption {(Color online) $t$ dependence for each $Q^2$-$x_B$ bin of the constant term $\kappa_{\rm LL}$ of the double-spin asymmetry. The pink two-dot-dashed curves are the calculations of the DSA for the Bethe-Heitler process alone. The curves show the predictions for the full $ep\gamma$ amplitude of four GPD models: i) VGG \cite{vgg} (red dashed), ii) KMM12 \cite{kreso} (cyan dotted), iii) GK \cite{kreso} (blue dash-dotted), and iv) GGL \cite{liuti} (orange dashed-three-dotted).}
\label{dsa_tdep_const}
\end{center}
\end{figure}

\begin{figure}
\begin{center}
\includegraphics[width=86mm]{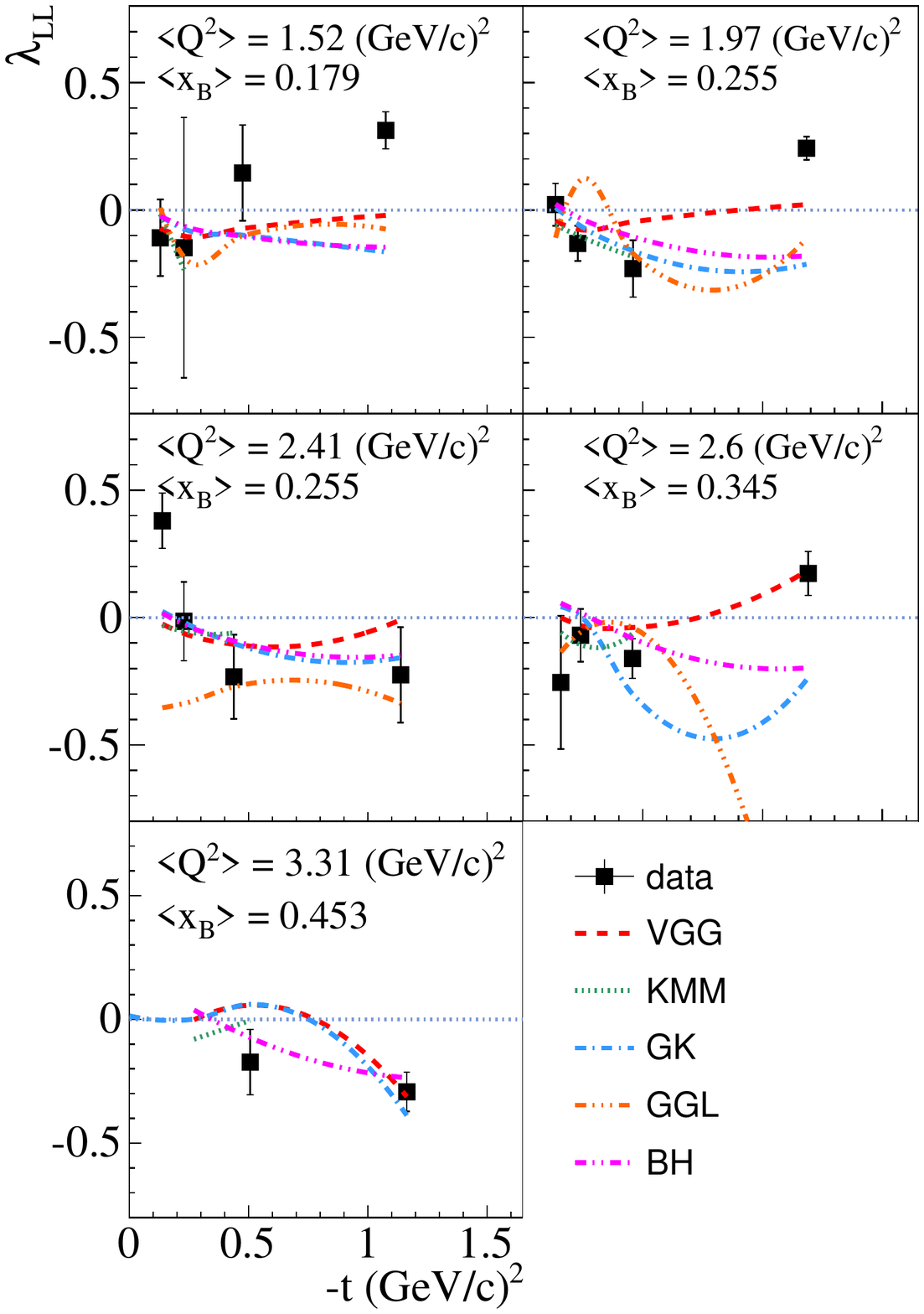}
\caption {(Color online) $t$ dependence for each $Q^2$-$x_B$ bin of the $\cos\phi$ term $\lambda_{\rm LL}$ of the double-spin asymmetry. The pink two-dot-dashed curves are the calculations of the DSA for the Bethe-Heitler process alone. The curves show the predictions for the full $ep\gamma$ amplitude of four GPD models: i) VGG \cite{vgg} (red dashed), ii) KMM12 \cite{kreso} (cyan dotted), iii) GK \cite{kreso} (blue dash-dotted), and iv) GGL \cite{liuti} (orange dashed-three-dotted).}
\label{dsa_tdep_cosphi}
\end{center}
\end{figure}

The double-spin asymmetry is plotted in Fig.~\ref{dsa_phi} as a function of $\phi$ for each bin in $-t$ (columns) and for each slice in the $Q^2$-$x_B$ space (rows). It is larger in magnitude than the single-spin asymmetries presented in the previous sections (around 0.6), seems rather flat as a function of $\phi$, and presents a slow decrease as a function of $-t$.
The data were fitted with the function
%
%
\begin{equation}\label{dsa_fit_func}
\frac{\kappa_{\rm LL} + \lambda_{\rm LL}\cos\phi}{1+\beta\cos\phi} ~,
\end{equation}
where the denominator parameter $\beta$ is the same as for the fits to the two single-spin asymmetries.
The two sets of fit parameters of the numerator, $\kappa_{\rm LL}$ and $\lambda_{\rm LL}$, are shown, as functions of $-t$ and for each $Q^2$-$x_B$ bin in Figs.~\ref{dsa_tdep_const} and \ref{dsa_tdep_cosphi}, respectively. The constant term dominates the asymmetry, while the $\cos\phi$ term of the numerator is compatible with zero for most kinematics. In Figs.~\ref{dsa_phi}, \ref{dsa_tdep_const}, and ~\ref{dsa_tdep_cosphi}, the three fit parameters of the double-spin asymmetries are compared to the four model predictions for DVCS+BH and to the calculations for BH only (green dot-dashed curve). It seems that Bethe-Heitler fully dominates the constant term, and all models - except for GGL, which misses both the magnitude and the $t$-dependence of this observable - predict this and correctly reproduce it. The best match for this term is provided by the VGG and GK models, which show sizeable differences only at the highest $-t$ values, where the DVCS contribution is expected to start to play a role. The models suggest a slight contribution from DVCS in the $\cos\phi$ term but the statistical precision of the data does not allow us to draw conclusions on which prediction provides the better fit. 

\section{Extraction of Compton Form Factors}
In recent years, various groups have developed and applied different procedures to extract Compton Form Factors from DVCS observables. The approach adopted here \cite{fitmick,michel2,mick_herve} is based on a local-fitting method at each given experimental $(Q^2, x_B,-t)$ kinematic point. In this framework, instead of four complex CFFs defined as in Eq.~\ref{def_cffs}, there are eight real CFFs defined as 
\begin{equation}
	F_{Re}(\xi,t) = \Re{\rm e}{\cal F}(\xi,t) 
\end{equation}
\begin{equation}
	F_{Im}(\xi,t) = -\frac{1}{\pi}\Im{\rm m}{\cal F}(\xi,t)=\left[ F(\xi,\xi,t)\mp F(-\xi,\xi,t) \right], 
\end{equation}
where the sign convention is the same as for Eq.~\eqref{dvcs-ampl}. These CFFs are the almost-free parameters - their values are allowed to vary within $\pm 5$ times the values predicted by the VGG model - that are extracted from DVCS observables using the well-established DVCS+BH theoretical amplitude. The BH amplitude is calculated exactly while the DVCS amplitude is taken at the QCD leading twist. The expression of these amplitudes can be found, for instance, in \cite{vgg}. 

\begin{figure}
\begin{center}
\includegraphics[width=86mm]{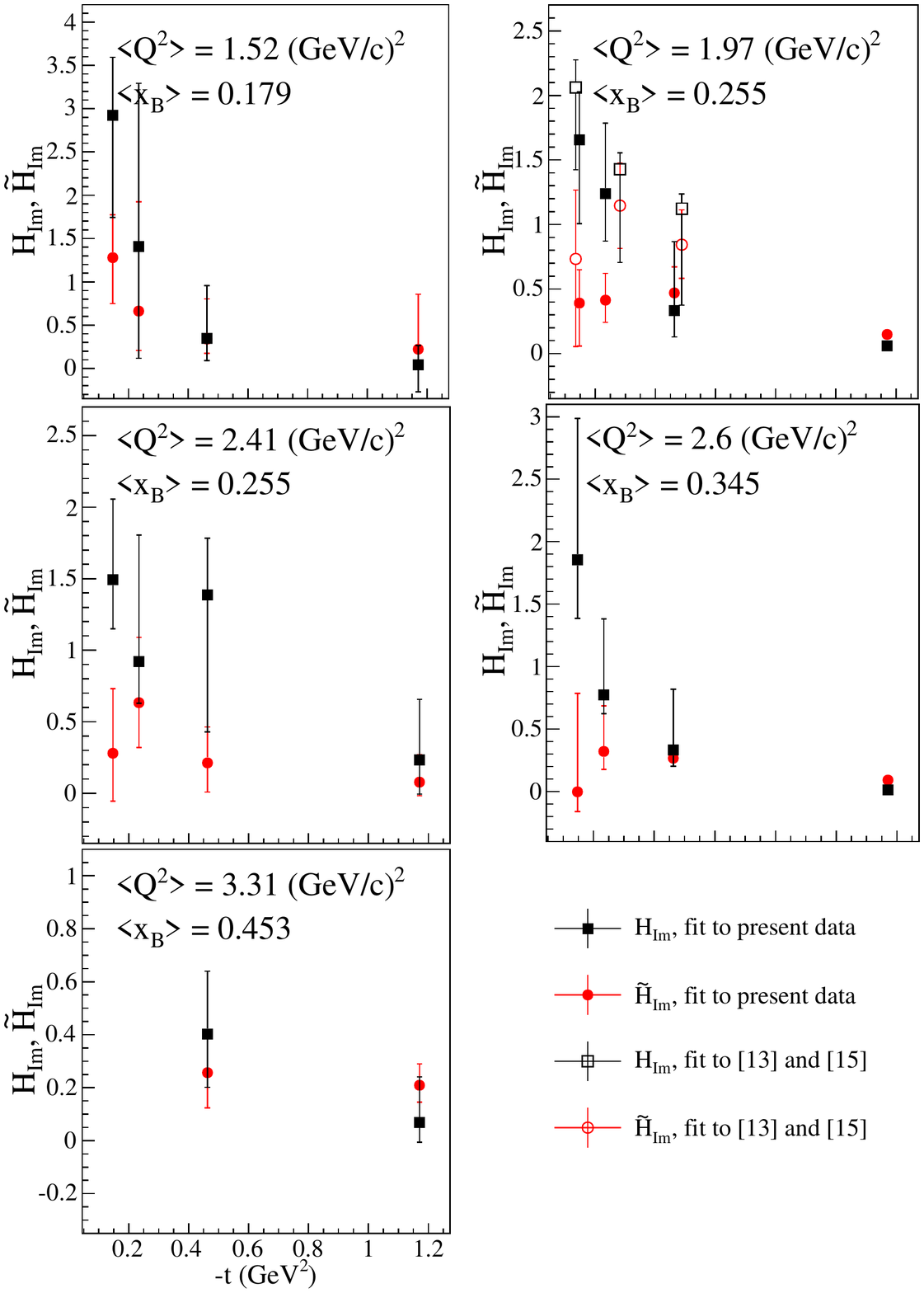}
\caption {(Color online) $t$ dependence for each $Q^2$-$x_B$ bin of $H_{Im}$ (black squares) and $\tilde{H}_{Im}$ (red circles). The full points are obtained by fitting the present data (TSA, BSA and DSA). The empty points were obtained by fitting the BSA results from \cite{fx} integrated over all values of $Q^2$ at $x_B \sim 0.25$, and the TSAs from \cite{shifeng}.}
\label{cff_comp}
\end{center}
\end{figure}

The three sets of asymmetries (BSA, TSA and DSA) for all kinematic bins were processed using this fitting procedure to extract the Compton Form Factors. 
In the adopted version of the fitter code, $\tilde{E}_{Im}$ is set to zero, as $\tilde{E}$ is assumed to be purely real - it is parametrized in the VGG model by the pion pole $(1/(t-m^2_{\pi}))$. Thus seven out of the eight real and imaginary parts of the CFFs are left as free parameters in the fit. 
Figure \ref{cff_comp} shows $H_{Im}$ (black full squares) and ${\tilde{H}}_{Im}$ (red full circles), which are obtained from the fit of the present data, as a function of $-t$ for each of our 5 $Q^2$-$x_B$ bins. These are the two CFFs that appear to be better constrained by the present results. Given that the size of the error bars reflects the sensitivity of the combination of observables to each CFF, it is evident that, as expected, our asymmetries are mostly sensitive to $\Im{\rm m}{\tilde{\cal H}}$. 

The results for $H_{Im}$ and ${\tilde{H}}_{Im}$ confirm what had been previously observed in a qualitative way by direct comparison of the $t$-dependence of our TSAs and BSAs in Section \ref{sec_results_tsa}: the $t$-slope of $\Im{\rm m}{\cal H}$ is much steeper than that of $\Im{\rm m}{\tilde{\cal H}}$, hinting at the fact that the axial charge (linked to $\Im{\rm m}{\tilde{\cal H}}$) might be more ``concentrated'' in the center of the nucleon than the electric charge (linked to $\Im{\rm m}{\cal H}$).
This effect seems stronger at the lowest values of $x_B$, while both CFFs tend to flatten out as $x_B$ increases. 
 
It is also interesting to compare the results obtained for the two equal-$x_B$ bins ($Q^2=1.97$ (GeV/$c$)$^2$ and $Q^2=2.41$ (GeV/$c$)$^2$): within the limits imposed by the size of the error bars and by the $Q^2$ lever arm (only 0.44 (GeV/$c$)$^2$), both sets of CFFs are compatible, at the 1-$\sigma$ level, which supports the validity of the scaling hypothesis. 

In Fig.~\ref{cff_comp} the values of $H_{Im}$ and ${\tilde{H}}_{Im}$  that were obtained \cite{fitmick3} using the same fitting code with the results from \cite{fx} for the beam-spin asymmetry and from \cite{shifeng} for the target-spin asymmetry, are also shown. Aside from the much larger kinematic coverage for the polarized-target observables made accessible by our data, in the kinematics where the previous extraction had been attempted, our data improve the precision of $\Im{\rm m}{\tilde{\cal H}}$. 

\section{Conclusions}
For the first time four-dimensional single-beam, single-target, and double (beam-target) spin asymmetries for deeply virtual Compton scattering were extracted over a large phase space at the same kinematics. This experiment used the CLAS detector in conjunction with the IC calorimeter and the Hall-B longitudinally polarized ${}^{14}$NH$_3$ target. 165 4-dimensional bins in $Q^2$, $x_B$, $-t$ and $\phi$, covering a wide kinematic range ($1<Q^2<5.2$ (GeV/$c$)$^2$, $0.12<x_B<0.6$, $0.08<-t<2$ (GeV/$c$)$^2$, $0<\phi<360^o$) were obtained for the three asymmetries, with systematic uncertainties largely below the statistical uncertainties. 
The $\phi$ dependence of the obtained asymmetries was studied. The dominance of the leading-order handbag mechanism is supported by the prevalence, especially at low $-t$, of the $\sin\phi$ term over higher $\sin n\phi$ components in both  single-spin asymmetries. The DSA is found to be mostly dominated by the constant term, which contains both BH and DVCS/BH interference, although in our kinematics the BH contribution is the strongest. These data bring important constraints to GPD parametrizations, especially for $H$ and $\tilde{H}$. Using one method among the various ones that are currently in development, these data allow us to extract the imaginary parts of the ${\tilde{\cal H}}$ and ${\cal H}$ Compton Form Factors and to gain insight, via their relative $t$ slopes, about the spatial distribution of the electric and axial charges in the proton.
The extraction of the Compton Form Factors will be further improved once the new CLAS results on DVCS cross sections will become available \cite{hyonsuk}. Furthermore, the extensive DVCS-devoted program planned at Jefferson Lab for the 12-GeV era will extend our knowledge of the Generalized Parton Distributions of the proton in terms of both kinematical coverage and statistical precision. 
\\
\\
We thank the staff of the Accelerator and the Physics Divisions and of the Target Group at Jefferson Lab for making the experiment possible. Special thanks to M. Guidal, F. Sabati\'e, S. Liuti, D. M\"{u}ller and K. Kumericki for the theoretical support. This work was supported in part by the U.S. Department of Energy and National Science Foundation, the French Centre National de la Recherche Scientifique and Commissariat \`a l'Energie Atomique, the French-American Cultural Exchange (FACE), the Italian Istituto Nazionale di Fisica Nucleare, the Chilean Comisi\'on Nacional de Investigaci\'on Cient\'ifica y Tecnol\'ogica (CONICYT), the National Research Foundation of Korea, and the UK Science and Technology Facilities Council (STFC).
The Jefferson Science Associates (JSA) operates the Thomas Jefferson National Accelerator Facility for 
the United States Department of Energy under contract DE-AC05-06OR23177. 
\clearpage
\begin{longtable*}{|c|c|c|c|c|c|c|c|c|}
\multicolumn{9}{c}%
{} \\
\hline
\multicolumn{1}{|c|}{$Q^2$} &
\multicolumn{1}{|c|}{$x_B$} &
\multicolumn{1}{|c|}{$-t$} &
\multicolumn{1}{|c|}{$\phi$} &
\multicolumn{1}{|c|}{BSA $\pm$ stat. $\pm$ syst.} &
\multicolumn{1}{|c|}{TSA $\pm$ stat. $\pm$ syst.} &
\multicolumn{1}{|c|}{$c_{A_{\rm UT}}$} &
\multicolumn{1}{|c|}{DSA $\pm$ stat. $\pm$ syst.} &
\multicolumn{1}{|c|}{$c_{A_{\rm LT}}$}\\
\multicolumn{1}{|c|}{(GeV/$c$)$^2$} &
\multicolumn{1}{|c|}{} &
\multicolumn{1}{|c|}{(GeV/$c$)$^2$} &
\multicolumn{1}{|c|}{(deg)} &
\multicolumn{1}{|c|}{} &
\multicolumn{1}{|c|}{} &
\multicolumn{1}{|c|}{} & 
\multicolumn{1}{|c|}{} &
\multicolumn{1}{|c|}{}\\
\hline
\endhead 
\hline
\endfoot
\caption{Values of the three asymmetries with their statistical and systematic uncertainties for each 4-dimensional bin. The values of the bin-by-bin transverse corrections for the TSA and DSA are also shown.}
\label{table_all_data}
\endlastfoot
\hline

1.68 & 0.194 & 0.11 & 25 & 0.37 $\pm$ 0.23$\pm$	0.01 & 0.44 $\pm$ 0.26 $\pm$ 0.02 & 0.0018 & 0.58 $\pm$ 0.33 $\pm$ 0.04 & 0.0116 \\  
1.68 & 0.190 & 0.12 & 60 & 0.34 $\pm$ 0.11$\pm$	0.03 & 0.12 $\pm$ 0.14 $\pm$ 0.02 & 0.0024 & 0.34 $\pm$ 0.16 $\pm$ 0.13 & 0.0070 \\  
1.58 & 0.186 & 0.13 & 92 & 0.23 $\pm$ 0.06$\pm$	0.01 & 0.18 $\pm$ 0.07 $\pm$ 0.01 & 0.0144 & 0.58 $\pm$ 0.08 $\pm$ 0.02 & 0.0072 \\  
1.54 & 0.178 & 0.13 & 128 & 0.13 $\pm$ 0.05$\pm$ 0.01 & 0.13 $\pm$ 0.06 $\pm$ 0.02 & 0.0184 & 0.56 $\pm$ 0.07 $\pm$ 0.02 & 0.0021 \\  
1.50 & 0.174 & 0.13 & 161 & 0.05 $\pm$ 0.05$\pm$ 0.01 & 0.07 $\pm$ 0.06 $\pm$ 0.01 & 0.0086 & 0.46 $\pm$ 0.07 $\pm$ 0.02 & -0.0024 \\  
1.51 & 0.173 & 0.13 & 198 & -0.05 $\pm$	0.05$\pm$ 0.00 & -0.08 $\pm$ 0.06 $\pm$	0.01 & -0.0086 & 0.51 $\pm$ 0.07 $\pm$ 0.01 & -0.0021 \\  
1.52 & 0.177 & 0.13 & 235 & -0.21 $\pm$	0.06$\pm$ 0.01 & -0.13 $\pm$ 0.07 $\pm$	0.02 & -0.0184 & 0.51 $\pm$ 0.08 $\pm$ 0.02 & 0.0018 \\  
1.60 & 0.185 & 0.13 & 266 & -0.28 $\pm$	0.06$\pm$ 0.01 & -0.18 $\pm$ 0.08 $\pm$	0.01 & -0.0138 & 0.44 $\pm$ 0.09 $\pm$ 0.02 & 0.0059 \\  
1.68 & 0.190 & 0.12 & 305 & -0.15 $\pm$	0.10$\pm$ 0.02 & -0.01 $\pm$ 0.12 $\pm$ 0.03 & -0.0017 & 0.46 $\pm$ 0.14 $\pm$ 0.04 & 0.0082 \\  

1.61 & 0.190 & 0.21 & 97 & 0.19 $\pm$ 0.19$\pm$	0.06 & 0.15 $\pm$ 0.21 $\pm$ 0.04 & 0.0193 & -0.05 $\pm$ 0.26 $\pm$ 0.10 & -0.0011 \\  
1.57 & 0.182 & 0.23 & 128 & 0.17 $\pm$ 0.10$\pm$ 0.03 & -0.01 $\pm$ 0.12 $\pm$ 0.03 & 0.0209 & 0.54 $\pm$ 0.14 $\pm$ 0.05 & 0.0034 \\  
1.51 & 0.179 & 0.23 & 161 & 0.09 $\pm$ 0.08$\pm$ 0.02 & 0.13 $\pm$ 0.09 $\pm$ 0.02 & 0.0111 & 0.64 $\pm$ 0.10 $\pm$ 0.06 & 0.0044 \\  
1.54 & 0.178 & 0.23 & 199 & -0.12 $\pm$	0.07$\pm$ 0.02 & 0.11 $\pm$ 0.09 $\pm$ 0.01 & -0.0085 & 0.60 $\pm$ 0.10 $\pm$ 0.03 & 0.0036 \\  
1.62 & 0.184 & 0.23 & 233 & -0.18 $\pm$	0.09$\pm$ 0.03 & -0.24 $\pm$ 0.11 $\pm$	0.01 & -0.0221 & 0.38 $\pm$ 0.13 $\pm$ 0.11 & 0.0022 \\  
1.64 & 0.191 & 0.22 & 263 & -0.12 $\pm$	0.17$\pm$ 0.05 & -0.41 $\pm$ 0.19 $\pm$ 0.08 & -0.0201 & 0.75 $\pm$ 0.21 $\pm$ 0.08 & 0.0055 \\  

1.63 & 0.188 & 0.49 & 96 & 0.39 $\pm$ 0.24$\pm$ 0.09 & 0.23 $\pm$ 0.27 $\pm$ 0.06 & 0.0193 & 0.81 $\pm$ 0.32 $\pm$ 0.14 & 0.0016 \\  
1.54 & 0.181 & 0.47 & 127 & 0.18 $\pm$ 0.11$\pm$ 0.05 & 0.14 $\pm$ 0.14 $\pm$ 0.04 & 0.0213 & 0.34 $\pm$ 0.16 $\pm$ 0.04 & 0.0023 \\  
1.36 & 0.173 & 0.49 & 162 & 0.21 $\pm$ 0.07$\pm$ 0.02 & 0.19 $\pm$ 0.08 $\pm$ 0.01 & 0.0117 & 0.50 $\pm$ 0.09 $\pm$ 0.05 & 0.0073 \\  
1.42 & 0.174 & 0.46 & 196 & -0.07 $\pm$ 0.08$\pm$ 0.02 & -0.03 $\pm$ 0.10 $\pm$ 0.01 & -0.0095 & 0.54 $\pm$ 0.11 $\pm$ 0.03 & 0.0071 \\  
1.60 & 0.181 & 0.43 & 234 & -0.22 $\pm$	0.12$\pm$ 0.02 & -0.17 $\pm$ 0.14 $\pm$ 0.04 & -0.0194 & 0.36 $\pm$ 0.16 $\pm$ 0.09 & 0.0020 \\  
1.56 & 0.185 & 0.51 & 261 & -0.04 $\pm$ 0.21$\pm$ 0.12 & -0.29 $\pm$ 0.24 $\pm$ 0.10 & -0.0218 & -0.02 $\pm$ 0.29 $\pm$ 0.12 & -0.0048 \\  

1.62 & 0.184 & 1.35 & 20 & 0.13 $\pm$ 0.16$\pm$	0.07 & 0.23 $\pm$ 0.19 $\pm$ 0.02 & 0.0016 & 0.36 $\pm$ 0.23 $\pm$ 0.07 & -0.0051 \\  
1.56 & 0.178 & 1.21 & 56 & 0.01 $\pm$ 0.14$\pm$ 0.05 & -0.05 $\pm$ 0.17 $\pm$ 0.07 & 0.0009 & 0.50 $\pm$ 0.20 $\pm$ 0.06 & -0.0017 \\  
1.46 & 0.178 & 1.02 & 92 & 0.12 $\pm$ 0.15$\pm$ 0.14 & 0.21 $\pm$ 0.20 $\pm$ 0.10 & 0.0029 & 0.08 $\pm$ 0.21 $\pm$ 0.18 & -0.0005 \\  
1.48 & 0.176 & 1.01 & 128 & 0.20 $\pm$ 0.19$\pm$ 0.05 & 0.36 $\pm$ 0.24 $\pm$ 0.04 & 0.0021 & 0.44 $\pm$ 0.27 $\pm$ 0.06 & 0.0062 \\  
1.36 & 0.170 & 1.00 & 164 & -0.09 $\pm$ 0.08$\pm$ 0.01 & 0.41 $\pm$ 0.09 $\pm$ 0.06 & 0.0034 & 0.12 $\pm$ 0.11 $\pm$ 0.05 & 0.0067 \\  
1.35 & 0.171 & 0.97 & 192 & 0.06 $\pm$ 0.14$\pm$ 0.11 & 0.07 $\pm$ 0.18 $\pm$ 0.07 & 0.0011 & 0.41 $\pm$ 0.19 $\pm$ 0.06 & 0.0095 \\  
1.40 & 0.171 & 1.07 & 239 & -0.22 $\pm$	0.17$\pm$ 0.07 & 0.09 $\pm$ 0.22 $\pm$ 0.03 & 0.0017 & 0.24 $\pm$ 0.23 $\pm$ 0.06 & 0.0048 \\  
1.41 & 0.178 & 1.03 & 266 & 0.04 $\pm$ 0.10$\pm$ 0.05 & -0.27 $\pm$ 0.12 $\pm$ 0.10 & -0.0034 & 0.30 $\pm$ 0.14 $\pm$ 0.08 & 0.0015 \\  
1.58 & 0.179 & 1.24 & 306 & 0.01 $\pm$ 0.14$\pm$ 0.09 & 0.04 $\pm$ 0.17 $\pm$ 0.10 & -0.0009 & 0.51 $\pm$ 0.20 $\pm$ 0.07 & -0.0016 \\  
1.62 & 0.183 & 1.35 & 340 & -0.11 $\pm$	0.18$\pm$ 0.12 & -0.20 $\pm$ 0.20 $\pm$ 0.10 & -0.0013 & 0.36 $\pm$ 0.24 $\pm$ 0.17 & -0.0051 \\  

2.03 & 0.255 & 0.13 & 19 & 0.03 $\pm$ 0.03$\pm$	0.01 & -0.03 $\pm$ 0.03 $\pm$ 0.01 & -0.0023 & 0.60 $\pm$ 0.04 $\pm$ 0.03 & 0.0328 \\  
1.99 & 0.250 & 0.14 & 52 & 0.17 $\pm$ 0.03$\pm$ 0.01 & 0.05 $\pm$ 0.04 $\pm$ 0.01 & 0.0027 & 0.60 $\pm$ 0.04 $\pm$ 0.03 & 0.0297 \\  
1.93 & 0.239 & 0.14 & 89 & 0.20 $\pm$ 0.03$\pm$ 0.01 & 0.05 $\pm$ 0.04 $\pm$ 0.01 & 0.0165 & 0.57 $\pm$ 0.05 $\pm$ 0.02 & 0.0179 \\  
1.91 & 0.235 & 0.14 & 124 & 0.16 $\pm$ 0.05$\pm$ 0.01 & 0.12 $\pm$ 0.06 $\pm$ 0.02 & 0.0247 & 0.63 $\pm$ 0.07 $\pm$ 0.02 & 0.0025 \\  
1.84 & 0.228 & 0.14 & 160 & 0.03 $\pm$ 0.06$\pm$ 0.01 & 0.20 $\pm$ 0.07 $\pm$ 0.01 & 0.0139 & 0.65 $\pm$ 0.08 $\pm$ 0.03 & -0.0105 \\  
1.89 & 0.231 & 0.14 & 199 & 0.02 $\pm$ 0.06$\pm$ 0.01 & -0.01 $\pm$ 0.07 $\pm$ 0.02 & -0.0116 & 0.59 $\pm$ 0.08 $\pm$ 0.02 & -0.0111 \\  
1.93 & 0.235 & 0.14 & 238 & -0.21 $\pm$	0.05$\pm$ 0.02 & -0.13 $\pm$ 0.06 $\pm$ 0.01 & -0.0246 & 0.50 $\pm$ 0.07 $\pm$ 0.02 & 0.0009 \\  
1.94 & 0.239 & 0.14 & 270 & -0.24 $\pm$	0.04$\pm$ 0.01 & -0.16 $\pm$ 0.04 $\pm$ 0.01 & -0.0176 & 0.63 $\pm$ 0.05 $\pm$ 0.02 & 0.0186 \\  
1.97 & 0.248 & 0.13 & 308 & -0.16 $\pm$	0.03$\pm$ 0.01 & -0.16 $\pm$ 0.03 $\pm$ 0.01 & -0.0041 & 0.56 $\pm$ 0.04 $\pm$ 0.03 & 0.0289 \\  
2.03 & 0.255 & 0.13 & 340 & -0.07 $\pm$	0.03$\pm$ 0.00 & -0.09 $\pm$ 0.04 $\pm$ 0.01 & -0.0007 & 0.55 $\pm$ 0.04 $\pm$ 0.03 & 0.0320 \\  

2.23 & 0.276 & 0.22 & 21 & 0.05 $\pm$ 0.05$\pm$	0.01 & -0.00 $\pm$ 0.06 $\pm$ 0.01 & -0.0037 & 0.58 $\pm$ 0.06 $\pm$ 0.02 & 0.0172 \\  
2.12 & 0.268 & 0.23 & 53 & 0.22 $\pm$ 0.04$\pm$ 0.01 & 0.14 $\pm$ 0.05 $\pm$ 0.02 & 0.0044 & 0.56 $\pm$ 0.06 $\pm$ 0.03 & 0.0169 \\  
2.01 & 0.251 & 0.23 & 91 & 0.27 $\pm$ 0.04$\pm$ 0.01 & 0.16 $\pm$ 0.05 $\pm$ 0.01 & 0.0246 & 0.49 $\pm$ 0.06 $\pm$ 0.01 & 0.0105 \\  
1.91 & 0.248 & 0.23 & 124 & 0.06 $\pm$ 0.06$\pm$ 0.01 & 0.06 $\pm$ 0.07 $\pm$ 0.01 & 0.0347 & 0.66 $\pm$ 0.07 $\pm$ 0.03 & 0.0060 \\  
1.79 & 0.246 & 0.24 & 161 & 0.03 $\pm$ 0.06$\pm$ 0.01 & 0.07 $\pm$ 0.07 $\pm$ 0.01 & 0.0185 & 0.55 $\pm$ 0.08 $\pm$ 0.02 & -0.0017 \\  
1.80 & 0.245 & 0.24 & 199 & 0.02 $\pm$ 0.06$\pm$ 0.01 & -0.03 $\pm$ 0.08 $\pm$ 0.02 & -0.0174 & 0.70 $\pm$ 0.08 $\pm$ 0.03 & 0.0004 \\  
1.97 & 0.242 & 0.23 & 235 & -0.11 $\pm$	0.06$\pm$ 0.01 & -0.18 $\pm$ 0.07 $\pm$ 0.01 & -0.0322 & 0.62 $\pm$ 0.08 $\pm$ 0.04 & 0.0041 \\  
2.01 & 0.248 & 0.23 & 270 & -0.31 $\pm$	0.04$\pm$ 0.01 & -0.21 $\pm$ 0.05 $\pm$ 0.01 & -0.0247 & 0.53 $\pm$ 0.06 $\pm$ 0.02 & 0.0109 \\  
2.14 & 0.268 & 0.23 & 307 & -0.28 $\pm$	0.04$\pm$ 0.02 & -0.15 $\pm$ 0.05 $\pm$ 0.02 & -0.0044 & 0.57 $\pm$ 0.06 $\pm$ 0.03 & 0.0167 \\  
2.21 & 0.277 & 0.22 & 340 & -0.05 $\pm$	0.04$\pm$ 0.01 & 0.00 $\pm$ 0.05 $\pm$ 0.01 & 0.0039 & 0.51 $\pm$ 0.06 $\pm$ 0.02 & 0.0167 \\  

2.34 & 0.286 & 0.40 & 26 & 0.22 $\pm$ 0.16$\pm$	0.01 & -0.17 $\pm$ 0.18 $\pm$ 0.01 & -0.0069 & 0.66 $\pm$ 0.22 $\pm$ 0.05 & 0.0039 \\  
2.20 & 0.276 & 0.45 & 56 & 0.21 $\pm$ 0.06$\pm$ 0.02 & 0.15 $\pm$ 0.08 $\pm$ 0.02 & 0.0075 & 0.43 $\pm$ 0.09 $\pm$ 0.04 & 0.0000 \\  
2.04 & 0.254 & 0.44 & 90 & 0.14 $\pm$ 0.06$\pm$ 0.01 & 0.29 $\pm$ 0.07 $\pm$ 0.04 & 0.0310 & 0.55 $\pm$ 0.08 $\pm$ 0.03 & 0.0022 \\  
1.82 & 0.255 & 0.45 & 127 & 0.08 $\pm$ 0.06$\pm$ 0.04 & 0.14 $\pm$ 0.08 $\pm$ 0.02 & 0.0422 & 0.51 $\pm$ 0.09 $\pm$ 0.02 & 0.0047 \\  
1.73 & 0.252 & 0.48 & 163 & 0.06 $\pm$ 0.04$\pm$ 0.01 & 0.12 $\pm$ 0.05 $\pm$ 0.01 & 0.0209 & 0.45 $\pm$ 0.06 $\pm$ 0.02 & 0.0056 \\  
1.74 & 0.254 & 0.47 & 197 & 0.02 $\pm$ 0.05$\pm$ 0.01 & -0.13 $\pm$ 0.06 $\pm$ 0.03 & -0.0211 & 0.48 $\pm$ 0.07 $\pm$ 0.02 & 0.0060 \\  
1.88 & 0.251 & 0.44 & 233 & -0.13 $\pm$	0.07$\pm$ 0.02 & -0.12 $\pm$ 0.09 $\pm$ 0.04 & -0.0380 & 0.44 $\pm$ 0.10 $\pm$ 0.03 & 0.0029 \\  
1.99 & 0.253 & 0.46 & 271 & -0.20 $\pm$	0.05$\pm$ 0.03 & -0.17 $\pm$ 0.06 $\pm$ 0.02 & -0.0310 & 0.50 $\pm$ 0.07 $\pm$ 0.03 & 0.0018 \\  
2.20 & 0.274 & 0.45 & 303 & -0.30 $\pm$	0.06$\pm$ 0.03 & -0.09 $\pm$ 0.07 $\pm$ 0.01 & -0.0066 & 0.41 $\pm$ 0.09 $\pm$ 0.04 & -0.0003 \\  
2.36 & 0.287 & 0.40 & 334 & -0.21 $\pm$	0.12$\pm$ 0.03 & -0.04 $\pm$ 0.14 $\pm$ 0.02 & -0.0042 & 0.34 $\pm$ 0.16 $\pm$ 0.02 & -0.0007 \\  

2.03 & 0.261 & 1.41 & 17 & -0.04 $\pm$ 0.05$\pm$ 0.02 & -0.08 $\pm$ 0.05 $\pm$ 0.01 & -0.0026 & 0.48 $\pm$ 0.06 $\pm$ 0.02 & -0.0031 \\  
1.98 & 0.258 & 1.12 & 54 & 0.10 $\pm$ 0.06$\pm$ 0.02 & 0.14 $\pm$ 0.07 $\pm$ 0.02 & 0.0060 & 0.34 $\pm$ 0.08 $\pm$ 0.03 & -0.0038 \\  
1.78 & 0.256 & 1.18 & 87 & 0.05 $\pm$ 0.07$\pm$ 0.03 & 0.20 $\pm$ 0.08 $\pm$ 0.04 & 0.0152 & 0.33 $\pm$ 0.09 $\pm$ 0.03 & 0.0005 \\  
1.82 & 0.261 & 1.13 & 128 & 0.14 $\pm$ 0.09$\pm$ 0.03 & 0.46 $\pm$ 0.12 $\pm$ 0.06 & 0.0188 & 0.26 $\pm$ 0.13 $\pm$ 0.02 & 0.0027 \\  
1.79 & 0.252 & 1.03 & 164 & 0.03 $\pm$ 0.05$\pm$ 0.01 & 0.16 $\pm$ 0.06 $\pm$ 0.02 & 0.0066 & 0.40 $\pm$ 0.06 $\pm$ 0.01 & 0.0062 \\  
1.78 & 0.254 & 1.03 & 195 & -0.03 $\pm$	0.07$\pm$ 0.02 & -0.09 $\pm$ 0.08 $\pm$ 0.03 & -0.0056 & 0.34 $\pm$ 0.09 $\pm$ 0.03 & 0.0055 \\  
1.84 & 0.257 & 1.23 & 238 & -0.05 $\pm$	0.08$\pm$ 0.02 & -0.30 $\pm$ 0.11 $\pm$ 0.04 & -0.0145 & 0.34 $\pm$ 0.12 $\pm$ 0.05 & 0.0039 \\  
1.75 & 0.257 & 1.14 & 270 & -0.02 $\pm$	0.05$\pm$ 0.02 & -0.26 $\pm$ 0.07 $\pm$ 0.04 & -0.0169 & 0.21 $\pm$ 0.07 $\pm$ 0.02 & -0.0018 \\  
2.01 & 0.260 & 1.12 & 307 & -0.07 $\pm$	0.07$\pm$ 0.01 & -0.20 $\pm$ 0.08 $\pm$ 0.02 & -0.0068 & 0.56 $\pm$ 0.09 $\pm$ 0.04 & 0.0005 \\  
2.03 & 0.261 & 1.40 & 343 & 0.00 $\pm$ 0.04$\pm$ 0.01 & 0.02 $\pm$ 0.05 $\pm$ 0.02 & 0.0018 & 0.36 $\pm$ 0.06 $\pm$ 0.02 & 0.0051 \\  

2.45 & 0.265 & 0.13 & 20 & -0.02 $\pm$ 0.04$\pm$ 0.01 & 0.02 $\pm$ 0.05 $\pm$ 0.01 & -0.0002 & 0.76 $\pm$ 0.06 $\pm$ 0.02 & 0.0271 \\  
2.36 & 0.253 & 0.14 & 53 & 0.07 $\pm$ 0.05$\pm$ 0.02 & 0.01 $\pm$ 0.05 $\pm$ 0.02 & 0.0018 & 0.62 $\pm$ 0.06 $\pm$ 0.04 & 0.0211 \\  
2.32 & 0.249 & 0.14 & 88 & 0.17 $\pm$ 0.05$\pm$ 0.01 & -0.01 $\pm$ 0.07 $\pm$ 0.01 & 0.0071 & 0.76 $\pm$ 0.07 $\pm$ 0.02 & 0.0125 \\  
2.30 & 0.246 & 0.14 & 125 & 0.20 $\pm$ 0.08$\pm$ 0.02 & 0.14 $\pm$ 0.10 $\pm$ 0.01 & 0.0109 & 0.61 $\pm$ 0.11 $\pm$ 0.03 & -0.0021 \\  
2.26 & 0.242 & 0.14 & 162 & 0.20 $\pm$ 0.10$\pm$ 0.03 & 0.25 $\pm$ 0.12 $\pm$ 0.05 & 0.0064 & 0.81 $\pm$ 0.13 $\pm$ 0.04 & -0.0117 \\  
2.27 & 0.242 & 0.14 & 199 & -0.07 $\pm$	0.09$\pm$ 0.04 & 0.07 $\pm$ 0.10 $\pm$ 0.04 & -0.0044 & 0.81 $\pm$ 0.11 $\pm$ 0.05 & -0.0116 \\  
2.28 & 0.243 & 0.14 & 237 & -0.29 $\pm$	0.07$\pm$ 0.02 & -0.17 $\pm$ 0.09 $\pm$ 0.02 & -0.0110 & 0.84 $\pm$ 0.09 $\pm$ 0.02 & -0.0007 \\  
2.32 & 0.248 & 0.14 & 271 & -0.22 $\pm$	0.06$\pm$ 0.02 & -0.15 $\pm$ 0.07 $\pm$ 0.01 & -0.0082 & 0.75 $\pm$ 0.07 $\pm$ 0.02 & 0.0124 \\  
2.36 & 0.253 & 0.14 & 307 & -0.14 $\pm$	0.04$\pm$ 0.01 & -0.10 $\pm$ 0.05 $\pm$ 0.01 & -0.0023 & 0.56 $\pm$ 0.06 $\pm$ 0.02 & 0.0206 \\  
2.45 & 0.265 & 0.14 & 340 & -0.01 $\pm$	0.05$\pm$ 0.02 & -0.07 $\pm$ 0.06 $\pm$ 0.02 & -0.0002 & 0.67 $\pm$ 0.07 $\pm$ 0.03 & 0.0265 \\  

2.57 & 0.281 & 0.22 & 24 & 0.03 $\pm$ 0.08$\pm$	0.02 & 0.03 $\pm$ 0.09 $\pm$ 0.03 & -0.0008 & 0.55 $\pm$ 0.10 $\pm$ 0.02 & 0.0130 \\  
2.50 & 0.269 & 0.23 & 53 & 0.11 $\pm$ 0.05$\pm$ 0.02 & 0.09 $\pm$ 0.06 $\pm$ 0.02 & 0.0031 & 0.68 $\pm$ 0.07 $\pm$ 0.05 & 0.0122 \\  
2.40 & 0.258 & 0.23 & 90 & 0.19 $\pm$ 0.06$\pm$ 0.02 & 0.07 $\pm$ 0.08 $\pm$ 0.01 & 0.0111 & 0.58 $\pm$ 0.09 $\pm$ 0.03 & 0.0076 \\  
2.35 & 0.252 & 0.23 & 126 & 0.16 $\pm$ 0.09$\pm$ 0.02 & 0.10 $\pm$ 0.11 $\pm$ 0.02 & 0.0150 & 0.55 $\pm$ 0.12 $\pm$ 0.03 & 0.0014 \\  
2.36 & 0.254 & 0.23 & 161 & 0.10 $\pm$ 0.10$\pm$ 0.03 & -0.18 $\pm$ 0.12 $\pm$ 0.02 & 0.0057 & 0.39 $\pm$ 0.14 $\pm$ 0.07 & -0.0045 \\  
2.37 & 0.254 & 0.23 & 199 & 0.02 $\pm$ 0.09$\pm$ 0.02 & 0.15 $\pm$ 0.11 $\pm$ 0.01 & -0.0059 & 0.70 $\pm$ 0.11 $\pm$ 0.06 & -0.0026 \\  
2.36 & 0.254 & 0.23 & 236 & -0.17 $\pm$	0.08$\pm$ 0.02 & -0.23 $\pm$ 0.10 $\pm$ 0.02 & -0.0160 & 0.63 $\pm$ 0.11 $\pm$ 0.03 & 0.0019 \\  
2.42 & 0.260 & 0.23 & 270 & -0.23 $\pm$	0.06$\pm$ 0.03 & -0.14 $\pm$ 0.08 $\pm$ 0.02 & -0.0118 & 0.61 $\pm$ 0.09 $\pm$ 0.04 & 0.0079 \\  
2.48 & 0.266 & 0.23 & 308 & -0.15 $\pm$	0.05$\pm$ 0.02 & -0.21 $\pm$ 0.06 $\pm$ 0.01 & -0.0039 & 0.59 $\pm$ 0.07 $\pm$ 0.02 & 0.0115 \\  
2.56 & 0.279 & 0.22 & 336 & -0.01 $\pm$	0.07$\pm$ 0.01 & -0.09 $\pm$ 0.08 $\pm$ 0.02 & 0.0001 & 0.49 $\pm$ 0.10 $\pm$ 0.03 & 0.0127 \\  

2.60 & 0.284 & 0.41 & 31 & -0.18 $\pm$ 0.22$\pm$ 0.09 & -0.01 $\pm$ 0.23 $\pm$ 0.06 & 0.0000 & 0.01 $\pm$ 0.31 $\pm$ 0.11 & -0.0045 \\  
2.55 & 0.275 & 0.44 & 52 & 0.17 $\pm$ 0.06$\pm$ 0.01 & 0.07 $\pm$ 0.07 $\pm$ 0.02 & 0.0055 & 0.54 $\pm$ 0.08 $\pm$ 0.05 & 0.0003 \\  
2.45 & 0.264 & 0.44 & 91 & 0.20 $\pm$ 0.07$\pm$ 0.02 & 0.04 $\pm$ 0.08 $\pm$ 0.04 & 0.0157 & 0.59 $\pm$ 0.09 $\pm$ 0.03 & 0.0022 \\  
2.37 & 0.254 & 0.44 & 122 & 0.23 $\pm$ 0.10$\pm$ 0.03 & 0.15 $\pm$ 0.12 $\pm$ 0.01 & 0.0188 & 0.69 $\pm$ 0.13 $\pm$ 0.03 & 0.0043 \\  
2.33 & 0.250 & 0.41 & 159 & -0.15 $\pm$	0.13$\pm$ 0.04 & -0.24 $\pm$ 0.16 $\pm$ 0.04 & 0.0063 & 0.59 $\pm$ 0.18 $\pm$ 0.04 & 0.0043 \\  
2.36 & 0.253 & 0.41 & 199 & -0.02 $\pm$	0.12$\pm$ 0.04 & 0.00 $\pm$ 0.14 $\pm$ 0.03 & -0.0079 & 0.59 $\pm$ 0.16 $\pm$ 0.04 & 0.0044 \\  
2.36 & 0.254 & 0.43 & 235 & -0.28 $\pm$	0.10$\pm$ 0.03 & -0.05 $\pm$ 0.12 $\pm$ 0.03 & -0.0184 & 0.71 $\pm$ 0.13 $\pm$ 0.04 & 0.0045 \\  
2.43 & 0.261 & 0.45 & 270 & -0.26 $\pm$	0.07$\pm$ 0.02 & -0.12 $\pm$ 0.09 $\pm$ 0.03 & -0.0161 & 0.60 $\pm$ 0.10 $\pm$ 0.02 & 0.0023 \\  
2.54 & 0.274 & 0.45 & 307 & -0.24 $\pm$	0.07$\pm$ 0.03 & -0.22 $\pm$ 0.08 $\pm$ 0.03 & -0.0065 & 0.43 $\pm$ 0.09 $\pm$ 0.05 & -0.0004 \\  
2.61 & 0.284 & 0.42 & 330 & -0.15 $\pm$	0.19$\pm$ 0.07 & -0.03 $\pm$ 0.18 $\pm$ 0.03 & 0.0004 & 0.59 $\pm$ 0.27 $\pm$ 0.14 & 0.0000 \\  

2.43 & 0.261 & 1.22 & 20 & -0.01 $\pm$ 0.08$\pm$ 0.02 & 0.08 $\pm$ 0.09 $\pm$ 0.01 & 0.0007 & 0.30 $\pm$ 0.11 $\pm$ 0.02 & -0.0068 \\  
2.45 & 0.265 & 1.06 & 52 & 0.14 $\pm$ 0.09$\pm$ 0.03 & -0.00 $\pm$ 0.11 $\pm$ 0.03 & 0.0034 & 0.60 $\pm$ 0.12 $\pm$ 0.02 & -0.0035 \\  
2.37 & 0.255 & 0.96 & 88 & 0.18 $\pm$ 0.20$\pm$ 0.12 & 0.21 $\pm$ 0.22 $\pm$ 0.05 & 0.0085 & 0.65 $\pm$ 0.26 $\pm$ 0.07 & 0.0004 \\    
2.43 & 0.261 & 0.96 & 273 & -0.31 $\pm$	0.17$\pm$ 0.05 & 0.09 $\pm$ 0.20 $\pm$ 0.08 & -0.0069 & 0.49 $\pm$ 0.23 $\pm$ 0.06 & 0.0006 \\  
2.44 & 0.262 & 1.08 & 308 & -0.16 $\pm$	0.10$\pm$ 0.04 & -0.04 $\pm$ 0.11 $\pm$ 0.03 & -0.0037 & 0.45 $\pm$ 0.13 $\pm$ 0.05 & 0.0045 \\  
2.46 & 0.265 & 1.23 & 340 & -0.04 $\pm$	0.07$\pm$ 0.02 & -0.09 $\pm$ 0.09 $\pm$ 0.01 & -0.0007 & 0.44 $\pm$ 0.10 $\pm$ 0.03 & 0.0059 \\  

2.55 & 0.319 & 0.16 & 19 & 0.13 $\pm$ 0.05$\pm$	0.02 & 0.07 $\pm$ 0.07 $\pm$ 0.01 & 0.0020 & 0.73 $\pm$ 0.07 $\pm$ 0.06 & 0.0546 \\  
2.50 & 0.319 & 0.16 & 50 & 0.25 $\pm$ 0.07$\pm$ 0.03 & 0.10 $\pm$ 0.09 $\pm$ 0.01 & 0.0045 & 0.60 $\pm$ 0.10 $\pm$ 0.06 & 0.0513 \\  
2.53 & 0.320 & 0.16 & 88 & 0.37 $\pm$ 0.14$\pm$ 0.08 & 0.16 $\pm$ 0.17 $\pm$ 0.03 & 0.0066 & 0.74 $\pm$ 0.19 $\pm$ 0.05 & 0.0464 \\  
2.55 & 0.317 & 0.16 & 240 & -0.13 $\pm$	0.22$\pm$ 0.01 & -0.19 $\pm$ 0.26 $\pm$ 0.05 & -0.0064 & 0.55 $\pm$ 0.29 $\pm$ 0.03 & 0.0345 \\  
2.59 & 0.321 & 0.16 & 273 & -0.12 $\pm$	0.15$\pm$ 0.05 & -0.24 $\pm$ 0.17 $\pm$ 0.07 & -0.0078 & 0.93 $\pm$ 0.19 $\pm$ 0.09 & 0.0479 \\  
2.53 & 0.319 & 0.16 & 309 & -0.21 $\pm$	0.07$\pm$ 0.02 & -0.11 $\pm$ 0.08 $\pm$ 0.01 & -0.0045 & 0.77 $\pm$ 0.09 $\pm$ 0.04 & 0.0528 \\  
2.55 & 0.318 & 0.16 & 342 & -0.12 $\pm$	0.06$\pm$ 0.02 & 0.02 $\pm$ 0.07 $\pm$ 0.01 & -0.0005 & 0.70 $\pm$ 0.08 $\pm$ 0.05 & 0.0534 \\  

2.72 & 0.338 & 0.24 & 19 & 0.08 $\pm$ 0.03$\pm$	0.00 & 0.10 $\pm$ 0.03 $\pm$ 0.01 & 0.0000 & 0.72 $\pm$ 0.04 $\pm$ 0.03 & 0.0398 \\  
2.68 & 0.338 & 0.24 & 51 & 0.20 $\pm$ 0.03$\pm$ 0.01 & 0.11 $\pm$ 0.04 $\pm$ 0.01 & 0.0050 & 0.69 $\pm$ 0.05 $\pm$ 0.03 & 0.0387 \\  
2.67 & 0.337 & 0.24 & 89 & 0.16 $\pm$ 0.05$\pm$ 0.02 & 0.15 $\pm$ 0.07 $\pm$ 0.02 & 0.0223 & 0.69 $\pm$ 0.07 $\pm$ 0.03 & 0.0293 \\  
2.43 & 0.337 & 0.25 & 122 & 0.13 $\pm$ 0.07$\pm$ 0.01 & 0.19 $\pm$ 0.08 $\pm$ 0.01 & 0.0373 & 0.72 $\pm$ 0.09 $\pm$ 0.04 & 0.0164 \\  
2.47 & 0.326 & 0.25 & 161 & 0.14 $\pm$ 0.12$\pm$ 0.01 & 0.25 $\pm$ 0.14 $\pm$ 0.03 & 0.0214 & 0.59 $\pm$ 0.15 $\pm$ 0.07 & -0.0042 \\  
2.55 & 0.322 & 0.25 & 201 & -0.07 $\pm$	0.09$\pm$ 0.03 & -0.25 $\pm$ 0.11 $\pm$ 0.06 & -0.0179 & 0.75 $\pm$ 0.12 $\pm$ 0.02 & -0.0020 \\  
2.55 & 0.338 & 0.24 & 238 & -0.20 $\pm$	0.07$\pm$ 0.02 & -0.18 $\pm$ 0.09 $\pm$ 0.01 & -0.0346 & 0.71 $\pm$ 0.09 $\pm$ 0.03 & 0.0142 \\  
2.73 & 0.337 & 0.24 & 271 & -0.26 $\pm$	0.06$\pm$ 0.03 & -0.28 $\pm$ 0.07 $\pm$ 0.02 & -0.0233 & 0.68 $\pm$ 0.07 $\pm$ 0.03 & 0.0278 \\  
2.69 & 0.336 & 0.24 & 309 & -0.26 $\pm$	0.03$\pm$ 0.02 & -0.16 $\pm$ 0.04 $\pm$ 0.01 & -0.0058 & 0.73 $\pm$ 0.05 $\pm$ 0.03 & 0.0385 \\  
2.68 & 0.338 & 0.24 & 341 & -0.04 $\pm$	0.03$\pm$ 0.00 & -0.08 $\pm$ 0.03 $\pm$ 0.01 & 0.0002 & 0.59 $\pm$ 0.04 $\pm$ 0.03 & 0.0395 \\  

2.81 & 0.356 & 0.44 & 20 & 0.09 $\pm$ 0.03$\pm$	0.01 & 0.04 $\pm$ 0.03 $\pm$ 0.01 & -0.0055 & 0.59 $\pm$ 0.04 $\pm$ 0.02 & 0.0123 \\  
2.74 & 0.346 & 0.46 & 51 & 0.20 $\pm$ 0.03$\pm$ 0.01 & 0.13 $\pm$ 0.03 $\pm$ 0.01 & 0.0049 & 0.66 $\pm$ 0.04 $\pm$ 0.02 & 0.0143 \\  
2.68 & 0.346 & 0.45 & 90 & 0.15 $\pm$ 0.05$\pm$ 0.02 & 0.18 $\pm$ 0.06 $\pm$ 0.02 & 0.0358 & 0.68 $\pm$ 0.06 $\pm$ 0.02 & 0.0135 \\  
2.24 & 0.349 & 0.47 & 126 & 0.07 $\pm$ 0.05$\pm$ 0.01 & 0.19 $\pm$ 0.06 $\pm$ 0.03 & 0.0636 & 0.53 $\pm$ 0.06 $\pm$ 0.02 & 0.0097 \\  
2.25 & 0.339 & 0.49 & 162 & 0.06 $\pm$ 0.06$\pm$ 0.01 & 0.25 $\pm$ 0.07 $\pm$ 0.03 & 0.0325 & 0.68 $\pm$ 0.08 $\pm$ 0.02 & 0.0061 \\  
2.23 & 0.339 & 0.49 & 202 & -0.13 $\pm$	0.06$\pm$ 0.02 & -0.15 $\pm$ 0.07 $\pm$ 0.02 & -0.0296 & 0.54 $\pm$ 0.08 $\pm$ 0.04 & 0.0022 \\  
2.43 & 0.347 & 0.47 & 232 & -0.06 $\pm$	0.06$\pm$ 0.01 & -0.21 $\pm$ 0.08 $\pm$ 0.03 & -0.0567 & 0.38 $\pm$ 0.09 $\pm$ 0.02 & 0.0034 \\  
2.72 & 0.344 & 0.47 & 271 & -0.22 $\pm$	0.05$\pm$ 0.02 & -0.26 $\pm$ 0.06 $\pm$ 0.02 & -0.0362 & 0.62 $\pm$ 0.06 $\pm$ 0.04 & 0.0118 \\  
2.77 & 0.347 & 0.45 & 309 & -0.21 $\pm$	0.03$\pm$ 0.01 & -0.18 $\pm$ 0.04 $\pm$ 0.01 & -0.0058 & 0.64 $\pm$ 0.04 $\pm$ 0.02 & 0.0137 \\  
2.77 & 0.355 & 0.45 & 340 & -0.15 $\pm$	0.02$\pm$ 0.01 & -0.07 $\pm$ 0.03 $\pm$ 0.01 & 0.0049 & 0.59 $\pm$ 0.03 $\pm$ 0.02 & 0.0129 \\  

2.59 & 0.352 & 1.30 & 20 & 0.02 $\pm$ 0.03$\pm$	0.00 & 0.04 $\pm$ 0.04 $\pm$ 0.01 & -0.0041 & 0.36 $\pm$ 0.04 $\pm$ 0.03 & -0.0051 \\  
2.59 & 0.348 & 1.09 & 52 & 0.13 $\pm$ 0.04$\pm$ 0.01 & 0.15 $\pm$ 0.05 $\pm$ 0.02 & 0.0088 & 0.39 $\pm$ 0.06 $\pm$ 0.02 & -0.0035 \\  
2.28 & 0.350 & 1.28 & 87 & 0.12 $\pm$ 0.06$\pm$ 0.02 & 0.09 $\pm$ 0.08 $\pm$ 0.06 & 0.0326 & 0.36 $\pm$ 0.08 $\pm$ 0.03 & -0.0008 \\  
2.27 & 0.353 & 1.09 & 129 & 0.08 $\pm$ 0.06$\pm$ 0.03 & 0.19 $\pm$ 0.08 $\pm$ 0.05 & 0.0375 & 0.30 $\pm$ 0.08 $\pm$ 0.04 & -0.0025 \\  
2.33 & 0.345 & 1.06 & 161 & -0.00 $\pm$	0.06$\pm$ 0.02 & 0.21 $\pm$ 0.08 $\pm$ 0.03 & 0.0188 & 0.52 $\pm$ 0.08 $\pm$ 0.02 & 0.0015 \\  
2.29 & 0.347 & 1.06 & 201 & -0.00 $\pm$	0.06$\pm$ 0.02 & -0.19 $\pm$ 0.08 $\pm$ 0.03 & -0.0184 & 0.52 $\pm$ 0.08 $\pm$ 0.02 & 0.0020 \\  
2.33 & 0.352 & 1.23 & 236 & -0.01 $\pm$ 0.06$\pm$ 0.02 & -0.25 $\pm$ 0.09 $\pm$ 0.06 & -0.0384 & 0.40 $\pm$ 0.09 $\pm$ 0.02 & 0.0003 \\  
2.25 & 0.347 & 1.23 & 269 & -0.04 $\pm$ 0.05$\pm$ 0.01 & -0.19 $\pm$ 0.06 $\pm$ 0.04 & -0.0356 & 0.30 $\pm$ 0.06 $\pm$ 0.02 & -0.0027 \\  
2.61 & 0.349 & 1.09 & 308 & -0.11 $\pm$	0.05$\pm$ 0.01 & -0.14 $\pm$ 0.06 $\pm$ 0.02 & -0.0086 & 0.48 $\pm$ 0.06 $\pm$ 0.03 & -0.0016 \\  
2.57 & 0.352 & 1.28 & 340 & -0.02 $\pm$ 0.03$\pm$ 0.00 & -0.07 $\pm$ 0.03 $\pm$ 0.01 & 0.0036 & 0.43 $\pm$ 0.04 $\pm$ 0.02 & -0.0034 \\  

3.21 & 0.414 & 0.27 & 19 & -0.24 $\pm$ 0.18$\pm$ 0.02 & 0.00 $\pm$ 0.00 $\pm$ 0.00 & 0.0000 & 0.90 $\pm$ 0.21 $\pm$ 0.02 & 0.0046 \\  
3.16 & 0.414 & 0.27 & 51 & 0.22 $\pm$ 0.21$\pm$ 0.09 & -0.04 $\pm$ 0.26 $\pm$ 0.08 & -0.0007 & 0.30 $\pm$ 0.29 $\pm$ 0.09 & 0.0070 \\  

3.36 & 0.443 & 0.51 & 17 & 0.07 $\pm$ 0.04$\pm$	0.01 & -0.02 $\pm$ 0.04 $\pm$ 0.01 & -0.0034 & 0.82 $\pm$ 0.05 $\pm$ 0.03 & 0.0398 \\  
3.41 & 0.444 & 0.49 & 52 & 0.21 $\pm$ 0.05$\pm$ 0.02 & 0.13 $\pm$ 0.06 $\pm$ 0.02 & 0.0066 & 0.77 $\pm$ 0.06 $\pm$ 0.04 & 0.0395 \\  
3.25 & 0.448 & 0.49 & 89 & 0.24 $\pm$ 0.07$\pm$ 0.03 & 0.27 $\pm$ 0.09 $\pm$ 0.04 & 0.0035 & 0.77 $\pm$ 0.10 $\pm$ 0.05 & 0.0397 \\  
2.84 & 0.430 & 0.53 & 124 & 0.12 $\pm$ 0.09$\pm$ 0.05 & 0.27 $\pm$ 0.11 $\pm$ 0.03 & 0.0548 & 0.64 $\pm$ 0.12 $\pm$ 0.07 & 0.0284 \\  
2.93 & 0.420 & 0.57 & 155 & -0.08 $\pm$ 0.24$\pm$ 0.08 & 0.07 $\pm$ 0.31 $\pm$ 0.08 & 0.0227 & 0.44 $\pm$ 0.30 $\pm$ 0.08 & 0.0040 \\  
2.94 & 0.421 & 0.58 & 207 & -0.44 $\pm$ 0.23$\pm$ 0.06 & -0.25 $\pm$ 0.28 $\pm$ 0.07 & -0.0273 & 0.50 $\pm$ 0.31 $\pm$ 0.07 & 0.0082 \\  
2.95 & 0.435 & 0.50 & 238 & -0.04 $\pm$ 0.11$\pm$ 0.05 & -0.18 $\pm$ 0.13 $\pm$ 0.05 & -0.0497 & 0.75 $\pm$ 0.14 $\pm$ 0.07 & 0.0309 \\  
3.33 & 0.440 & 0.48 & 271 & -0.08 $\pm$ 0.08$\pm$ 0.02 & -0.24 $\pm$ 0.10 $\pm$ 0.01 & -0.0322 & 0.82 $\pm$ 0.11 $\pm$ 0.04 & 0.0381 \\  
3.46 & 0.445 & 0.50 & 310 & -0.28 $\pm$ 0.05$\pm$ 0.02 & -0.23 $\pm$ 0.06 $\pm$ 0.01 & -0.0879 & 0.68 $\pm$ 0.06 $\pm$ 0.04 & 0.0364 \\  
3.35 & 0.443 & 0.52 & 342 & -0.08 $\pm$ 0.03$\pm$ 0.01 & -0.16 $\pm$ 0.04 $\pm$ 0.01 & -0.0010 & 0.77 $\pm$ 0.04 $\pm$ 0.03 & 0.0392 \\  

3.37 & 0.465 & 1.12 & 18 & 0.06 $\pm$ 0.03$\pm$	0.01 & 0.08 $\pm$ 0.04 $\pm$ 0.01 & -0.0064 & 0.51 $\pm$ 0.04 $\pm$ 0.02 & 0.0015 \\  
3.46 & 0.465 & 1.13 & 52 & 0.21 $\pm$ 0.05$\pm$ 0.02 & 0.16 $\pm$ 0.07 $\pm$ 0.03 & 0.0095 & 0.60 $\pm$ 0.07 $\pm$ 0.03 & 0.0071 \\  
3.18 & 0.459 & 1.27 & 89 & -0.00 $\pm$ 0.07$\pm$ 0.02 & 0.15 $\pm$ 0.10 $\pm$ 0.05 & 0.0507 & 0.55 $\pm$ 0.10 $\pm$ 0.04 & 0.0078 \\  
3.13 & 0.456 & 1.15 & 126 & 0.11 $\pm$ 0.06$\pm$ 0.01 & 0.04 $\pm$ 0.08 $\pm$ 0.05 & 0.0558 & 0.55 $\pm$ 0.08 $\pm$ 0.02 & 0.0007 \\  
3.30 & 0.434 & 1.18 & 159 & 0.07 $\pm$ 0.11$\pm$ 0.09 & 0.12 $\pm$ 0.15 $\pm$ 0.04 & 0.0241 & 0.43 $\pm$ 0.15 $\pm$ 0.07 & -0.0116 \\  
3.11 & 0.430 & 1.09 & 203 & -0.10 $\pm$ 0.13$\pm$ 0.03 & 0.00 $\pm$ 0.16 $\pm$ 0.06 & -0.0219 & 0.78 $\pm$ 0.16 $\pm$ 0.06 & -0.0021 \\  
3.16 & 0.455 & 1.25 & 236 & -0.06 $\pm$ 0.07$\pm$ 0.01 & -0.30 $\pm$ 0.10 $\pm$ 0.06 & -0.0644 & 0.46 $\pm$ 0.09 $\pm$ 0.03 & -0.0026 \\  
3.12 & 0.454 & 1.35 & 269 & -0.02 $\pm$ 0.06$\pm$ 0.02 & -0.21 $\pm$ 0.09 $\pm$ 0.05 & -0.0541 & 0.56 $\pm$ 0.08 $\pm$ 0.03 & 0.0083 \\  
3.49 & 0.467 & 1.10 & 310 & -0.31 $\pm$ 0.06$\pm$ 0.04 & -0.25 $\pm$ 0.07 $\pm$ 0.03 & -0.0119 & 0.77 $\pm$ 0.08 $\pm$ 0.05 & 0.0115 \\  
3.36 & 0.464 & 1.15 & 342 & -0.07 $\pm$ 0.03$\pm$ 0.01 & -0.08 $\pm$ 0.04 $\pm$ 0.01 & 0.0063 & 0.51 $\pm$ 0.04 $\pm$ 0.02 & 0.0014 \\
\hline  
\multicolumn{9}{c}{}
\end{longtable*}

\end{document}